\documentclass{article}[jheppub]
\usepackage{bm}
\usepackage{amsmath}
\usepackage{amsfonts}
\usepackage{jheppub}
\usepackage{url}
\usepackage[utf8x]{inputenc}
\newcommand{\nn}{\nonumber}
\newcommand{\da}{{\dot{\alpha}}}
\newcommand{\db}{{\dot{\beta}}}
\newcommand{\beq}{\begin{equation}}
\newcommand{\eeq}{\end{equation}}
\newcommand{\beqn}{\begin{eqnarray}}
\newcommand{\eeqn}{\end{eqnarray}}
\newcommand\one{\bm 1}
\newcommand\two{\bm 2}

\newdimen\eqskip
\newdimen\txtskip
\eqskip=20pt\def\bP{\bm{P}}

\def\boldsigma{{\bm{\sigma}}}

\def\slsh{\rlap{$\;\!\!\not$}}     
\def\cA{{\cal A}}
\def\qb{\bar{q}}
\def\Qb{{\bar{Q}}}

\def\mt{m}
\def\spa#1.#2{\langle#1#2\rangle}
\def\s#1.#2{s_{#1#2}}
\def\spb#1.#2{\left[#1#2\right]}
\def\spab#1.#2.#3{\left\langle#1|#2|#3\right]}
\def\spba#1.#2.#3{\left[#1|#2|#3\right\rangle}
\def\spaa#1.#2.#3.#4{\left\langle#1|#2|#3|#4\right\rangle}
\def\spaba#1.#2.#3.#4{\left\langle#1|#2|#3|#4\right\rangle}
\def\spbb#1.#2.#3.#4{\left[#1|#2|#3|#4\right]}
\def\spbab#1.#2.#3.#4{\left[#1|#2|#3|#4\right]}
\def\spaabb#1.#2.#3.#4.#5{\left\langle#1|#2|#3|#4|#5\right]}
\def\spabab#1.#2.#3.#4.#5{\left\langle#1|#2|#3|#4|#5\right]}
\def\spababa#1.#2.#3.#4.#5.#6{\left\langle#1|#2|#3|#4|#5|#6\right\rangle}
\def\spbaba#1.#2.#3.#4.#5{\left[#1|#2|#3|#4|#5\right\rangle}
\def\spbabab#1.#2.#3.#4.#5.#6{\left[#1|#2|#3|#4|#5|#6\right]}
\def\Athree{A_{3}}
\def\Afour{A_{4}}
\def\Afive{A_{5}}
\def\Asix{A_{6}}
\def\bra#1{\langle #1|}
\def\ket#1{|#1 \rangle}

\def\braket#1{\langle #1 \rangle}

\def\Tr{\operatorname*{Tr}}

\author[a]{John M. Campbell,}
\emailAdd{johnmc@fnal.gov}

\author[b]{R. Keith Ellis,}
\emailAdd{keith.ellis@durham.ac.uk}

\affiliation[a]{Fermilab, PO Box 500, Batavia IL 60510-5011, USA}
\affiliation[b]{Institute for Particle Physics Phenomenology, Durham University, Durham, DH1 3LE, UK}

\preprint{FERMILAB-PUB-23-459-T, IPPP/23/47}
\title{Top tree amplitudes for higher order calculations}

\abstract{We present compact analytic results for tree-level amplitudes
  containing a $t \bar{t}$ pair accompanied by up to four massless partons,
  $t \bar{t}gg$,
  $t \bar{t}ggg$,  
  $t \bar{t}gggg$,
  $t \bar{t}q\bar{q}$,
  $t \bar{t}q\bar{q}g$,
  $t \bar{t}q\bar{q}gg$
  and $t\bar{t}q\bar{q}q^\prime \bar{q}^\prime$.
The results, obtained using BCFW on-shell recursion, are based both on previous
published results and on the new calculations performed in this paper.
These amplitudes are sufficient to calculate the production of a $t\bar{t}$
pair and zero, one, or two light parton jets, with the option to include the tree-level
decays $t \to b \nu e^+$ and $\bar{t} \to \bar{b}  e^- \bar{\nu}$ efficiently.
Our results are part of the NNLO corrections to $t \bar{t}$ production
including the decay correlations for on-shell top quarks.
}
\begin{document}
\maketitle

\section{Introduction}

The calculation of tree-graph amplitudes for massless particles is
radically simplified by the exploitation of spinor
methods~\cite{DeCausmaecker:1981jtq,Berends:1981uq,Kleiss:1985yh,Xu:1986xb,Mangano:1990by,Dixon:1996wi}.
However it is less widely appreciated that even in the presence of masses, spinor techniques can lead to
compact expressions for tree-graph amplitudes.

This has recently been demonstrated for amplitudes containing a
$t \bar{t}$-pair and $(n-2)$ gluons~\cite{Ochirov:2018uyq} where beautiful results
have been obtained for all $n\geq 4$ for two particular helicity combinations.
The two cases comprise the amplitude with all gluons with identical helicity,
and the amplitude with one opposite-helicity gluon color-adjacent to one
of the quarks. In a second paper results have been provided for amplitudes involving
two massive quark-antiquark pairs and an arbitrary number of identical helicity gluons~\cite{Lazopoulos:2021mna}.
These relations are proved using
Britto-Cachazo-Feng-Witten (BCFW) recursion~\cite{Britto:2004ap,Britto:2005fq}.
The other required amplitudes for $t\bar{t}ggg$ and some additional amplitudes for $t\bar{t}gggg$
can be obtained using Bern-Carrasco-Johansson
(BCJ) relations~\cite{Bern:2008qj,Bern:2019prr}. 

Automatic procedures to calculate tree (and one-loop) graphs are
available~\cite{Maltoni:2002qb,Gleisberg:2008fv,Actis:2016mpe,Buccioni:2019sur}. 
Nevertheless it seemed opportune to apply the theoretical results described above for the concrete
case of $t \bar{t}$+jets, supplementing the results given in ref.~\cite{Ochirov:2018uyq} with
explicit expressions for $t \bar{t}q\bar{q}$ and the five- and six-parton amplitudes,
$t \bar{t}q\bar{q}g$, $t \bar{t}q\bar{q}gg$
and $t\bar{t}q\bar{q}q^\prime \bar{q}^\prime$.
This is particularly useful
because amplitude expressions allow the inclusion of the tree-level decay of the top quark~\cite{Kleiss:1988xr}.

The BCFW technique allows the
iterative construction of higher point amplitudes
starting from three-point amplitudes evaluated at complex momenta for both massless and
massive~\cite{Badger:2005jv,Badger:2005zh,Ozeren:2006ft,Huang:2012gs,Ochirov:2018uyq,Lazopoulos:2021mna} amplitudes.
Since the amplitudes are constructed using on-shell results, they are free of the redundant gauge degrees of
freedom which are present in a normal quantum field theory calculation. Our results are presented using the formalism of
Arkani-Hamed, Huang and Huang (AHH)~\cite{Arkani-Hamed:2017jhn}, who have extended the spinor-helicity formalism
for massless particles.  In their formalism the covariance properties of the amplitudes under little group rotations
are made manifest by the addition of a $SU(2)$ little group index.  The resulting Spin-spinors carry an $SU(2)$ index ($I$)
in addition to the Lorentz group SL$(2,\mathbb{C})$ indices, $(\alpha, \da)$.
From the point of view of the amplitude program, which asserts that amplitudes calculated recursively using on-shell
ingredients are more fundamental than their quantum field theory analogues, the extension to massive particles is
an important and necessary step.

Our aim in this paper is more prosaic; we want to investigate the
benefits for top quark physics of analytic tree-level amplitudes calculated
using BCFW techniques.  The work of BCFW and BCJ has shown that full
amplitudes can be calculated from a limited number of ingredients.
At low perturbative order
analytic results can be computationally more efficient (see for
example ref.~\cite{Lazopoulos:2021mna}) than results based on off-shell
Berends-Giele recursion~\cite{Berends:1987me}, which is often the
automatic procedure of choice for the calculation of tree graphs.  These
amplitudes will be incorporated in MCFM~\cite{MCFM}, exploiting the possibility of
including the tree-level decay of the top quark with decay
correlations at essentially zero cost~\cite{Kleiss:1988xr}.  Finally,
we note that compact low order tree-graph results can be useful
ingredients for loop calculations via unitarity, see for example
refs.~\cite{Bern:1995db,Budge:2020oyl}.

\subsection{Plan of the paper}
Section~\ref{spin-spinor} gives an introduction to the massless and massive spinor formalism,
following the method of AHH for the massive case. 
Section~\ref{sec:3parton}  addresses the definition of color-ordered primitives
and the BCJ relations between them.
The basic 3-parton building blocks for the BCFW recursion are also presented here.
Section~\ref{sec:4parton} illustrates the use of BCFW recursion for the calculation of
$A_4(\one,3_g^+,4_g^+,\two_{\Qb})$ and $A_4(\one,3_g^+,4_g^-,\two_{\Qb})$, and 
presents a full set of results for the 4-parton amplitudes.
Section \ref{BCFWexample} uses the results of the previous two sections to
calculate the 5-parton amplitude $A_5(\one,\two_{\Qb},3_q^-,4_{\qb}^+,5_g^+)$ to further
illustrate the application of BCFW techniques.
Sections~\ref{sec:5parton} and \ref{sec:6parton}
present the results for all the 5-parton and 6-parton amplitudes.
Both sections contain a 
description of the color decomposition of the amplitude, the form of the squared amplitude after summing over colors,
a complete set of results for the subamplitudes in terms of massless and massive spinors for all helicity 
combinations of the massless particles and a description of the BCJ relations between the sub-amplitudes, if
applicable.
In section~\ref{classic} we give an explicit representation of the Spin-spinors that is closely connected
to the Kleiss-Stirling method~\cite{Kleiss:1986qc} and review the implementation of tree-level top-quark decay.
In section~\ref{conclusions} we draw some conclusions.
Appendix~\ref{SpinorReview} derives the results needed for the calculation in the Spin-spinor formalism
and appendix~\ref{sec:Melia} gives an alternative color decomposition for $Q\bar Q q\bar q g g$ amplitudes.
\section{Spin-spinor formalism}
\label{spin-spinor}

In this section we will introduce the essence of the Spin-spinor formalism of
Arkani-Hamed, Huang and Huang (AHH)~\cite{Arkani-Hamed:2017jhn}.
A more detailed exposition of this formalism is given in
refs.~\cite{Ochirov:2018uyq,Christensen:2018zcq,Christensen:2019mch,Lazopoulos:2021mna}.
Appendix~\ref{SpinorReview} gives a
detailed derivation of the results that we will need for our calculation.

\subsection{Massless partons}
We consider a spinor state $|p\rangle_\beta$ which is a solution to
the massless Weyl equation $p^{\da \beta}\, |p\rangle_\beta=0$ where
$p^{\da \beta} = p^\mu\sigma_{\mu}^{\da \beta}$ is derived from
the four-momentum $p^\mu$ of the particle.
The indices $\da$ and $\beta$ are the SL$(2,\mathbb{C})$ Lorentz group indices
that are normally superfluous in the angle and square bracket formalism,
but we sometimes find it convenient to retain them here.
Since the particle is massless, $p^{\da \beta}$
is a rank one matrix and is expressible as
\beq \label{bispinor1}
p^{\da \beta}=|p]^\da \langle p|^{\beta},\;\;
\eeq
which is clearly invariant under the little group rescaling
\beq
\langle p|^{\beta} \to t\, \langle p|^{\beta}\,,\quad\quad
|p]^\da  \to \frac{1}{t} |p]^\da \, .
\eeq
The spinors are in the fundamental representation of the group SL$(2,\mathbb{C})$ and
the particular components are indicated by indices $\da,\beta$.
For a massless particle we can go to a frame in which
the momentum is directed along the $z$ direction $p=(E,0,0,E)$. The little group is thus the
group of rotations in the $x,y$ plane, namely $SO(2)\equiv U(1)$.The great utility of the spinor formalism
derives from the fact that amplitudes are directly functions of spinor helicity variables.

\subsection{Massive partons}
The extension of this formalism to massive particles notes that the little group in this case can be
deduced in the rest frame of the particle. In the rest frame the little group is the set of rotations in
$3$ dimensions, namely $SO(3)\equiv SU(2)$. Amplitudes can now be expressed in terms of Spin-spinors
which transform as a direct product of the $SU(2)$ spin group tensor and the SL$(2,\mathbb{C})$ Lorentz group.
These Spin-spinors are denoted by
$|\bm{p}^I\rangle^\beta$ and $|\bm{p}^I]^\da$.
In the angle and square bracket notation
combination rules for the dotted and undotted SL$(2,\mathbb{C})$ indices are mandated by the
angle and square brackets, so they can be dropped in spinor products.
Amplitudes involving massive particles, with momenta $p_1$ and $p_2$
are naturally expressed in terms of spinor products such as $\spa{\bm{1}^I}.{\bm{2}_J}$, $\spb{\bm{1}^I}.{\bm{2}_J}$
since these spinor products reflect the little group transformation properties of the amplitudes themselves.

The Spin-spinors so defined satisfy a number of relations that are necessary to perform the BCFW recursion.
These identities are,
\beq
   \begin{aligned}
   \ket{\bm{p}^I}_{\;\!\!\alpha}\;\![\bm{p}_I|_{\db}& = +p_{\alpha\db} \\
   |\bm{p}^I]^{\da}\:\!\bra{\bm{p}_I}^{\beta}& =-p^{\da\beta} \\
   \ket{\bm{p}^I}_{\;\!\!\alpha} \bra{\bm{p}_I}^{\beta} & =+m\:\!\delta_\alpha^\beta \\
   |\bm{p}^I]^{\da}\;\:\!\![\bm{p}_I|_{\db} & = -m\:\!\delta^{\da}_{\db}
   \end{aligned} \qquad \quad
   \begin{aligned}
   \ket{\bm{p}_I}_{\;\!\!\alpha}\;\![\bm{p}^I|_{\db}& = -p_{\alpha\db} \\
   |\bm{p}_I]^{\da}\:\!\bra{\bm{p}^I}^{\beta}& =+p^{\da\beta} \\
   \ket{\bm{p}_I}_{\;\!\!\alpha} \bra{\bm{p}^I}^{\beta} & =-m\:\!\delta_\alpha^\beta \\
   |\bm{p}_I]^{\da}\;\:\!\![\bm{p}^I|_{\db} & = +m\:\!\delta^{\da}_{\db}
   \end{aligned} \qquad \quad
   \begin{aligned}
   p^{\da\beta} \ket{\bm{p}^I}_{\;\!\!\beta} & = -m |\bm{p}^I]^{\da} \\
   p_{\alpha\db} |\bm{p}^I]^{\db} & = -m \ket{\bm{p}^I}_{\;\!\!\alpha} \\
   \bra{\bm{p}^I}^{\alpha} p_{\alpha\db} & = +m [\bm{p}^I|_{\db} \\
   [\bm{p}^I|_{\da} p^{\da\beta} & = +m \bra{\bm{p}^I}^{\beta} 
   \end{aligned} \qquad \quad
\label{identities}
\eeq
The derivation of these relations is presented in Appendix~\ref{SpinorReview}. The SU(2) indices
are raised and lowered using the two-dimensional totally antisymmetric tensor $\epsilon^{IJ}$.

We adopt the convention that,\footnote{This has been discussed at length in
refs.~\cite{Ochirov:2018uyq,Lazopoulos:2021mna} where explicit spinors obeying these relations can be found.}
\beq
|-p\rangle = -|p\rangle\, ,\quad\quad
|-p] = |p]\, .
\label{eq:signflip}
\eeq
In this paper we calculate amplitudes will all momenta outgoing.
With this convention we have that
\beq
\gamma_\mu p^\mu +m =
\left(\begin{matrix}
  |-p_I]^\da [p^I|_\db & \; |-p_I]^\da \langle p^I|^\beta\\
  |-p_I\rangle_\alpha [p^I|_\db & \; |-p_I\rangle_\alpha \langle p^I|^\beta
\end{matrix}\right) \,,
\eeq
which shows that sewing together amplitudes in the BCFW method,
where one line must perforce have a negative momentum, reproduces the numerator of the massive fermion propagator.

Armed with the basic results for the 3-point vertices involving massive and massless particles
we can construct higher point tree-level amplitudes using BCFW recursion. In addition we can 
illustrate the BCJ relations between the analytical results that we calculate.

\section{Color and counting of primitives}
\label{sec:3parton}
It is well known that for the case of pure gluon scattering, the color-trace decomposition into
color-ordered primitives is~\cite{Mangano:1990by}, 
\beq
    {\cal A}_n(1_g,2_g,\ldots n_g)= \!\!
      \sum_{\sigma \in S_{n-1}(\{2,\dots,n\})} \!\!\!
      \Tr\!\big( t^{C_1} t^{C_{\sigma(2)}} \dots t^{C_{\sigma(n)}} \big)
      \bar{A}_n(1,\sigma(2),\dots,\sigma(n)) \,,
      \label{color-trace}
      \eeq
where the sum is over $(n-1)!$ primitives, since the cyclicity of the trace allows one to fix the first argument.   
This decomposition has the disadvantage that the color coefficients are not all linearly independent. Consequently
the color-ordered sub-amplitudes are not the minimal set. Indeed for the pure gluon case, the color sub-amplitudes
defined in Eq.~(\ref{color-trace}) are related by the Kleiss-Kuijf~(KK)
relations~\cite{Kleiss:1988ne}, which reduce
the number of independent primitives to $(n-2)!$. It was subsequently observed by
Del Duca, Dixon and Maltoni~\cite{DelDuca:1999rs} that the color decomposition
\beq
    {\cA}_{n}(1_g,2_g,\ldots n_g) = \!\!
      \sum_{\sigma \in S_{n-2}(\{3,\dots,n\})} \!\!\!
      \big( T^{C_{\sigma(3)}} \dots T^{C_{\sigma(n)}}
      \big)_{C_1 C_2}
      A_n(1,2,\sigma(3),\dots,\sigma(n)) \,,
\eeq
where $T$ are $SU(3)$ matrices in the {\bf adjoint} representation,
contains only linearly independent color structures and automatically reduces the number of independent
color-subamplitudes to $(n-2)!$.

In this paper we will be dealing with amplitudes with one or more quark lines. For the case of
one quark line, the trace representation
\beq
    {\cA}_{n}(\one_Q,\two_{\Qb},3_g \ldots n_g) = \!\! = \!\!
      \sum_{\sigma \in S_{n-2}(\{3,\dots,n\})} \!\!\!
      \big( T^{C_{\sigma(3)}} \dots T^{C_{\sigma(n)}}
      \big)_{x_1 x_2}
      A_n(\one_Q,\two_{\Qb},\sigma(3),\dots,\sigma(n)) \,,
\eeq
is free of further relations of the KK type.
In addition to pure gluon processes, color decompositions for processes involving one quark line have
been considered in ref.~\cite{DelDuca:1999rs}.
In the case where we consider more than one quark line the equivalent 
color decompositions have been given in refs.~\cite{Ellis:2011cr,Johansson:2015oia,Melia:2013bta,Melia:2013epa,Melia:2015ika}.
Table~\ref{KKtable} presents the number of 
primitive color sub-amplitudes after application of all KK-type relations, for the top pair
production amplitudes that we consider in this paper.

\begin{table}
\begin{center}
\begin{tabular}{|c|c|c|c|}
  \hline
  $n$             & $k$                 & $n_g$          & $n_P$\\
  $\#$ of partons & $\#$ of quark pairs & $\#$ of gluons & $\#$ of primitives\\
  \hline
  4 & 1 & 2 & 2\\
  5 & 1 & 3 & 6\\
  6 & 1 & 4 & 24\\
  4 & 2 & 0 & 1\\
  5 & 2 & 1 & 3\\
  6 & 2 & 2 & 12\\
  6 & 3 & 0 & 4\\
  \hline
\end{tabular}
\caption{Number of independent primitive amplitudes $n_P=(n-2)!/k!$, after imposition of KK-type
constraints. $n$ is the total number of partons, and $k$ is the number of
distinguishable quark pairs.} 
\label{KKtable}
\end{center}
\end{table}

\subsection{Relations between the kinematic part of tree amplitudes}
Bern, Carrasco and Johansson (BCJ) have discovered additional relations
obeyed by amplitudes involving external gluons.  The quark-gluon BCJ
relations, for one or more massive quark lines, are given by the general
formula~\cite{Bern:2008qj,Johansson:2015oia,Bern:2019prr},
\beq
\sum_{i=2}^{n-1} \Big( \sum_{j=2}^{i} s_{jn} - m _j^2 \Big) A_n(\one_Q, \two_{\Qb},
\dots i,n_g,i+1, \dots ,n-1) = 0 \,.
\label{fBCJmassive}
\eeq
where particle $n$ is strictly a gluon, while the remaining $(n-1)$ particles can be of any type: quark/antiquark/gluon.
Table \ref{AfterBCJ} gives results for the number of primitives after imposition of BCJ relations.
\begin{table}
\begin{center}
\begin{tabular}{|c|c|c|c|}
  \hline
  $n$             & $k$                 & $n_g$          & $n_P$\\
  $\#$ of partons & $\#$ of quark pairs & $\#$ of gluons & $\#$ of primitives\\
  \hline
  4 & 1 & 2 & 1 \\
  5 & 1 & 3 & 2 \\
  6 & 1 & 4 & 6 \\
  4 & 2 & 0 & 1 \\
  5 & 2 & 1 & 2 \\
  6 & 2 & 2 & 6 \\
  6 & 3 & 0 & 4 \\
  \hline
\end{tabular}
\caption{Number of independent primitive amplitudes $n_P$ after imposition of KK
  and BCJ constraints. $n$ is the total number of partons, and $k$ is the number of
  distinguishable quark pairs. For $k=1$ this is $n_P=(n-3)!$, while for $k \geq 2$
  $n_P= (n-3)! \, (2k-2)/k!$. Adapted from ref.~\cite{Johansson:2015oia}.}
\label{AfterBCJ}
\end{center}
\end{table}

\subsection{Three parton amplitudes}
In this section we provide the basic building blocks for 3-point amplitudes.  These are necessary in order
to start the BCFW recursion. For the $ggg$ process we have,
\beq
\cA_3(1_g^-,2_g^-,3_g^+)= g \tilde{f}^{{C_1} {C_2} {C_3} }\; \frac{\spa1.2^3}{\spa2.3\,\spa3.1} =
-i g \big(\Tr\{t^{C_1} t^{C_2} t^{C_3}\}-\Tr\{t^{C_1} t^{C_3} t^{C_2}\})\; \frac{\spa1.2^3}{\spa2.3\,\spa3.1}\, .
\eeq
For the $Q g \Qb$ process we have,
\beq
\cA_3(\one_{x_1},3_C,\two_{x_2}) = g \, (t^C)_{x_1 x_2} \, A_3(\one_Q,3,\two_{\Qb})
\eeq
where,
\beqn
 -i  A_3(\one_Q,3^+,\two_{\Qb}) &=&- 
  \frac{\big( [\one 3]\braket{q \two}+\braket{\one q}[3 \two] \big)}{\braket{q 3}} \nn \\
&=&- 
    \frac{\big( \spab{\one}.{\one}.3 \braket{q \two} -\braket{\one q}\spba{3}.\two.\two \big)}{m \braket{q 3}} 
   =- \frac{\braket{\one \two} \langle q|\one|3]}{m\braket{q 3}}\, ,
   \label{eq:qgpa}\\
-i A_3(\one_Q,3^-,\two_{\Qb}) &=& - 
      \frac{\big( \braket{\one 3}[q \two] + [\one q] \braket{3 \two} \big)}{[3\;\!q]}\nn \\
    &=& - \frac{\big( [\one|\one| 3\rangle [q \two] - [\one q]\spab{3}.{\two}.{\two} \big)}{m [3 q]}
    = - \frac{[\one \two] \bra{3}\one|q]}{m[3 q]} \, .
    \label{eq:qgma}
\eeqn
These formula require complex on-shell kinematics and $q$ is an arbitrary light-like momentum.
The first form in Eqs.~(\ref{eq:qgpa}) and~(\ref{eq:qgma}) is valid for both massless and massive quarks.
The application to massless quarks however, requires picking out the term with the right little group scaling, dependent
on the desired helicities of the massless quarks. 
The last form in Eqs.~(\ref{eq:qgpa}) and~(\ref{eq:qgma}),
obtained using the equation of motion from Eq.~(\ref{identities}), is the most compact expression
for massive fermions~\cite{Ochirov:2018uyq}.
The SU(3) color matrices in the fundamental representation are normalized such that,
\beq
    {\rm Tr} \{t^C t^D\}=\delta^{CD},\;\;\; [t^A, t^B]=i \tilde{f}^{ABC} t^C,~{\rm where}~\tilde{f}^{ABC}=\sqrt{2} f^{ABC}\, .
\eeq

Much of the concision of the expressions which we present in the following is due to the notation which we have chosen.
We employ a notation in which slashed momenta can denote either
$\boldsigma.p$ or $\bar{\boldsigma}.p$ depending on the spinor string in which it appears. Moreover we can drop the slash
inside the spinor sandwiches. 
Momenta $p_j$ are mostly represented by the symbol $j$ alone.
Thus,
\beqn
\spab{i}.{\slsh{p}_{j}}.{l} &\equiv&\spab{i}.{p_{j}}.{l}
\equiv \spab{i}.{\bar{\boldsigma}\cdot p_{j}}.{l}
\equiv \langle i|^\alpha\, (p_j)_{\alpha \db} |l]^\db
\equiv \spab{i}.{j}.{l}\, ,\nn\\
\spab{i}.{\slsh{p}_{jk}}.{l} &\equiv&\spab{i}.{p_{jk}}.{l} \equiv \spab{i}.{\bar{\boldsigma}\cdot p_{jk}}.{l}\equiv \spab{i}.{(j+k)}.{l}~{\rm where}~ p_{jk}=p_{j}+p_{k}\, .
\eeqn
More complicated spinor strings are defined in a similar way.
In these expressions $p_i,p_l$ are light-like momenta, whereas $p_j,p_k$ are not necessarily light-like.
In the angle and square bracket notation, the SL$(2,\mathbb{C})$ indices $\alpha,\db$ are superfluous; they are shown above for completeness only.
The momenta of massive quarks are always denoted
in boldface. The covariance properties of the amplitudes under little group transformations
are manifested by the SU(2) indices $I$ and $J$ of the external massive particles.
These external indices are never summed. In practice, these indices will not be displayed,
and their presence in the formula should be understood. In practice it is useful to consider the SU(2) indices of the
outgoing massive quarks to be in the raised position, (transforming as an SU(2) doublet),
whereas the index of the outgoing antiquark is in the lower position, (transforming as an SU(2) anti-doublet).
\beqn
\langle \one| &\equiv \langle \one^I|\, \quad\quad [\one| &\equiv [\one^I| \nn \\
|\two\rangle &\equiv |\two_I\rangle\, \quad\quad |\two] &\equiv |\two_I]
\eeqn
As such the indices are in the right positions to apply the identities given in Eq.~(\ref{identities}) which involve
sums over SU(2) indices with one index up and the other down.
The SU(2) indices $I$ and $J$ run over the values $1$ and $2$. We follow the Einstein notation that repeated indices are summed.

\section{Four parton amplitudes}
\label{sec:4parton}
\subsection{One quark pair, two gluon amplitudes}
\subsubsection{Color algebra}
The color decomposition for a tree-level amplitude with
$Q\bar{Q}$ + $(n-2)$-gluons is,
\beq \label{QQn-2g}
  \cA_n(1_{x_1},3_{C_3},\ldots,n_{C_n},2_{x_2})
 \ =\  g^{n-2} \sum_{\sigma\in S_{n-2}}
 (t^{C_{\sigma(3)}}\ldots t^{C_{\sigma(n)}})_{x_1 x_2}
  A_n(1_{Q},\sigma(3),\ldots,\sigma(n),2_{\Qb})\ ,
\eeq
where $S_{n-2}$ is the permutation group on $n-2$ elements,
and $A_n$ are the tree-level partial amplitudes.

For the case at hand, $n=4$, the square of the amplitude summed over colors of quarks and gluons is,
\beqn
  \sum_{C_3,C_4,x_1,x_2}\, |\cA_4(1_{x_1},3_{C_3},4_{C_4},2_{x_2})|^2&=&
g^4\, V \Big\{ N \Big[\big|\Afour(1_{Q},3_g,4_g,2_{\Qb})\big|^2+\big|\Afour(1_{Q},4_g,3_g,2_{\Qb})\big|^2\Big] \nn\\
&-& \frac{1}{N} \big|\Afour(1_{Q},3_g,2_{\Qb},4_g)\big|^2\Big\} \, ,
\eeqn
where $V=N^2-1$ and the expression for the subleading color amplitude is given by a sum of the two leading color amplitudes,
\beq \label{subleading}
\Afour(1_{Q},3_g,2_{\Qb},4_g)=\Afour(1_{Q},3_g,4_g,2_{\Qb})+\Afour(1_{Q},4_g,3_g,,2_{\Qb})
\eeq

\subsubsection{Results for one quark pair + two gluon amplitudes}
\begin{figure}[t]
\centering
\includegraphics[width=0.25\textwidth,angle=270]{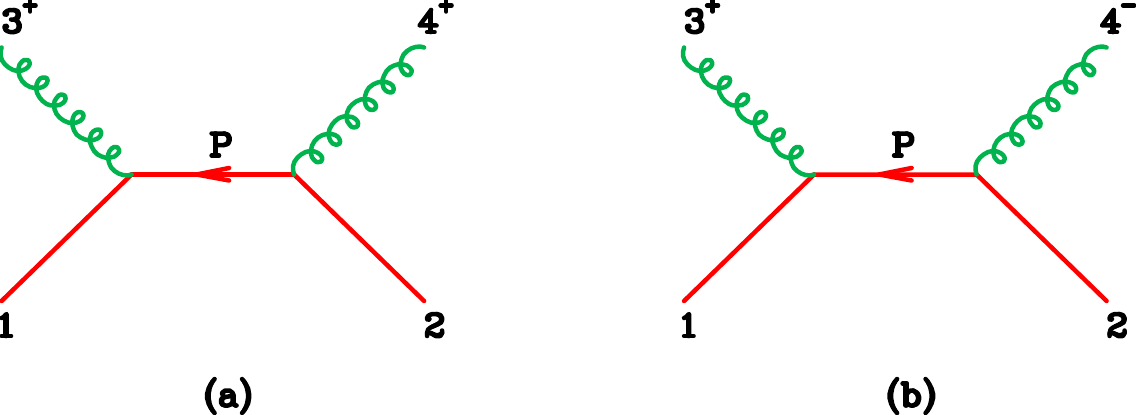}
\caption{Diagrams for BCFW recursion for $A_4(\one,3_g,4_g,\two_{\Qb})$}
\label{fig:QQbgg}
\end{figure}
We can now calculate the one quark pair + two gluon amplitudes using the 3-parton amplitudes given in
Eqs.~(\ref{eq:qgpa}),~(\ref{eq:qgma})
by BCFW recursion. As usual for the choice of the BCFW shift momentum,
\beq
|\hat{j}]=|j]-z |i],\quad\quad
|\hat{i}>=|i\rangle +z |j\rangle,
\eeq
the helicities of the marked particles can take the values,
$(h_i, h_j) = (+,-) , (+,+) , (-,-)$ but not $(h_i, h_j) = (-,+)$, in order that the amplitude as a function
of $z$ vanishes as $z \to \infty$~\cite{Britto:2005fq}.
The four-parton amplitudes are then obtained from,
\beq
A_4(\one_Q,3_g,4_g,\two_{\Qb}) = A_3(\one,\hat{3}_g, -\bP) \, \frac{i}{\spab3.\one.3} \, A_3(\bP,\hat{4}_g,\two_{\Qb}) \,.
\eeq

The relevant diagram for the calculation of $A_4(\one_Q,3_g^+,4_g^+,\two_{\Qb})$ is shown in Fig.~\ref{fig:QQbgg}(a).
Taking $i=3$ and $j=4$ we have,
\beq
|\hat{3}\rangle = |3\rangle +z |4\rangle\, ,\quad\quad |\hat{4}]=|4]-z |3]\, ,\quad\quad P=p_1+\hat{p}_3\, .
\eeq
The onshell condition on the intermediate quark line $P^2-m^2=\spba{3}.{\one}.{\hat{3}}=0$ determines that,
\beq
z=-\frac{\spba{3}.\one.{3}}{\spba{3}.\one.{4}}\, , \quad\quad
\spa{x}.{\hat{3}} =\frac{\spba3.\one.x}{\spba3.\one.4} \spa4.3\, , \quad\quad
\spb{x}.{\hat{4}}=\frac{\spbab{x}.{(3+4)}.{\one}.3}{\spba3.\one.4}=-\frac{\spbab{x}.{(\one+\two)}.{\one}.3}{\spba3.\one.4}\, .
\eeq
From Eq.~(\ref{eq:qgpa}) using Eq.~(\ref{eq:signflip}) we have that, 
\beqn
    A_3(\one,\hat{3}^+_g, -\bP) &=& -i \frac{\langle{\one|-\bP}\rangle \spab{q}.\one.{3}}{m \spa{q}.{\hat{3}}} 
               =  -i \frac{\spa\one.\bP \spab4.\one.3}{m \spa3.4}\, , \\
    A_3(\bP,\hat{4}^+_g,\two_{\Qb}) &=& i \frac{\spa{\bP}.{\two} \spab{q}.{\two}.{\hat{4}}} {m \spa{q}.{4}} 
= -i \frac{\spa{\bP}.{\two} \spabab{3}.{\two}.{(\one+\two)}.{\one}.3} {m \spa{3}.4 \spba3.\one.4}
= -i \frac{\spa{\bP}.{\two} \spb3.4 m} {\spba3.\one.4}\, .
\eeqn
For clarity, when a momentum has a negative sign we introduce an additional vertical line in the spinor products,
e.g.~$\spa{i}.j=\spa{i|}.j$.
Therefore the answer by BCFW is,
\beq
-i A_4(\one_Q,3_g^+,4_g^+,\two_{\Qb}) = -\frac{\spb3.4}{\spa3.4} \frac{1}{\spab3.\one.3} \spa\one.\bP \spa{\bP}.{\two}
= m \frac{\spb3.4}{\spa3.4} \frac{1}{\spab3.\one.3} \spa\one.\two \,,
\eeq
where we have used the relation, c.f. Eq.~(\ref{identities}),
\beq \label{identity1}
\spa\one.\bP \spa{\bP}.{\two} =\langle{\one^J|\bP_I}\rangle \spa{\bP^I}.{\two_K} = -m \spa{\one^J}.{\two_K} \, .
\eeq

For the calculation of $A_4(\one_Q,3_g^+,4_g^-,\two_{\Qb})$, shown in Fig.~\ref{fig:QQbgg}(b), we use the same shift
and require the amplitude,
\beqn
A_3(\bP,\hat{4}^-_g,\two_{\Qb}) &=&
-i \frac{(\spa{\bP}.{4}\spb{q}.\two+\spb{\bP}.{q}\spa4.{\two})} {\spb{\hat{4}}.{q}} 
=-i \frac{(\spa{\bP}.{4}\spb3.\two+\spb{\bP}.3\spa4.{\two})} {\spb4.3} \, .
\eeqn
Therefore the answer by BCFW is
\beqn
-i A_4(\one_Q,3_g^+,4_g^-,\two_{\Qb})
&=& -\frac{\spab4.\one.3}{ m \spab3.\one.3 s_{34} }
\spa{\one}.{\bP} (\spa{\bP}.{4}\spb3.\two+\spb{\bP}.3\spa4.{\two}) \nn\\
&=&  \frac{\spab4.\one.3}{ \spab3.\one.3 s_{34} }
(\spa{\one}.{4}\spb3.\two+\spb{\one}.{3}\spa4.{\two})\, ,
\eeqn
where we have used Eq.~(\ref{identity1}) with a suitable choice of arguments.
Additionally we have $\ket{\bm{p}_I}_{\alpha}[\bm{p}^I|_{\db}=  -p_{\alpha\db}$, (see  Appendix~\ref{SpinorReview}) so that, 
\beq
\spa\one.\bP \spb{\bP}.{3} =\langle{\one^J \bP_I\rangle \, [\bP^I}{3}] =-\spab{\one^J}.{\bP}.{3} =
-m \spb{\one^J}.{3} \, .
\eeq

Summarizing, the two primitive leading-color amplitudes are given by,
\beqn 
\label{eq:2q2gapp}
-i \Afour(\one_{Q},3_g^+,4^+_g,\two_{\Qb})&=& m \frac{\spb3.4}{\spa3.4} \frac{\spa{\one}.{\two}}{(s_{13}-m^2)} \, ,\\
-i \Afour(\one_{Q},3_g^+,4^-_g,\two_{\Qb})&=& 
   \frac{\spab4.\one.3 \big( [\one 3] \braket{4\two} + \braket{\one 4} [3\two] \big)}{(s_{13}-m^2)s_{34}}\, .
\label{eq:2q2gamp}
\eeqn
From Eqs.~(\ref{eq:2q2gapp},\ref{eq:2q2gamp}) the remaining helicity combinations can be obtained by charge conjugation
and line reversal,
\beqn
-i \Afour(\one_{Q},3_g^-,4^-_g,\two_{\Qb})&=&  i \Afour(\two_{Q},4_g^+,3^+_g,\one_{\Qb})|_{\spa{}.{} \leftrightarrow \spb{}.{}}
= m \frac{\spa3.4}{\spb3.4} \frac{\spb{\one}.{\two}}{(s_{13}-m^2)} \, ,\nn\\
-i \Afour(\one_{Q},3_g^-,4^+_g,\two_{\Qb})&=& i \Afour(\two_{Q},4_g^+,3^-_g,\one_{\Qb}) =
 \frac{\bra{3}\one|4] \big( \braket{\one 3} [4\two] + [\one 4] \braket{3\two} \big)}{(s_{13}-m^2)s_{34}} \, .
\label{eq:2q2gb}
\eeqn
The amplitudes in Eqs.~(\ref{eq:2q2gapp},\ref{eq:2q2gamp}) and~(\ref{eq:2q2gb}) clearly satisfy the BCJ relation,
\beq
(s_{13}-m^2) \Afour(\one_{Q},3_g,4_g,\two_{\Qb})=(s_{14}-m^2) \Afour(\one_{Q},4_g,3_g,\two_{\Qb}) \, .
\eeq

The subleading color amplitudes can be obtained from Eq.~(\ref{eq:2q2gb}) using the relation Eq.~(\ref{subleading}).
Applying Eq.~(\ref{subleading}) and simplifying we have,
\beqn
-i \Afour(\one_{Q},3^+_g,\two_{\Qb},4_g^+) & = & m \frac{\spb3.4^2 \spa{\one}.{\two}}{(s_{13}-m^2)(s_{14}-m^2)}\, , \nn \\
-i \Afour(\one_{Q},3_g^-,\two_{\Qb},4_g^+) &=&-                                          
 \frac{\bra{3}\one|4] \big( \braket{\one 3} [4\two] + [\one 4] \braket{3\two} \big)}{(s_{13}-m^2)(s_{14}-m^2)}\, .
\eeqn
\subsection{Two quark pair amplitude}
\subsubsection{Color algebra}
We now write down the amplitude for 4 quarks, where 1 and 2 have mass $m$, and 3 and 4 are massless,
\beq
  \cA_4(\one_{x_1},\two_{x_2},3_{x_3}^{h_3},4_{x_4}^{h_4})  = g^2 (t^C)_{x_1,x_2} \, (t^C)_{x_3,x_4} \; \Afour(\one_Q,\two_{\Qb},3_q^{h_3},4_{\qb}^{h_4})\, .
 \eeq
The result for the amplitude squared summed over colors is,
\beq
\sum_{x_1,x_2,x_3,x_4}   |\cA_4(\one_{x_1},\two_{x_2},3^{h_3}_{x_3},4^{h_4}_{x_4})|^2=g^4 V \, \big| \Afour(\one_Q,\two_{\Qb},3_q^{h_3},4_{\qb}^{h_4})\big|^2 \, .
\eeq
\subsubsection{Result for two quark pair amplitude}
The result for two quark pair amplitude is simply given by,
\beq \label{twoquarkpair}
  -i \Afour(\one_Q,\two_{\Qb},3_q^-,4_{\qb}^+)= \frac{\spa\one.3\spb4.\two+\spb\one.4\spa3.\two}{s_{34}}\, .
\eeq
The primitive amplitude with opposite helicities of the massless quarks is obtained by exchanging labels 3 and 4.  

\section{Example of BCFW recursion}
\label{BCFWexample}
In this section we illustrate the calculation of the 5-parton amplitude 
using BCFW recursion exploiting the amplitudes presented in sections \ref{sec:3parton} and \ref{sec:4parton}.
As an example we calculate one of the amplitudes for one massive quark pair, one massless quark pair
and a gluon, $\cA_5(\one,\two_{\Qb},3_q^-,4_{\qb}^+,5_g^+)$.
For the BCFW shift we take $i=5,j=4$ so that, 
\beq
\ket{\hat{5}}\equiv \ket{5}+z\, \ket{4},\;\;\;\;
|\hat{4}] \equiv |4]-z\, |5]
\eeq

\subsection{Residue at $z_{15}$}
\begin{figure}[t]
\centering
\includegraphics[width=0.25\textwidth,angle=270]{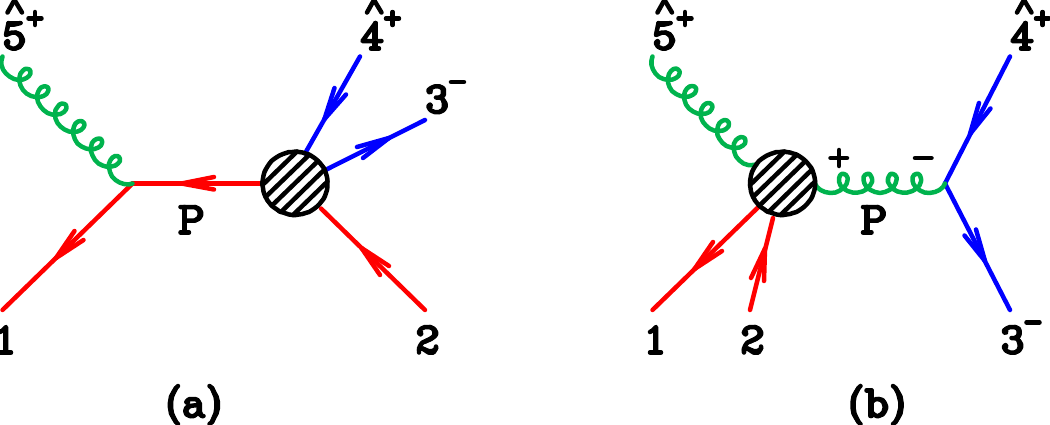}
\caption{Diagrams for BCFW recursion for $A(\one_Q,\two_{\Qb},3_q^-,4_{\qb}^+,5_g^+)$}
\label{fig:BCFW}
\end{figure}
For the diagram in Fig.~\ref{fig:BCFW}(a) we have that, 
\beq
\left. \Afive(\one_Q,\two_{\Qb},3_q^-,4_{\qb}^+,5_g^+) \right|_{z_{15}}=
-\Athree(\one_Q,\hat{5}^+_g,-\bP_{\Qb}) \frac{i}{\spab5.\one.5} \Afour(\bP_Q,\two_{\Qb},3_q^-,\hat{4}_{\qb}^+)\, ,
\eeq
where $P = p_1 + \hat{p}_5$ and the overall sign is due to the definition of the color decomposition, see Eq.~(\ref{colorcoeffs4q1g}).
For $i=5$ and $j=4$ the onshell condition for the massive intermediate quark is
$\bra{5}\one|5]+z_{15}\bra{4}\one|5]=0$.
The shifted spinors are,
\beq
\label{BCFWshift1}\!\!\!
   z_{15}  = -\frac{\bra{5}\one|5]}{\bra{4}\one|5]}\, , \quad~\,
   |\hat{5}\rangle =-\frac{|\one|5|4\rangle}{[5|\one|4\rangle} , \quad~\,
   |{\hat{4}}]\!= \frac{|(4+5)|\one|5]}{\langle 4|\one|5]} \,.
\eeq
For the amplitude on the left hand side of Fig.~\ref{fig:BCFW}(a) we use the expression given in Eq.~(\ref{eq:qgpa}) 
with Eq.~(\ref{eq:signflip}) and we choose $\langle q|=\langle 4|$. For the amplitude on the right hand side
of Fig.~\ref{fig:BCFW}(a) we use Eq.~(\ref{twoquarkpair}),
\beqn
\Athree(\one_Q,\hat{5}^+_g,-\bP_{\Qb})&=& -i
\frac{\langle \one| -\bP\rangle \spab{q}.{\one}.5}{m \spa{q}.{\hat{5}}}
= i \frac{\spa{\one}.{\bP} \spab{4}.{\one}.5}{m \spa{4}.{5}} \\ 
\Afour(\bP_Q,\two_{\Qb},3_q^-,\hat{4}_{\qb}^+) &=&
i \frac{(\spa{\bP}.3 \spb\hat{4}.\two + \spb{\bP}.{\hat{4}} \spa3.\two)}{s_{3\hat 4}}
=i \frac{(\spb{\bP}.{\hat{4}} \spa3.\two-\spa{\bP}.3 \spb{\two}.{\hat{4}})}{\spa4.3 \spb{3}.{\hat{4}}}\\
&=& i \frac{(\spbab{\bP}.{(4+5)}.{\one}.{5} \spa3.\two-\spa{\bP}.3 \spbab{\two}.{(4+5)}.{\one}.{5})}{\spa4.3 \spbab{3}.{(4+5)}.{\one}.{5}}\eeqn
Now using the relations for massive spinors in Eq.~(\ref{identities}),
\beqn
\spa\one.{\bP} \spa {\bP}.{x}   &=& \spa{\one^J}.{\bP_I} \spa {\bP^I}.{x}  = -m \spa{\one^J}.{x} \, ,\nn \\
\spa\one.{\bP} \spb {\bP}.{x} &=& \spa{\one^J}.{\bP_I} \spb {\bP^I}.{x}= -\spab{\one^J}.{\bP}.{x}
\eeqn
we thus have,
\beqn
\spa{\one}.{\bP} \spbab{\bP}.{(4+5)}.\one.5 &=& -\spabab{\one}.{(\one+\hat{5})}.{(4+5)}.{\one}.{5} \nn \\
&=& -m\spbab{\one}.{(4+5)}.{\one}.{5}
-\spa{\one}.{\hat{5}} \spbab{5}.{(4+5)}.{\one}.{5} \nn \\
&=& m\spbab{5}.{\one}.{(4+5)}.{\one}
+m \spba{\one}.{5}.{4} \spb{5}.{4}
\eeqn
Hence,
\beqn
\frac{\spa{\one}.{\bP}}{m} \Afour(\bP_Q,\two_{\Qb},3_q^-,\hat{4}_{\qb}^+)&=&
  i \Big( \big(\spbab{5}.{\one}.{(4+5)}.{\one}+\spba{\one}.{5}.{4} \spb{5}.{4}\big) \spa3.{\two}
  +\spa{\one}.3 \spbab{\two}.{(4+5)}.{\one}.5 \Big)\nn\\
  &\times& \frac{1}{\spa4.3 \spbab3.{(4+5)}.{\one}.5}\, .
\eeqn
So the final result is,
\beqn
\Afive(\one_Q,\two_{\Qb},3_q^-,4_{\qb}^+,5_g^+) \big|_{z_{15}}
&=& \frac{i}{\spab5.\one.5} \frac{\spab4.{\one}.5}{\spa4.5} \frac{1}{\spa4.3 \spbab3.{(4+5)}.{\one}.5} \nn\\
&\times&
\Big( \big(\spbab{5}.{\one}.{(4+5)}.{\one}+\spb{\one}.{5}\spa{5}.{4} \spb{5}.{4}\big) \spa3.{\two}
    +\spa{\one}.3 \spbab{\two}.{(4+5)}.{\one}.5 \Big) \nn\\
&=& \frac{i}{\spab5.\one.5} \frac{\spab4.{\one}.5}{\spa4.5} \frac{1}{\spa3.4 \spbab3.{(4+5)}.{\one}.5} \nn\\
&\times&
    \Big( \big(
    \spbab{\one}.{(4+5)}.{\one}.{5}
    +\spb{\one}.{5}\spa{4}.{5} \spb{5}.{4}\big) \spa3.{\two}
    +\spa{\one}.3 \spbab{5}.{\one}.{(4+5)}.{\two}
    \Big) \nn\\
\label{eq:bcfwres1}
\eeqn

\subsection{Residue at $z_{34}$}
For the second diagram, Fig.~\ref{fig:BCFW}(b),
\beq
\left. \Afive(\one_Q,\two_{\Qb},3_q^-,4_{\qb}^+,5_g^+) \right|_{z_{34}} =
-\Afour(\one_Q,\hat{5}^+_g,P_g^+,\two_{\Qb}) \frac{-i}{s_{34}} \Athree(3_q^-,-P_g^-,\hat 4_{\qb}^+) \, ,
\eeq
where $P=p_3+\hat{p}_4$ and overall sign is because of the definition of the color amplitude.
For $i=5$ and $j=4$ the onshell condition for the intermediate gluon line 
is $(\spb3.4-z_{34}\spb3.5)\spa4.3=0$.
The shifted spinors in this case are,
\beq
\label{BCFWshift2}\!\!\!
   z_{34}  = \frac{\spb3.4}{\spb3.5}\, ,\quad
   |\hat{5}\rangle =-\frac{|(4+5)|3]}{\spb3.5} \, ,\quad
     |{\hat{4}}] = |3]\frac{\spb4.5}{\spb3.5} , \quad~\,
    \langle P|= -\frac{[5|(3+4)|}{\spb3.5}\, , \quad
    |P]= |3]\, .
\eeq
Inserting the amplitudes from Eqs.~(\ref{eq:2q2gapp}) and~(\ref{eq:qgma}) gives,
\beqn
\label{LHS2}
\Afour(\one_Q,\hat{5}^+_g,P_g^+,\two_{\Qb}) &=& i m \frac{\spb{P}.{5}}{\spa{P}.{\hat{5}}} \frac{\spa\one.\two}{\spba{5}.\one.{\hat{5}}}
=-im \frac{\spb3.5^2}{s_{345}}  \frac{\spa\one.\two \spb3.5}{\spbab{3}.{(4+5)}.\one.{5}} \, \\
\label{RHS2}
\Athree(3_q^-,-P_g^-,\hat 4_{\qb}^+)&=&  (-i) \frac{\spa3.{|-\!P} \spb q.{\hat 4}}{\spb{-P}.q}= i \frac{\spa3.P \spb{5}.{4}}{\spb{P}.5}
= -i \frac{\spa3.4 \spb4.5^2}{\spb3.5^2} \, .
\eeqn
since,
\beq
\spb{P}.5 = \spb3.5\,,\quad\quad \spa{P}.{\hat{5}}=\frac{\spbab5.(3+4).(4+5).3}{\spb3.5^2} = -\frac{s_{345}}{\spb3.5}\,,\quad\quad
\spba{5}.\one.{\hat{5}}=\frac{\spbab3.{(4+5)}.\one.{5}}{\spb3.5}
\eeq
Inserting the results from Eqs.~(\ref{LHS2}) and (\ref{RHS2}) gives,
\beqn \label{eq:bcfwres2}
\Afive(\one_Q,\two_{\Qb},3_q^-,4_{\qb}^+,5_g^+) \big|_{z_{34}} &=&
(-1) \times (-im) \frac{\spb3.5^2}{s_{345}}  \frac{\spa\one.\two \spb3.5}{\spbab{3}.{(4+5)}.\one.{5}} \times \frac{i}{\spa3.4 \spb3.4}
\times (-i) \frac{\spa3.4 \spb4.5^2}{\spb3.5^2} \nn\\
&=& i \frac{m \spb3.5 \spb4.5^2 }{s_{345}}  \frac{\spa\one.\two}{\spb3.4 \spbab{3}.{(4+5)}.\one.{5}}
\eeqn

The diagram in Fig.~\ref{fig:BCFW}(b) with the opposite helicity of gluon exchanged vanishes, because 
the amplitude on the right hand side is proportional to,
\beq
\Athree(3_q^-,-P_g^+,4_{\qb}^+) = i \frac{\spa3.q [\hat{4}|{-P}]}{\langle{-P}|q\rangle}
\eeq
and from Eq.~(\ref{BCFWshift2}) both $[\hat{4}|$ and $|-P]$ are proportional to $|3]$.
Thus the sum of the two contributions given in Eqs.~(\ref{eq:bcfwres1}) and~(\ref{eq:bcfwres2}) gives
the total result for $\Afive(\one_Q,\two_{\Qb},3_q^-,4_{\qb}^+,5_g^+)$,
\beqn
\label{eq:q3mq4pg5pbcfw}
 -i \Afive(\one_Q, \two_{\Qb},3^-_q,4^+_{\qb},5^+_g) &=&
   \frac{1}{\spbab3.(4+5).\one.5} \* \Big[ \frac{m \* \spa\one.\two \* \spb3.5 \* \spb4.5^2}{s_{345} \* \spb3.4} \\
 &+&\frac{1}{\spab5.\one.5} \* \frac{\spab4.\one.5}{\spa4.5 \* \spa3.4} \* \Big(
  \spa\one.3 \* \spbab5.\one.(4+5).\two+\spbab\one.(4+5).\one.5 \* \spa3.\two 
   + \spb\one.5 \* \spa3.\two \* s_{45} \Big) \Big] \, .\nn
\eeqn
 
We note that, at this stage, the amplitude appears to contain an unphysical pole
represented by the overall factor $\spbab3.(4+5).\one.5$. 
This is also the case for the result given in ref.~\cite{Lazopoulos:2021mna}, which is presented in
a slightly different form but with which this agrees (after taking the limit in
which one quark pair is massless). By using the
equations of motion and applying Schouten identities one can demonstrate explicitly
that this pole is not present.  The result for this amplitude presented below in
section~\ref{sec:5parton} has been simplified in this way.

\section{Five parton amplitudes}
\label{sec:5parton}
\subsection{One quark pair + 3 gluon amplitudes}
\subsubsection{Color algebra}
The general color decomposition is given by Eq.~(\ref{QQn-2g}).
For $n=5$ we have,
\beq \label{QQ3g}
  \cA_5(\one_{x_1},3_{C_3},4_{C_4},5_{C_5},\two_{x_2})
 \ =\  g^3 \sum_{\sigma\in S_3}
 (t^{C_{\sigma(3)}} t^{C_{\sigma(4)}} t^{C_{\sigma(5)}})_{x_1 x_2}
  \Afive(\one_Q,\sigma(3)_g,\sigma(4)_g,\sigma(5)_g,\two_{\Qb})\ ,
\eeq
Squaring the amplitude and summing over colors
we obtain the following expression~\cite{Mangano:1987kp},
\beq
  \sum_{\rm colors} |\cA(\one_{x_1},3_{C_3},4_{C_5},5_{C_5},\two_{x_2})|^2 =
  g^6\,\frac{(N^2-1)}{N^2} \sum_{j=0}^2 N^{2j} \sum_{\sigma\in S_{3}} \, H_j(3,4,5) \, .
\eeq
Introducing the compact notation, 
\beq
(3,4,5)=\Afive(\one_Q,3_g,4_g,5_g,\two_{\Qb})
\eeq
we get
\beqn
H_2(3,4,5) &=& |(3,4,5)|^2 \,,\\
H_1(3,4,5) &=& -(3,4,5)^*\, [2 (3,4,5) + (3,5,4) + (4,3,5) - (5,4,3)] \,, \label{eq:H1} \\
H_0(3,4,5) &=& (3,4,5)^* \sum_{\sigma\in S_3} (i,j,k) \,.
\eeqn
Note that ref.~\cite{Mangano:1987kp} contains a typographical error in the sign of the final term
in the expression for $H_1(3,4,5)$ which we have corrected in Eq.~(\ref{eq:H1}).

\subsection{Results for one quark pair + 3 gluon amplitudes}
The amplitude with gluons of all positive helicity is taken directly from ref.~\cite{Ochirov:2018uyq},
\beq
-i \Afive(\one,3_g^+,4_g^+,5_g^+,\two) =
-m \*\frac{\spb4.5\* \spab4.\one.3+\spab3.\one.3\* \spb3.5}{\spa3.4\* \spa4.5}
  \* \frac{\spa\one.\two}{\spab3.\one.3\* \spab5.\two.5}
 \,  .
 \\
\eeq
 The amplitude for gluon 3 of negative helicity is also given in ref.~\cite{Ochirov:2018uyq}.  However
it contains an unphysical pole that can, with suitable application of Schouten
identities and momentum conservation, be removed.  
The final simplified result is,
\beqn
&& -i \Afive(\one,3_g^-,4_g^+,5_g^+,\two)=
 \frac{1}{\spa3.4\* \spa4.5\* \spab3.\one.3\* \spab5.\two.5} \nn \\
&\times& \Bigg\{ \Bigg[
 \frac{m\* \spa3.4\* \spb4.5\* \big(\spab3.\one.4\* s_{345} - \spab3.\one.3\* \spb4.5\* \spa5.3 \big)}{s_{345}\* \spb3.4}
 -\frac{\spaa3.\one.(4+5).3\* \spab3.\two.5\* \spab3.1.3}{m\* s_{345}} \Bigg] \* \spa\one.\two \nn \\
&+&\frac{ \big( \spab3.\two.5\* \spab3.\one.3 - m^2\* \spa3.4\* \spb4.5 \big)}{m} \*\spa\one.3 \* \spa3.\two \Bigg\}
\,  .
\eeqn
 
The position of the negative helicity gluon can be moved to the other end of the string
through the line-reversal relation,
\beq \label{line-reversal}
\Afive(\one_q,3_g^+,4_g^+,5_g^-,\two_{\qb}) = \Afive(\two_q,5_g^-,4_g^+,3_g^+,\one_{\Qb}) \,.
\eeq

The final amplitude we need, with the negative helicity gluon in the middle of the
string, is fixed by the previous two equations through the BCJ relation,
\beqn
  (s_{13}+s_{34}-m^2)  \, \Afive(\one_Q,4_g^+,3_g^-,5_g^+,\two_{\Qb}) &=&  \nn \\
   (s_{23}-m^2)  \, \Afive(\one_Q,4_g^+,5_g^+,3_g^-,\two_{\Qb})
 - (s_{13}-m^2)\, \Afive(\one_Q,3_g^-,4_g^+,5_g^+,\two_{\Qb}) \, .
\eeqn
Forming the appropriate combination and manipulating to remove the spurious pole we
find,
\beqn
&& -i \Afive(\one_q,3_g^+,4_g^-,5_g^+,\two_{\qb}) = 
\frac{\spa\one.4\* \spa4.\two\* \spab4.\one.3\* \spab4.\two.5}{m\* \spa3.4\* \spa4.5\* \spab3.\one.3\*\spab5.\two.5}\nn \\
&+& \frac{\spa\one.\two}{\spa3.5\* \spab3.\one.3\*\spab5.\two.5\* s_{345}}
\* \Big(m\* \frac{\spb3.5}{\spb4.5}
   \* \Big[\spa4.3\* \spb3.5\* \spab4.\two.4 - s_{345}\* \spab4.\two.5\Big]
          - \frac{\spab4.\one.3\* \spab4.\two.4\* \spaa4.\one.\two.4}{m\* \spa4.5}\Big)\nn \\
       &+& \frac{\spa\one.\two}{\spa3.5\* \spab5.\two.5\* s_{345}}
       \* \Big(\frac{m\* \spa3.5\* \spb3.5^3}{\spb4.3\* \spb4.5}  
       + \frac{\spaa4.\one.\two.4^2}{m\* \spa4.3\* \spa4.5}\Big)
\,  .
\eeqn
 
The remaining amplitudes are obtained through a simple operation,
\beq
\Afive(\one,3_g^{-h_3},4_g^{-h_4},5_g^{-h_5},\two) = 
 - \Afive(\one,3_g^{h_3},4_g^{h_4},5_g^{h_5},\two) |_{\spa{}.{} \leftrightarrow \spb{}.{}} \,,
\eeq
in an obvious notation where $\spa{}.{} \leftrightarrow \spb{}.{}$ denotes the interchange of angle
and square brackets.

\subsection{Two quark pairs and one gluon}
\subsubsection{Color algebra for two quark pairs and one gluon}
In the case of two
quark pairs and one gluon the possible color structures are the following,
\beqn \label{QQqqgcolorstructure}
  \cA_5(\one_{x_1}, \two_{x_2},3_{x_3},4_{x_5},5_{C_5}) &=& g^3
  \Big(-(t^{C_5})_{x_1 x_4}  \delta_{x_3 x_2} A^{(1)}
  -(t^{C_5})_{x_3 x_2}  \delta_{x_1 x_4} A^{(2)} \nn \\
  &-&\frac{1}{N}(t^{C_5})_{x_1 x_2}  \delta_{x_3 x_4} A^{(3)}
  -\frac{1}{N}  (t^{C_5})_{x_3 x_4}  \delta_{x_1 x_2} A^{(4)}\Big) \, .
\eeqn
Squaring and summing over colors we find ($V=N^2-1$)
\beqn
\sum_{\rm colors} |\cA_5(\one_{x_1}, \two_{x_2},3_{x_3},4_{x_5},5_{C_5})|^2
&=& \frac{V}{N}\Big[(N^2-1) \* \left(|A^{(1)}|^2+|A^{(2)}|^2\right) \nn\\
&& - A^{(1)} A^{(2) *} -A^{(2)} A^{(1) *} -A^{(3)} A^{(4) *} -A^{(4)} A^{(3) *}
\nn \\
&& + |A^{(1)}+A^{(2)}+A^{(3)}+A^{(4)}|^2 \Big]\, .
\eeqn
Note that the term on the final line will not contribute since this
combination of subamplitudes is identically zero.
Imposing this condition we obtain,
\beqn
\sum_{\rm colors} |\cA_5(\one_{x_1}, \two_{x_2},3_{x_3},4_{x_5},5_{C_5})|^2
& =& V N \,\Big[|A^{(1)}|^2+|A^{(2)}|^2\nn\\
   &+&\frac{1}{N^2} \Big(|A^{(3)}|^2+|A^{(4)}|^2
    -2 \big|(A^{(1)}+A^{(2)})\big|^2\Big) \Big] \, .
\eeqn

Using the Melia basis, as described in ref.~\cite{Johansson:2015oia}, this
should be written in terms of three independent primitives as,
\beqn
  \cA_5(\one_{x_1}, \two_{x_2},3_{x_3},4_{x_5},5_{C_5}) &=& g^3 \Big(
   A_{12345} \, C_{12345} + A_{12354} \, C_{12354} + A_{12534} \, C_{12534} 
  \Big)  \, ,
\eeqn
where the color coefficients are given by,
\beqn
C_{12345} &=&
-(t^{C_5} t^{D})_{x_1 x_2} (t^{D})_{x_3 x_4} \nn\\
C_{12534} &=&
-(t^{D} t^{C_5})_{x_1 x_2} (t^{D})_{x_3 x_4} \nn\\
C_{12354} &=& 
-(t^{C_5} t^{D})_{x_1 x_2} (t^{D})_{x_3 x_4}
	       -(t^{D})_{x_1 x_2} (t^{D} t^{C_5})_{x_3 x_4} \,  .
\label{colorcoeffs4q1g}
\eeqn
Performing the color algebra and comparing we can thus identify,
\beqn
A^{(1)} &=& A_{12345}, \; 
A^{(2)} = A_{12534}, \;
A^{(3)} = -A_{12345} - A_{12354} - A_{12534}, \;
A^{(4)} = A_{12354}
\label{eq:A1sign}
\eeqn
These clearly satisfy the constraint alluded to above,
\beq
   A^{(1)}+A^{(2)}+A^{(3)}+A^{(4)}=0
\label{eq:ampconstraint} \,.
\eeq

\subsubsection{BCJ relations}
We can further reduce the set of primitives by using the
kinematic-algebra basis that also accounts for BCJ relations between the
amplitudes.  In that case we fix quark 3 to be in position 3 and find the result
in terms of two amplitudes,
\beq
p_{2}\cdot p_{5} \; A_{12534}-p_{1}\cdot p_{5} \; A_{12345}-p_{14}\cdot p_{5} \; A_{12354} =0\, ,
\eeq
\beqn
  \cA_5(\one_{x_1}, \two_{x_2},3_{x_3},4_{x_4},5_{C_5}) &=& g^3 \bigg[
   A_{12345} \Big( C_{12345} + C_{12534} \, \frac{p_1\cdot p_5}{p_2\cdot p_5} \Big) \nn
   \\
 && \;\; + A_{12354} \Big( C_{12354} + C_{12534} \, \frac{p_{14}\cdot p_5}{p_2\cdot p_5} \Big)
  \bigg]\, .
\eeqn

\subsection{Results for two quark pairs + 1 gluon amplitudes}
Manipulating the result derived in the previous section in Eq.~(\ref{eq:q3mq4pg5pbcfw}) to remove
the unphysical pole, we find the amplitude for a positive helicity gluon,
\beqn
  -i A(\one_Q, \two_{\Qb},3^-_q,4^+_{\qb},5^+_g) &=&
\frac{1}{m\* \spa4.5} \* \Bigg[
    -\frac{\spa\one.3\* \spa 3.\two\*\spab4.\one.5}{\spa3.4 \* \spab5.\one.5} \nn \\
    &+&\frac{\spa\one.\two}{s_{34}} \* \Big( \frac{\spab3.\two.4 \* \spab4.\one.5}{\spab5.\one.5}
    +\frac{\spaba3.\two.(3+5).4 \* \spb4.5}{s_{345}}\Big)\Bigg]
\,  .
\eeqn
 The corresponding result for a negative helicity gluon after removal of the unphysical pole is,
\beqn
  -i A(\one_Q, \two_{\Qb},3^-_q,4^+_{\qb},5^-_g) &=&
\frac{1}{m\* \spb4.5}\* \Bigg[  
       -\frac{ \spb\one.4\* \spb4.\two\* \spab5.\one.4}{\spab5.\one.5\* \spb3.4} \nn \\	      
      &+&\frac{\spb\one.\two}{s_{34}}\* \Big( \frac{\spa3.5 \* \spbab4.(3+5).\two.4}{s_{345}}
        -\frac{\spab3.\two.4\* \spab5.\one.4}{\spab5.\one.5} \Big)
\Bigg]
\,  .
\eeqn
 
\subsubsection{Relationship to other amplitudes}
The other leading color amplitude is related to the one given above through
charge conjugation,
\beq
  A(\one_Q, \two_{\Qb},5^-_g,3^-_q,4^+_{\qb}) =
  A(\two_Q, \one_{\Qb},4^-_q,3^+_{\qb},5^+_g)
   \big._{\spa{}.{} \leftrightarrow \spb{}.{}} \, .
\eeq
The subleading color amplitude $A(\one_Q, \two_{\Qb},3_q,5_g,4_{\qb})$
can be obtained by forming a combination
with amplitudes in which the heavy quark and antiquark are interchanged,
\beqn
  A(\one_Q, \two_{\Qb},3^-_q,5_g,4^+_{\qb}) &=&
  A(\two_Q, \one_{\Qb},3^-_q,4^+_{\qb},5_g)
 +A(\one_Q, \two_{\Qb},5_g,3^-_q,4^+_{\qb}) \, .
\eeqn
Simplifying this combination we find,
\beq
-i \Afive(\one_Q, \two_{\Qb},3^-_q,5^+_g,4^+_{\qb})=
\frac{1}{m\* \spa3.5\* \spa4.5}\* \Bigg[ 
  \spa\one.\two\* \Big( \frac{\spab3.1.4}{\spb3.4}
  - \frac{\spb4.5\* \spaba5.(3+4).\one.3}{\spb3.4 \* s_{345}} \Big) - \spa\one.3\* \spa3.\two \Bigg]
\,  .
\eeq
 \beq
-i \Afive(\one_Q, \two_{\Qb},3^-_q,5^-_g,4^+_{\qb})=
-\frac{1}{m\* \spb3.5\* \spb4.5}
  \* \Bigg[ \spb\one.\two\* \Big( \frac{\spab3.1.4}{\spa3.4}
  - \frac{\spa3.5\* \spbab5.(3+4).\one.4}{\spa3.4\* s_{345}} \Big) + \spb\one.4\* \spb4.\two\Bigg]
\,  .
\eeq
 Together with the identities in Eq.~(\ref{eq:A1sign}), all amplitudes needed to
construct the full squared matrix element for this process are at hand.

\section{Six parton amplitudes}
\label{sec:6parton}

\subsection{One quark pair + 4 gluon amplitudes}
\subsubsection{Color structure}
Here we describe the color structure for one massive quark pair + 4 gluon amplitudes.
The form of the expansion into color-ordered primitives is taken from Eq.~(\ref{QQn-2g}).
\beq \label{QQ4g}
\cA_6(1_{x_1},3_{C_3},4_{C_4},5_{C_5},6_{C_6},2_{x_2})
 \ =\  g^4 \sum_{\sigma\in S_4}
 (t^{C_{\sigma(3)}}t^{C_{\sigma(4)}}t^{C_{\sigma(5)}}t^{C_{\sigma(6)}})_{x_1 x_2}
 \Asix(1_{Q},\sigma(3),\sigma(4),\sigma(5),\sigma(6),2_{\Qb})\ ,
\eeq
where $S_{4}$ is the permutation group on $4$ elements,
and $A_6$ are the tree-level partial amplitudes.

Squaring the amplitude and summing over colors
we obtain the following expression~\cite{Mangano:1987kp},
\beq
  \sum_{\rm colors} |\cA_6(\one_Q,3_{C_3},4_{C_4},5_{C_5},6_{C_6},\two_{\Qb})|^2 =
  g^8\, \frac{(N^2-1)}{N^3} \sum_{j=0}^3\, N^{2j} \sum_{\sigma\in S_{4}} \, H_j(3,4,5,6) \, .
\eeq
Introducing the compact notation for the color-ordered primitives $A$, 
\beq
(3,4,5,6)=\Asix(\one_Q,3_g,4_g,5_g,6_g,\two_{\Qb})
\eeq
we get,
\beqn
H_3(3,4,5,6) &=& |(3,4,5,6)|^2,\nn \\
H_2(3,4,5,6) &=& (3,4,5,6)^*\, [-3 (3,4,5,6) - (3,4,6,5) - (3,5,4,6)\nn \\
&-&(4,3,5,6) + (3,6,5,4) + (5,4,3,6) + (5,6,3,4)\nn \\
&+&(5,6,4,3) + (6,4,5,3) + (6,5,3,4)], \nn \\
H_1(3,4,5,6) &=& (3,4,5,6)^*\, [M(3,4,5,6) -M(6,5,4,3)]\nn \\
M(3,4,5,6)&=&3 (3,4,5,6)+2 (3,4,6,5)+2 (3,5,4,6)+2 (4,3,5,6)+(3,5,6,4)\nn \\
&+&(3,6,4,5) +(4,3,6,5) + (4,5,3,6) + (5,3,4,6), \nn \\
H_0(3,4,5,6) &=& -(3,4,5,6)^*\, \sum_{\sigma\in S_{4}} (i,j,k,l).
\eeqn

\subsubsection{Results for $\Asix(\one_Q,3_g,4_g,5_g,6_g,\two_{\Qb})$}
The all-plus helicity result is taken from Ochirov~\cite{Ochirov:2018uyq},
\beqn
 -i A(\one_Q,3_g^+,4_g^+,5_g^+,6_g^+,\two_{\Qb}) &=&
  m \* \spa\one.\two \* \frac{[3| \big(\slsh{p}_{13} \slsh{p}_{4} + s_{13}-m^2\big)
  \* \big(\slsh{p}_{134} \slsh{p}_5 + s_{134}-m^2)\big) | 6]}{
  (\spab3.\one.3\* (s_{134}-m^2)\* \spab6.2.6\* \spa3.4\* \spa4.5\* \spa5.6)}
  \nonumber \\
  &=&
  -m \* \spa\one.\two \* \frac{\mt^2\* \spa4.5\* \spb3.4\* \spb5.6
  +(s_{134}-\mt^2)\*\spbab6.(3+4+5).\one.3}{
  (\spab3.\one.3\* (s_{134}-m^2)\* \spab6.2.6\* \spa3.4\* \spa4.5\* \spa5.6)} 
\,  . \eeqn

The result with one negative helicity adjacent to the massive quark is also taken from ref~\cite{Ochirov:2018uyq},
\beqn
  && -i \Asix(\one,3_g^-,4_g^+,5_g^+,6_g^+,\two)=
   \Bigg[
      \frac{\spaba3.\one.(4+5+6).3}{m\* s_{3456}\* \spaba3.\one.(3+4+5).6}
      \*\Big(\spa\one.\two\*\spaba3.\one.(4+5+6).3-\spa\one.3\* \spa3.\two \* s_{3456}\Big) \nn \\
     & +&m\*\spa4.5\*\spaba3.\one.4.3\* \frac{\spab3.4.6\*\spab6.\two.6+\spab3.4.5\* \spab5.\two.6}{
           s_{34} \*(s_{134}-m^2)\* \spab6.\two.6\* \spaba3.\one.3.4\*\spaba3.\one.(3+4).5}
           \*(\spa\one.\two\* \spaba3.\one.4.3-\spa\one.3\* \spa3.\two\*s_{34}) \nn \\
      &+&m\*\frac{\spa5.6\* \spaba3.\one.(4+5).3 \*\spab3.(4+5).6}{
            s_{345}\*\spab6.\two.6\* \spaba3.\one.(3+4).5\* \spaba3.\one.(3+4+5).6}
           \*\big(\spa\one.\two\* \spaba3.\one.(4+5).3-\spa\one.3\* \spa3.\two\*s_{345}\big)
      \Bigg]\* \frac{1}{\spa3.4\*\spa4.5\*\spa5.6} 
\,  .
\nonumber \\
\eeqn
 
The unphysical poles present in this result can be removed, at the expense of generating a slightly longer
expression,
\beqn
  && -i \Asix(\one,3_g^-,4_g^+,5_g^+,6_g^+,\two)=
 -\frac{1}{\spa4.5\* \spa5.6} \* \Bigg[ \nn \\ &&
         \Big(\spa\one.\two\* \frac{\spaba3.\one.(4+5).3}{s_{345}}- \spa\one.3\* \spa3.\two \Big)
        \* \frac{\mt}{(s_{134}-\mt^2)} \* \Big(
           \frac{\spa3.5\* \spb5.6}{\spa3.4 \* \spab6.2.6}
         + \frac{\spb4.5\* \spab5.2.6}{\spab3.1.3 \* \spab6.2.6}
         + \frac{\spb4.6}{\spab3.1.3} \Big) \nn \\ &&
       +\frac{\spa\one.3\* \spa3.\two \* \spab3.2.6}{\spa3.4 \* \spab6.2.6 \* \mt}
       -\frac{\spa\one.\two}{s_{3456}\* \spa3.4\* \spab6.2.6} \* \Big(
           \frac{\spa3.6\* \spab3.(4+5).6^2\* \mt}{s_{345}}
         - \frac{\spaba3.\one.\two.3\* \spab3.2.6}{\mt} \Big) \nn \\ &&
       +\frac{\spa\one.\two\* \spa3.4\* \spa4.5\* \spb4.5\* \spab3.1.4\* \mt}{\spab3.1.3 \* s_{34} \*s_{345} \* (s_{134}-m^2)} \* \Big(
           \frac{\spb4.5\* \spab5.2.6}{\spab6.2.6}
         + \spb4.6 \Big)
	 \Bigg]
\,  .
\eeqn
 
A complete set of amplitudes can be generated after specifying the results for four other helicity
combinations.  The first corresponds to a single gluon of negative helicity but in a different position
in the string,
\beqn
  && -i \Asix(\one,3_g^+,4_g^-,5_g^+,6_g^+,\two)=
       - \frac{m \* \spa\one.\two \* \spb5.6 \* \spab4.\one.3^2 \* \spab4.(\one+3).5^2}{
           \spa3.4 \* \spab3.\one.3 \* \spab6.\two.6 \* \spbab5.(3+4).\one.3 \* \spaba6.\two.(5+6).4 \* (s_{134}-m^2)}
 \nonumber \\ &&
       + \frac{\spa\one.\two \* \spb5.6 \* \spab4.\one.3 \*
            \spaba4.\one.\two.4 \* \spaba4.\two.(4+5).6
          }{
           \spa3.6 \* \spa4.5 \* \spab3.\one.3 \* \spaba6.\two.(5+6).4 \* m \* s_{56}\* s_{3456} } 
       + \frac{\spa\one.\two \* \spa4.6 \* \spb5.6 \* \spaba4.\one.\two.4^2}{
           \spa3.4 \* \spa3.6 \* \spa4.5 \* \spaba6.\two.(5+6).4 \* m \* s_{56} \* s_{3456}}
 \nonumber \\ &&
       - \frac{m \* \spa\one.\two \* \spb5.6 \* \spab4.(5+6).3}{
           \spa3.6 \* \spa4.5 \*\spab3.\one.3 \* \spaba6.\two.(5+6).4 \* s_{56} \* s_{456}}  \*  \Big(
            \frac{\spa3.4 \* \spab4.(5+6).3 \* \spaba4.\two.(4+5).6}{s_{3456}}
          + \spa4.6 \* \spaba4.\two.(5+6).4
          \Big)
 \nonumber \\ &&
       + \frac{m \* \spa\one.\two \* \spa3.4 \* \spb3.5^4 \* \spbab3.\one.(3+4+5).6}{
           \spb4.5 \* \spab6.\two.6 \* \spab6.(4+5).3 \* \spbab5.(3+4).\one.3 \* s_{34} \* s_{345}}
       + \frac{m \* \spa\one.\two \* \spa4.6 \* \spb5.6 \* \spab4.(5+6).3^3}{
           \spa4.5 \* \spab6.(4+5).3 \* \spaba6.\two.(5+6).4 \* s_{56} \* s_{456} \* s_{3456}}
 \nonumber \\ &&
       + \frac{\spa\one.4 \* \spa4.\two \* \spb5.6 \* \spab4.\one.3}{
           \spa3.4 \* \spa4.5 \* \spab3.\one.3 \* \spab6.\two.6 \* m \* s_{56}} \* \Big(
	   \spab4.\two.6 + \frac{m^2 \* \spa4.5 \* \spb6.5}{(s_{134}-m^2)}
           \Big)
\,.
\eeqn
 
The other three amplitudes contain two negative-helicity gluons and are given by,
\beqn
&& -i \,\Asix(\one_q,3_g^-,4_g^-,5_g^+,6_g^+,\two_{\qb}) = 
 \nn \\ &&
 \frac{\mt^2\* \spa3.4\* \spb5.6\* \spab4.(\one+3).5}{\spab3.\one.3\* \spab6.\two.6\* \spbab3.\one.(3+4).5\* \spaba6.\two.(5+6).4}\* \Big(
  \frac{\spab\two.(\two+5+6).\one\* \spab4.(\one+3).5}{(s_{134}-\mt^2)}
 +  \spb\one.5\* \spa4.\two
 \Big)
 \nn \\ &&
 + \frac{\spaba3.\one.(5+6).4}{\spa4.5\* \spa5.6\* \spab3.\one.3\* \spab6.(4+5).3\* \mt\* s_{456}}\* \Big(
  \frac{\spb\one.\two\* \spababa4.(5+6).\two.\one.(5+6).4}{s_{3456}}
 - \spab4.(5+6).\one\* \spab4.(5+6).\two
 \Big)
 \nn \\ &&
 + \frac{\mt\* \spa3.4\* \spa4.6}{\spa4.5\* \spa5.6\* \spab3.\one.3\* \spab6.(4+5).3\* \spaba6.\two.(5+6).4}\* \Big(
  \spb\one.\two\* \spaba4.\two.(5+6).4
 - \spab4.(5+6).\one\* \spab4.(5+6).\two
 \Big)
 \nn \\ &&
 - \frac{\spbab5.(3+4).\two.6}{\spb3.4\* \spb4.5\* \spab6.\two.6\* \spab6.(4+5).3\* \mt\* s_{345}}\* \Big(
  \frac{\spa\one.\two\* \spbabab5.(3+4).\one.\two.(3+4).5}{s_{3456}}
 + \spab\one.(3+4).5\* \spab\two.(3+4).5
 \Big)
 \nn \\ &&
 + \frac{\mt\* \spb3.5\* \spb5.6}{\spb3.4\* \spb4.5\* \spab6.\two.6\* \spab6.(4+5).3\* \spbab5.(3+4).\one.3}\* \Big(
  \spa\one.\two\* \spbab5.(3+4).\one.5
 - \spab\one.(3+4).5\* \spab\two.(3+4).5
 \Big)
 \nn \\ &&
 - \frac{\mt\* \spa3.4^2\* \spab4.(5+6).3\* \spb\one.\two}{\spa4.5\* \spa5.6\* \spab3.\one.3\* \spab6.(4+5).3\* s_{3456}}
 - \frac{\mt\* \spa\one.\two\* \spab6.(3+4).5\* \spb5.6^2}{\spb3.4\* \spb4.5\* \spab6.\two.6\* \spab6.(4+5).3\* s_{3456}}
\,  .
\eeqn
 \beqn
&& -i\* \Asix(\one_q,3_g^+,4_g^-,5_g^+,6_g^-,\two_{\qb}) = 
 \nn \\ &&
   \frac{\spab4.\one.3\* \spab4.(\one+3).5^2\* \spab6.\two.5}{
   \spa3.4\* \spb5.6\* \spab3.\one.3\* \spab6.\two.6\* \spbab5.(3+4).\one.3\* \spaba6.\two.(5+6).4\* (s_{134}-\mt^2)}
 \nn \\ &&
 \times \Big(
    - \spb\one.3\* \spa6.\two\* \spa4.6\* \spb5.6
    + \spa\one.4\* \spb5.\two\* \spab6.(4+5).3
    + \spa\one.4\* \spb3.\two\* \spab6.\two.5
    + \spb\one.3\* \spa4.\two\* \spab6.\two.5
    \Big) \nn \\ &&
 - \frac{\spa\one.4\* \spb5.\two\* \spa4.6\* \spab4.\one.3\* \spab4.(\one+3).5\* \spab6.\two.5}{
    \spa3.4\* \spb5.6\* \spab3.\one.3\* \spab6.\two.6\* \spaba6.\two.(5+6).4\* (s_{134}-\mt^2)}
 \nn \\ &&
 + \frac{\mt\* \spa4.6^4}{
    \spa4.5\* \spa5.6\* \spab3.\one.3\* \spaba6.\two.(5+6).4\* s_{456}}\* \Big(
      \frac{\spab4.\one.3\* \spb\one.\two}{\spa3.4}
    - \spb\one.3\* \spb3.\two
    \Big)
 \nn \\ &&
 + \frac{\mt\* \spb3.5^4}{\spb3.4\* \spb4.5\* \spab6.\two.6\* \spbab5.(3+4).\one.3\* s_{345}} \* \Big(
      \frac{\spa\one.\two\* \spab6.\two.5}{\spb5.6}
    - \spa\one.6\* \spa6.\two\Big)
 \nn \\ &&
 - \frac{\spb3.5^4\* \spab6.\one.3}{\spb3.4\* \spb4.5\* \spab6.(4+5).3\* \spbab5.(3+4).\one.3\* \mt\* s_{345}}
  \* \Big(\frac{\spa\one.\two\* \spaba6.\two.\one.6}{s_{3456}} - \spa\one.6\* \spa6.\two\Big)
 \nn \\ &&
 + \frac{\spa4.6^4\* \spab6.\two.3}{\spa4.5\* \spa5.6\* \spab6.(4+5).3\* \spaba6.\two.(5+6).4\* \mt\* s_{456}}\* \Big(
      \frac{\spb\one.\two\* \spbab3.\one.\two.3}{s_{3456}}
    + \spb\one.3\* \spb3.\two
    \Big)
 \nn \\ &&
 - \frac{\mt\* \spa\one.\two\* \spab6.(3+4).5\* \spb3.5^4}{\spb3.4\* \spb4.5\* \spb5.6\* \spbab5.(3+4).\one.3\* s_{345}\* s_{3456}}
 - \frac{\mt\* \spa4.6^4\* \spb\one.\two\* \spab4.(5+6).3}{\spa3.4\* \spa4.5\* \spa5.6\* \spaba6.\two.(5+6).4\* s_{456}\* s_{3456}}
\,  .
\eeqn
 \beqn
&& -i \, \Asix(\one_q,3_g^+,4_g^-,5_g^-,6_g^+,\two_{\qb}) = 
 \nn \\ &&
 -\frac{\mt^2\* \spa4.5\* \spab4.\one.3\* \spab4.(\one+3).6\* \spb\one.3\* \spaba4.(\one+3).(4+5).\two}{
  \spa3.4\* \spab3.\one.3\* \spab6.\two.6\* \spbab5.(3+4).\one.3\* \spaba6.\two.(5+6).4\* (s_{134}-\mt^2)} 
 \nn \\ &&
 - \frac{\mt^3\* \spab4.\one.3 
 \* \Big(
   \spa\one.4\* \spaba4.(\one+3).(4+5).\two\* \spa4.5\* \spb3.6
  + \spa4.5^2\* \spab4.(\one+3).5\* \spb\one.3\* \spb6.\two
  \Big)}{
  \spa3.4\* \spab3.\one.3\* \spab6.\two.6\* \spbab5.(3+4).\one.3\* \spaba6.\two.(5+6).4\* (s_{134}-\mt^2)}
 \nn \\ &&
 - \frac{\mt^4\* \spa\one.4\* \spb6.\two\* \spa4.5^2\* \spb3.5\* \spab4.\one.3}{\spa3.4\* \spab3.\one.3\* \spab6.\two.6\* \spbab5.(3+4).\one.3\* \spaba6.\two.(5+6).4\* (s_{134}-\mt^2)}
 \nn \\ &&
 + \frac{\mt\* \spab4.\one.3\* \spab4.(\one+3).6\* \Big(
   \spa\one.4\* \spaba4.(\one+3).(4+5).\two\* \spb3.6
  + \spa4.5\* \spab4.(\one+3).5\* \spb\one.3\* \spb6.\two
  \Big)}{\spa3.4\* \spb5.6\* \spab3.\one.3\* \spbab5.(3+4).\one.3\* \spaba6.\two.(5+6).4\* (s_{134}-\mt^2)}
 \nn \\ &&
 - \frac{\mt^2\* \spa\one.4\* \spb6.\two\* \spa4.5\* \spab4.\one.3}{
  \spa3.4\* \spab3.\one.3\* \spbab5.(3+4).\one.3\* \spaba6.\two.(5+6).4\* (s_{134}-\mt^2)} \* \Big(
   \spab4.\one.3
  - \frac{\spab4.(\one+3).6\* \spb3.5}{\spb5.6}
  \Big)
 \nn \\ &&
 + \frac{\spab4.\one.3\* \spab4.(\one+3).6\* \Big(
    \spa\one.4\* \spb6.\two\* \spab4.\one.3\* \spab6.\two.6
  + \spab4.(\one+3).6\* \spb\one.3\* \spaba4.(\one+3).(4+5).\two
  \Big)}{
  \spa3.4\* \spb5.6\* \spab3.\one.3\* \spbab5.(3+4).\one.3\* \spaba6.\two.(5+6).4\* (s_{134}-\mt^2)}  
 \nn \\ &&
 + \frac{\mt\* \spa4.5^3}{\spa5.6\* \spab3.\one.3\* \spaba6.\two.(5+6).4\* s_{456}}\* \Big(
   \frac{\spab4.\one.3\* \spb\one.\two}{\spa3.4}
  - \spb\one.3\* \spb3.\two
  \Big)
 - \frac{\mt\* \spa\one.\two\* \spb3.6^3}{\spb3.4\* \spb4.5\* \spb5.6\* \spab6.\two.6\* s_{3456}}
 \nn \\ &&
 - \frac{\spa\one.\two\* \spbabab3.(4+5).\two.\one.(4+5).3^2}{
   \spb3.4\* \spb4.5\* \spab6.\two.6\* \spab6.(3+4).5\* \spab6.(4+5).3\* \mt\* s_{345}\* s_{3456}}
 \nn \\ &&
 - \frac{\spa\one.\two\* \spbab3.\one.(4+5).3\* \spbab5.\two.(4+5).3\* \spbabab3.(4+5).\two.\one.(4+5).3}{
   \spb3.4\* \spb4.5\* \spab6.\two.6\* \spab6.(3+4).5\* \spab6.(4+5).3\* \spbab5.(3+4).\one.3\* \mt\* s_{3456}}
 \nn \\ &&
 + \frac{\mt\* \spa\one.\two\* \spb3.5\* \spb3.6\*  \Big(
   \spb3.5\* \spbab6.\two.(4+5).3\* s_{3456}
  - s_{345}\* \spb3.6\* \spbab5.\two.(4+5).3
  \Big)}{
   \spb3.4\* \spb4.5\* \spb5.6\* \spab6.\two.6\* \spab6.(3+4).5\* \spbab5.(3+4).\one.3\* s_{3456}}
 \nn \\ &&
 - \frac{(\spa\one.4\* \spb3.4+\spa\one.5\* \spb3.5)\* (\spa\two.4\* \spb3.4+\spa\two.5\* \spb3.5)\* \spbab3.\one.(4+5).3\* \spbab6.\two.(4+5).3
  }{\spb3.4\* \spb4.5\* \spab6.\two.6\* \spab6.(4+5).3\* \spbab5.(3+4).\one.3\* \mt\* s_{345}}
 \nn \\ &&
 + \frac{\mt^2\* \spa4.5\* \spab4.\one.3}{
   \spab6.\two.6\* \spbab5.(3+4).\one.3\* \spaba6.\two.(5+6).4\* (s_{134}-\mt^2)}
   \* \Big(
   \spb\one.3\* \spa4.\two\* \spb3.6  - \frac{\spb\one.6\* \spa4.\two\* \spab4.\one.3}{\spa3.4}
   \Big)
 \nn \\ &&
 + \frac{\spa4.5^3\* \spab6.\two.3}{
   \spa5.6\* \spab6.(4+5).3\* \spaba6.\two.(5+6).4\* \mt\* s_{456}}\*  \Big(
   \frac{\spb\one.\two\* \spbab3.\one.\two.3}{s_{3456}}
  + \spb\one.3\* \spb3.\two
  \Big)
 \nn \\ &&
 + \frac{\spab4.\one.3\* \spab4.(\one+3).6}{\spa3.4\* \spb5.6\* \spbab5.(3+4).\one.3\* \spaba6.\two.(5+6).4\* (s_{134}-\mt^2)}
   \* \Big(\spb\one.6\* \spa4.\two\* \spab4.\one.3 + \spb\one.3\* \spa4.\two\* \spa4.3\* \spb3.6 \Big)
 \nn \\ &&
 - \frac{\mt\* \spa4.5^3\* \spb\one.\two\* \spab4.(5+6).3}{\spa3.4\* \spa5.6\* \spaba6.\two.(5+6).4\* s_{456}\* s_{3456}}
\,  .
\label{eq:g3pg4mg5mg6p}
\eeqn
 
We note that both $\Asix(\one_Q,3_g^-,4_g^-,5_g^+,6_g^+,\two_{\Qb})$
and $\Asix(\one_Q,3_g^+,4_g^-,5_g^+,6_g^-,\two_{\Qb})$ are symmetric
under the relation,
$1 \leftrightarrow 2$, $3 \leftrightarrow 6$, $4 \leftrightarrow 5$,
$\langle\rangle \leftrightarrow []$.  Similarly,
$\Asix(\one_Q,3_g^+,4_g^-,5_g^-,6_g^+,\two_{\Qb})$ is anti-symmetric under
the relation,
$1 \leftrightarrow 2$, $3 \leftrightarrow 6$, $4 \leftrightarrow 5$.
These relations can be understood from the charge conjugation properties of
these color-ordered amplitudes. 

\subsubsection{Rules for obtaining remaining amplitudes}
Amplitudes with opposite gluon helicities are related by complex conjugation,
\beq
\Asix(\one^{-I},3_g^{-h_3},4_g^{-h_4},5_g^{-h_5},6_g^{-h_6},\two^{-J}) = 
 - \Asix(\one^{I},3_g^{h_3},4_g^{h_4},5_g^{h_5},6_g^{h_6},\two^{J}) |_{\spa{}.{} \leftrightarrow \spb{}.{}}
\eeq
In addition we have line reversal,
\beq
\Asix(\one^{-I},3_g^{h_6},4_g^{h_5},5_g^{h_4},6_g^{h_3},\two^{-J}) = 
 - \Asix(\two^{J},6_g^{h_3},5_g^{h_4},4_g^{h_5},3_g^{h_6},\one^{I}) 
\label{eq:linereversal}
\eeq
Starting from gluon helicities $(-,+,+,+)$ and $(+,-,+,+)$ this allows us to
compute $(+,+,+,-)$ and $(+,+,-,+)$.

We note that we could have used
the 6-point BCJ relations~\cite{Bern:2008qj} to reduce
the number of helicity combinations that we have to compute.  In our
labeling, but suppressing gluon subscripts, the simplest relation is,
\beqn
\label{eq:BCJ1}
&&\Asix(\one_Q,3,4,5,6,\two_{\Qb})=\Big(
   \Asix(\one_Q,4,3,5,6,\two_{\Qb})(s_{23}+s_{36}+s_{35}-m^2) \\ && \quad
 + \Asix(\one_Q,4,5,3,6,\two_{\Qb})(s_{23}+s_{36}-m^2)
 + \Asix(\one_Q,4,5,6,3,\two_{\Qb})(s_{23}-m^2) \Big) / (s_{13}-m^2) \nn
\eeqn
In this equation the position of gluon 4 on the right-hand side is
fixed, immediately following the quark 1. This allows the helicities $(+,-,+,+)$
and $(-,+,-,-)$ to be obtained from $(-,+,+,+)$ and $(+,-,-,-)$ respectively.
By using Eq.~\eqref{eq:linereversal} the same relation allows the combinations $(+,+,-,+)$
and $(-,-,+,-)$ to be determined. 

In similar fashion, Eq.~\eqref{eq:BCJ1} could be used to obtain the
amplitude $(-,+,+,-)$ from the results for helicities $(+,+,-,-)$ and $(+,-,+,-)$, and
similarly for $(+,-,-,+)$.  We have chosen to instead compute this amplitude directly,
c.f. Eq.~\eqref{eq:g3pg4mg5mg6p}.  We also note that there are further BCJ relations, for example,
\beqn
\Asix(\one_Q,3,4,5,6,\two_{\Qb})&=-&\Big(
   \Asix(\one_Q,5,3,4,6,\two_{\Qb}) s_{35} (s_{24}+s_{46}-m^2)
 + \Asix(\one_Q,5,3,6,4,\two_{\Qb}) s_{35} (s_{24}-m^2) \nn \\ && \quad
 + \Asix(\one_Q,5,6,3,4,\two_{\Qb}) (s_{35}+s_{36}) (s_{24}-m^2) \nn \\ && \quad
 + \Asix(\one_Q,5,4,3,6,\two_{\Qb}) (s_{23}+s_{36}-m^2) (s_{134}+s_{45}-m^2) \nn \\ && \quad
 + \Asix(\one_Q,5,4,6,3,\two_{\Qb}) (s_{23}-m^2) (s_{134}+s_{45}-m^2) \nn \\ && \quad
 + \Asix(\one_Q,5,6,4,3,\two_{\Qb}) (s_{23}-m^2) (s_{134}+s_{45}+s_{46}-m^2) \nn \\ && \quad
 \Big) / \left[ (s_{13}-m^2) (s_{134}-m^2) \right]\, .  \label{eq:BCJ2}
\eeqn
In this equation gluon 5 always
appears immediately after quark 1 on the right-hand side, so it could also be used to obtain $(+,+,-,+)$
and $(-,-,+,-)$ directly.  We find it simpler and more efficient to simply present a complete set
of helicities without appealing to the BCJ relations.  However we have checked that they are satisfied
by the analytic formulae given above.

\subsection{Two quark pair + two gluon amplitudes}
\subsubsection{Color structure for two quark pairs and two gluons}

The case of two
quark pairs and two gluons is the most complicated set of amplitudes that we calculate.
The expectation from Table~\ref{KKtable} is that we there will be 
12 primitive amplitudes, reducing to six after imposition of BCJ relations.
The possible color structures are the following,
\beqn
  &&\cA_6(\one_Q, \two_{\Qb},3_q,4_{\qb},5_g,6_g) = g^4\nn \\
  &\times& \Big((t^{C_5} t^{C_6} )_{x_1 x_4}  \delta_{x_3 x_2} A^{(1)}
  +\delta_{x_1 x_4} (t^{C_5} t^{C_6})_{x_3 x_2}  A^{(2)}
  +(t^{C_5})_{x_1 x_4}  \, (t^{C_6} )_{x_3 x_2} A^{(3)}\nn \\
  &+&\frac{1}{N}(t^{C_5}\, t^{C_6})_{x_1 x_2}  \delta_{x_3 x_4} A^{(4)}
  +\frac{1}{N}  (t^{C_5}\, t^{C_6})_{x_3 x_4}  \delta_{x_1 x_2} A^{(5)}
+ \frac{1}{N}(t^{C_5}\,  )_{x_1 x_2}  (t^{C_6} )_{x_3 x_4} A^{(6)}\nn \\
  &+&(t^{C_6} t^{C_5} )_{x_1 x_4}  \delta_{x_3 x_2} B^{(1)}
    +\delta_{x_1 x_4} (t^{C_6} t^{C_5})_{x_3 x_2}  B^{(2)}
  +(t^{C_6})_{x_1 x_4}  \, (t^{C_5} )_{x_3 x_2} B^{(3)}\nn \\
  &+&\frac{1}{N}(t^{C_6}\, t^{C_5})_{x_1 x_2}  \delta_{x_3 x_4} B^{(4)}
      +\frac{1}{N}  (t^{C_6}\, t^{C_5})_{x_3 x_4}  \delta_{x_1 x_2} B^{(5)}
   +\frac{1}{N}(t^{C_6}\,  )_{x_1 x_2}  (t^{C_5} )_{x_3 x_4} B^{(6)}\Big)
\eeqn
We find that, numerically, these are related in analogous fashion to the
one gluon case,
\beq
\sum_{i=1}^6 \left( A^{(i)}+B^{(i)} \right) = 0 \,.
\label{eq:4q2grelation}
\eeq
Squaring the amplitude and summing over colors we have,
\beqn
&&\sum_{colors} |\cA_6|^2= g^8 V\Big[(N^2-1) \* (|A^{(1)}|^2+|A^{(2)}|^2+|A^{(3)}|^2+|B^{(1)}|^2+|B^{(2)}|^2+|B^{(3)}|^2) \nn\\
       	       	                             &+&|A^{(4)}|^2+|A^{(5)}|^2+|A^{(6)}|^2+|B^{(4)}|^2+|B^{(5)}|^2+|B^{(6)}|^2\nn\\
&+&A^{(1)}\*(A^{(6)}+A^{(5)}+A^{(4)}+A^{(2)}+B^{(2)}-B^{(1)})^*+B^{(1)}\*(B^{(6)}+B^{(5)}+B^{(4)}+B^{(2)}+A^{(2)}-A^{(1)})^*\nn\\
&+&A^{(2)}\*(B^{(6)}+B^{(1)}-B^{(2)}+A^{(5)}+A^{(4)}+A^{(1)})^*+B^{(2)}\*(A^{(6)}+A^{(1)}-A^{(2)}+B^{(5)}+B^{(4)}+B^{(1)})^*\nn\\
&+&A^{(3)}\*(B^{(6)}+B^{(5)}+B^{(3)}+A^{(6)}+A^{(4)})^*+B^{(3)}\*(A^{(6)}+A^{(5)}+A^{(3)}+B^{(6)}+B^{(4)})^*\nn\\
&+&A^{(4)}\*(A^{(3)}+A^{(2)}+A^{(1)})^*+B^{(4)}\*(B^{(3)}+B^{(2)}+B^{(1)})^*\nn\\
&+&A^{(5)}\*(A^{(2)}+A^{(1)}+B^{(3)})^*+B^{(5)}\*(B^{(2)}+B^{(1)}+A^{(3)})^*\nn\\
&+&A^{(6)}\*(B^{(3)}+B^{(2)}+A^{(3)}+A^{(1)})^*+B^{(6)}\*(A^{(3)}+A^{(2)}+B^{(3)}+B^{(1)})^*\nn\\
&+&\frac{2}{N^2} |A^{(1)}+A^{(2)}+A^{(3)}+B^{(1)}+B^{(2)}+B^{(3)}|^2\nn\\
&-&\frac{1}{N^2} (|A^{(5)}+B^{(5)}-A^{(4)}-B^{(4)}|^2+|A^{(6)}-B^{(6)}|^2)\Big] \,.
\eeqn
We can also use a decomposition in terms of color-ordered amplitudes~\cite{Johansson:2015oia}, similar to
the Melia basis,\footnote{
The actual basis proposed by Melia~\cite{Melia:2013xok} contains the same number of color subamplitudes,
but is one in which the color factor for each subamplitude can be written as a single
term that can be easily read off from a representative Feynman diagram.  This alternative
decomposition is given in Appendix~\ref{sec:Melia}.}
\beqn
  \cA_6(\one_Q, \two_{\Qb},3^{h_3}_q,4^+_{\qb},5^+_g,6^+_g) &=& g^4 \Big(
   A_{125634} \, C_{125634} + A_{125364} \, C_{125364} + A_{125346} \, C_{125346} \nn \\
&&\quad+ A_{123564} \, C_{123564}+ A_{123546} \, C_{123546} + A_{123456} \, C_{123456}
  \Big) \nn \\
&& + (5 \leftrightarrow 6) \,.
\eeqn
The color coefficients are given by,
\beqn
C_{123456} &=& \left( t^{C_6} t^{C_5} t^B \right)_{x_1 x_2} \left( t^B \right)_{x_3 x_4} \nn \\
C_{123546} &=& \left( t^{C_6} t^{C_5} t^B \right)_{x_1 x_2} \left( t^B \right)_{x_3 x_4}
             - \left( t^{C_6} t^B \right)_{x_1 x_2} \left(t^B t^{C_5} \right)_{x_3 x_4} \nn \\
C_{123564} &=& \left( t^{C_6} t^{C_5} t^B \right)_{x_1 x_2} \left( t^B \right)_{x_3 x_4}
             - \left( t^{C_6} t^B \right)_{x_1 x_2} \left(t^B t^{C_5} \right)_{x_3 x_4} \nn \\
           &&  - \left( t^{C_5} t^B \right)_{x_1 x_2} \left(t^B t^{C_6} \right)_{x_3 x_4}
               + t^B_{x_1 x_2} \left (t^B t^{C_5} t^{C_6} \right)_{x_3 x_4} \nn \\ 
C_{125346} &=&\left( t^{C_6} t^{B} t^{C_5} \right)_{x_1 x_2} \left( t^B \right)_{x_3 x_4} \nn \\
C_{125364} &=& \left( t^{C_6} t^B t^{C_5} \right)_{x_1 x_2} \left( t^B \right)_{x_3 x_4}
             - \left( t^B t^{C_5} \right)_{x_1 x_2} \left(t^B t^{C_6} \right)_{x_3 x_4} \nn \\
C_{125634} &=& \left( t^B t^{C_6} t^{C_5} \right)_{x_1 x_2} \left( t^B \right)_{x_3 x_4}
\label{eq:4q2gcolfacs}
\eeqn
and similarly for $5 \leftrightarrow 6$.  Performing the color algebra allows the two
decompositions to be related as follows,
\beqn
A^{(1)}&=&A_{123465},\;\;
A^{(2)}=A_{126534},\;\;
A^{(3)}=A_{126345} \nn \\
A^{(4)}&=&-A_{126534}-A_{126354}-A_{126345}-A_{123654}-A_{123645}-A_{123465} \nn \\
A^{(5)}&=&-A_{123564},\;\;
A^{(6)}=A_{123654}+A_{123645}+A_{125364}+A_{123564} \nn \\
B^{(1)}&=&A_{123456},\;\;
B^{(2)}=A_{125634},\;\;
B^{(3)}=A_{125346} \nn \\
B^{(4)}&=&-A_{125634}-A_{125364}-A_{125346}-A_{123564}-A_{123546}-A_{123456} \nn \\
B^{(5)}&=&-A_{123654},\;\;
B^{(6)}=A_{126354}+A_{123654}+A_{123564}+A_{123546}
\eeqn
In this basis the relationship between the subamplitudes in Eq.~(\ref{eq:4q2grelation})
is demonstrated explicitly.  The existence of this relationship demonstrates
that this basis is overcomplete.  This can be understood by noting
that the color coefficients in Eq.~(\ref{eq:4q2gcolfacs}) are not independent.  By explicitly 
evaluating them we observe that,
\beq
C_{125364}+C_{126354}-C_{123546}-C_{123645} = 0 \,.
\eeq

We could also employ the BCJ basis, in which the quark $3$ always appears
in position three~\cite{Johansson:2015oia}.  The remaining six color subamplitudes
can be expressed in terms of these through the BCJ relations,
\beqn
&0=&
-A_{126354} \spab6.\two.6
   + (s_{16}+s_{46}+s_{56}-m^2) \, A_{123654}
   + (s_{16}+s_{46}-m^2) \, A_{123564}
   + \spab6.\one.6 \, A_{123546}
    \nn
\\
&0=&
-A_{126345} \spab6.\two.6
   + (s_{16}+s_{46}+s_{56}-m^2) \, A_{123645}
   + (s_{16}+s_{56}-m^2) \, A_{123465}
   + \spab6.\one.6 \, A_{123456}
    \nn
\\
&0=&
-A_{126534} \spab6.\two.6 (s_{256}-m^2)
          - \spab6.\one.6 \, (s_{256}+s_{35}-m^2) \, A_{123546}
          - s_{36} \, (s_{15}+s_{45}-m^2) \, A_{123654} \nn \\ && \quad
          - (s_{16}+s_{46}-m^2) \, (s_{35}+s_{256}-m^2) \, A_{123564}
          - \spab5.\one.5 \, (s_{36}+s_{46}) \, A_{123465} \nn \\ && \quad
          - \spab5.\one.5 \, s_{36} \, A_{123645}
          - \spab6.\one.6 \, (s_{35}+s_{45}+s_{256}-m^2) \, A_{123456}   \, .
\eeqn
A further three relations can be obtained from these by interchanging the
gluon labels $5$ and $6$.
We have chosen not to make use of 
the BCJ relations to determine all amplitudes from this smaller set.
However we have checked that they are fulfilled numerically.

\subsubsection{Complete set of amplitudes}

We specify the amplitudes with light quark helicity assignments $(3_q^-, 4_{\bar q}^+)$.  The opposite
helicity amplitudes, $(3_q^+, 4_{\bar q}^-)$ are obtained by complex conjugation. 

Five color-ordered subamplitudes that are used to construct our complete set of amplitudes
are given in sections
~\ref{sec:qagg3456} ($A_{123456}$), 
~\ref{sec:qagg3564} ($A_{123564}$), 
~\ref{sec:qagg5346} ($A_{125346}$), 
~\ref{sec:qagg3546} ($A_{123546}$), 
and~\ref{sec:qagg6354} ($A_{126354}$). 
The amplitude $A_{125634}$ is computed from $A_{123456}$ by performing the operation
$1 \leftrightarrow 2$, $3 \leftrightarrow 4$, $5 \leftrightarrow 6$,
$\langle\rangle \leftrightarrow []$.  Finally, the operation $5 \leftrightarrow 6$ is used
to generate the full set of 12 amplitudes.

\subsubsection{Results for $\Asix(\one_Q,\two_{\Qb},3_q,4_{\qb},5_g,6_g,)$}
\label{sec:qagg3456}
\beqn
&&-i \Asix(\one,\two,3_q^-,4_{\qb}^+,5_g^-,6_g^-)=
-\frac{1}{m\*\spb3.4\*\spb4.5\*\spb5.6}\nn \\
       &\times&\Big[
     \frac{\spb\one.\two}{\spab3.(4+5).6} \* \Big\{
      \frac{\spb4.6\*\spab3.\two.4\*\spbab4.(5+6).\one.4}{\spbab4.(5+6).\one.6}
     -\frac{\spab3.(5+6).4\*\spbab4.\one.\two.4}{s_{3456}}\Big\}
-\frac{\spb\one.4\*\spb4.\two\*\spab6.\one.4}{\spab6.\one.6} \Big]\nn\\
       &-&\frac{m}{s_{34}}\*\Big[\frac{\spb\one.\two\*\spa3.5}{\spabab5.(3+4).(\one+\two).\one.6}
    \* \Big\{\frac{\spa3.5^2\*\spb3.4\*s_{3456}}{s_{345}\*\spab3.(4+5).6}
        -\frac{\spa5.6\*\spbab4.(3+5).\two.4}{\spb4.5\* \spab6.\one.6}
        -\frac{\spa5.6\*\spb3.4\*\spb4.6\*\spab3.\two.4}{
	   \spb4.5\*\spbab4.(5+6).\one.6} \Big\}\nn\\
       &-&\frac{\spb\one.\two\*\spa5.6\*\spab3.\two.4\*\spab5.(\one+6).4}{
	 \spab6.\one.6\*\spbab4.(5+6).\one.6\*(s_{156}-m^2)}
        - \frac{\spb\one.4\*\spb4.\two\*\spa3.4\*\spa5.6}{
	 \spb5.6\*\spab6.\one.6\*(s_{156}-m^2)}\Big]
\,  .
\eeqn
 \beqn
&&-i \Asix(\one,\two,3_q^-,4_{\qb}^+,5_g^+,6_g^+)=
\frac{1}{m\*\spa4.5\*\spa5.6\*\spab6.\one.6\*s_{34}}\nn \\
  &\times&\Big[
  (\spa\one.\two\* \spab3.\two.4 +\spa\one.3\* \spa3.\two\* \spb3.4) \*
   \frac{\big(\spabab4.\one.(5+6).\one.6+\spab4.\one.6\*s_{56}\big)}{(s_{156}-m^2)} \nn\\
& -& 
 \frac{m^2 \* \Big(
 \spa\one.\two\*\spa5.6\*\spab6.\one.6\*\spab3.(4+5).6^2\*\spab4.(3+5).6\*\spb3.4 \Big)}{
 \spabab5.(3+4).(\one+\two).\one.6\*s_{345}\*s_{3456}}\nn \\
 &+&\spa\one.\two\*(\spa4.5\*\spabab3.\two.(5+6).\one.6 +\spa4.3\* \spab3.\two.3\*\spab5.\one.6 )
 \* \frac{\spbab4.(5+6).\one.6}{\spabab5.(3+4).(\one+\two).\one.6}
 \Big]
\,  .
\eeqn
 \beqn
&& -i \Asix(\one,\two,3_q^-,4_{\qb}^+,5_g^-,6_g^+)=
\frac{1}{m\* \spb3.4\* \spa3.4\* \spa5.6} \nn\\
    &\times&\Big[
       \frac{\big(\spb\one.\two\* \spab5.\one.6 +\spb\one.6\* \spb6.\two\* \spa5.6\big)\* \spa3.5\* \spab5.\two.4\* \spab5.\one.6}{
      \spab6.\one.6\* \spabab5.(3+4).(\one+\two).\one.6} \nn\\
       &-& \frac{\big(\spb\one.\two\* \spab5.\one.6 +\spb\one.6\* \spb6.\two\* \spa5.6\big)
        \* \spa3.5\* \spab3.\two.4\* \spab5.\one.6^2\* \spb3.4}{
       \spab6.\one.6\* \spabab5.(3+4).(\one+\two).\one.6\* \spbab4.(5+6).\one.6}\nn\\
       &+&\frac{\spab5.\one.6\* \spab5.(\one+6).4}{\spab6.\one.6\* \spbab4.(5+6).\one.6\* (s_{156}-m^2)}\nn\\
         &\times& \Big((\spb\one.\two\* \spab5.\one.6+\spb\one.6\* \spb6.\two\* \spa5.6)\* \spab3.\two.4
         +\spa3.4\* (\spa5.6\* \spb\one.6\* \spb4.\two\* \spb4.6-\spb\one.4\* \spb4.\two\* \spab5.\one.6)\Big)\Big]\nn\\
   &+&\frac{\spb4.6^3}{m\* \spb3.4\* \spb4.5\* \spb5.6}\* \Big[
        (\spa\one.\two\* \spab3.\two.4+\spa\one.3\* \spa3.\two\* \spb3.4)
	  \* \frac{\spab4.\one.6\* \spb4.6+\spab5.\one.6\* \spb5.6}{\spab3.(4+5).6\* \spbab6.\one.(5+6).4\* s_{456}}\nn\\
       &-&  \frac{\spa\one.\two\* \spab3.\two.6}{\spab3.(4+5).6\* s_{3456}}
       -\frac{\spa\one.\two\* \spab3.\two.3}{s_{3456}\* s_{456}}\Big]\nn\\
     &+& \frac{\spa3.5^3 \* \spbab6.\one.\two.6}{\spa3.4\* \spa4.5\* \spab3.(4+5).6\* s_{345}\* s_{3456}\* \spabab5.(3+4).(\one+\two).\one.6} \nn\\
       &\times& \Big[(\spa\one.5\* \spb6.\two+\spb\one.6\* \spa5.\two)\* \spab4.(3+5).6
          -(\spa\one.4\* \spb6.\two+\spb\one.6\* \spa4.\two)\* \spab5.(3+4).6\Big]
\,  .
\eeqn
 \beqn
&& -i \Asix(\one,\two,3_q^-,4_{\qb}^+,5_g^+,6_g^-)=
\spa3.4\*\spb3.5\*(\spab6.\one.4\*\spab6.(3+4).5-\spab6.\one.5\*\spab6.(3+5).4)\nn\\
 &&\* \frac{(
   (\spa\one.6\*\spb5.\two+\spb\one.5\*\spa6.\two)\*\spab6.(3+5).4
  -(\spa\one.6\*\spb4.\two+\spb\one.4\*\spa6.\two)\*\spab6.(3+4).5)}{
   \spab6.(4+5).3\* \spabab6.\one.(\one+\two).(3+4).5\*s_{345}\*s_{3456}\*s_{34}}\nn \\
&+&\frac{1}{m \* s_{34}} \* \Big[
  \frac{\spa3.6\*\spa4.6\*\spb\one.\two\*\spb3.4\*\spaba3.\two.(4+5).6}{\spa4.5\*\spa5.6\*s_{456}\*s_{3456}}
\nn \\
&-&\frac{(\spa\one.\two\*\spab6.\one.5+\spa\one.6\*\spa6.\two\*\spb5.6) \* \spab3.\two.5\* \spab6.\one.5\*\spb4.5}{
  \spab6.\one.6\*\spabab6.\one.(\one+\two).(3+4).5\*\spb5.6}
 \nn\\
&+&\frac{(\spa\one.\two\*\spab6.\one.5+\spa\one.6\*\spa6.\two\*\spb5.6 )\* \spab6.\one.5\*\spb4.5}{
  \spab6.\one.6\*\spabab6.\one.(\one+\two).(3+4).5\*\spb5.6\*\spaba4.(5+6).\one.6}
 \* \spa3.4\*\spab3.\two.3\*\spab6.\one.5
\nn\\
&-&  \big((\spa\one.\two\*\spab3.\two.4+\spa\one.3\*\spa3.\two\*\spb3.4)\*\spab6.\one.5 
  -\spa\one.6\*\spa3.\two\*\spa3.6\*\spb3.4\*\spb5.6+\spa\one.6\*\spa6.\two\*\spab3.\two.4\*\spb5.6\big) \nn\\
&\times& \frac{\spab6.\one.5\*\spab4.(\one+6).5}{\spb5.6\*\spab6.\one.6\*\spaba4.(5+6).\one.6\*(s_{156}-m^2)}
 \nn\\
&+&\frac{\spa3.6\*\spa4.6\*\spb\one.\two\*\spb3.4\*\spaba6.\two.(4+5).6}{
  \spa4.5\*\spa5.6\*\spab6.(4+5).3\*s_{3456}}
 \nn\\
  &-&\frac{\spba\one.(4+5).6\*\spba\two.(4+5).6\*\spa3.4\*\spb3.4\*\spa4.6\*\spaba6.\one.(4+5).6}{
 \spa4.5\*\spa5.6\*s_{456}\*\spab6.(4+5).3\*\spaba6.\one.(5+6).4}
 \nn\\
&-&\frac{\spb\one.\two\*\spa4.6^2\*\spb3.4\*\spaba3.\two.(4+5).6\*\spaba6.\one.(4+5).6}{
 \spa4.5\*\spa5.6\*s_{456}\*\spab6.(4+5).3\*\spaba6.\one.(5+6).4}
 \Big]
\,  .
\eeqn
 
\subsubsection{Results for $\Asix(\one_Q,\two_{\Qb},3_q,5_g,6_g,4_{\qb})$}
\label{sec:qagg3564}
\beqn
-i \Asix(\one,\two,3_q^-,5_g^-,6_g^-,4_{\qb}^+)&=&
-\frac{1}{m \* \spb3.5 \* \spb4.6 \* \spb5.6}  \*  \Big(
   \frac{\spb\one.\two \* \spbab4.\one.\two.4}{s_{3456}}
  +\spb\one.4 \* \spb4.\two \Big)
\,  .
\eeqn
 \beqn
-i \Asix(\one,\two,3_q^-,5_g^+,6_g^+,4_{\qb}^+)&=&
\frac{1}{m \* \spa3.5 \* \spa4.6 \* \spa5.6}  \*  \Big(
  \frac{\spa\one.\two \* \spaba3.\one.\two.3}{s_{3456}}
 +\spa\one.3 \* \spa3.\two \Big)
\,  .
\eeqn
 \beqn
&&-i \Asix(\one,\two,3_q^-,5_g^-,6_g^+,4_{\qb}^+)=
\frac{1}{m \* \spab4.(5+6).3}  \*  \Big[
 \nonumber \\ &&
         \frac{\spb3.6}{ \spb3.5 \* \spb5.6\* s_{356}}  \*  \Big(
            \frac{\spa\one.\two \* \spab4.(3+5).6 \* \spbab4.\one.(3+5).6}{s_{3456}}
	     + \spab\one.(3+5).6 \* \spab\two.(3+5).6 \Big)
 \nonumber \\ &&
       - \frac{\spa4.5}{\spa4.6 \* \spa5.6\* s_{456}}  \*  \Big(
          - \frac{\spb\one.\two \* \spaba5.(4+6).\one.3 \* \spab5.(4+6).3}{s_{3456}}
	   + \spab5.(4+6).\one \* \spab5.(4+6).\two \Big)
 \nonumber \\ &&
       + \frac{1}{s_{3456}} \* \Big( \frac{\spa\one.\two \* \spb3.6 \* \spbab6.\one.(3+5).6}{\spb3.5 \* \spb5.6}
         - \frac{\spb\one.\two \* \spa4.5 \* \spaba5.\one.(4+6).5}{ \spa4.6 \* \spa5.6} \Big)
        \Big]
\,  .
\eeqn
 \beqn
-i \Asix(\one,\two,3_q^-,5_g^+,6_g^-,4_{\qb}^+)&=&
\frac{\spa3.6^3}{m \* \spa3.5 \* \spa5.6 \* \spab3.(5+6).4\* s_{356}}  \*  \Big(
      \frac{\spb\one.\two \* \spbab4.\one.\two.4}{s_{3456}}
    + \spb\one.4 \* \spb4.\two \Big)
 \nonumber \\ &&
 -\frac{\spb4.5^3}{m \* \spb4.6 \* \spb5.6 \* \spab3.(5+6).4\* s_{456}}  \*  \Big(
      \frac{\spa\one.\two \* \spaba3.\one.\two.3}{s_{3456}}
    + \spa\one.3 \* \spa3.\two \Big)
\,  .
\eeqn
 
We note that the $(5^-,6^-)$ amplitude can be obtained from the
$(5^+,6^+)$ one by the operation,
$1 \leftrightarrow 2$, $3 \leftrightarrow 4$, $5 \leftrightarrow 6$,
$\langle\rangle \leftrightarrow []$.  On the other hand, the
$(5^-,6^+)$ and $(5^+,6^-)$ amplitudes are symmetric under this operation,
although this is not manifest in the forms given above.
This can be understood from the charge conjugation properties of this
color-ordered amplitude. 

\subsubsection{Results for $\Asix(\one_Q,\two_{\Qb},5_g,3_q,4_{\qb},6_g)$}
\label{sec:qagg5346}
\beqn
&& -i \Asix(\one,\two,5_g^-,3_q^-,4_{\qb}^+,6_g^-)=
-\frac{\spb\one.4\* \spb4.\two\* \spab5.\two.3\* \spab6.\one.4}{m\* \spb3.4\* \spb3.5\* \spb4.6\* \spab5.\two.5\* \spab6.\one.6} \nn \\
&+& \frac{\spb\one.\two}{m\* \spb3.4\* \spb3.5\* \spb4.6} \* \Big[
         -\frac{\spab5.\two.3\* \spab6.\one.4\* \spabab5.(\two+3).(3+5).\two.4}{\spab5.\two.5\* \spab6.\one.6\* \spaba4.(3+5).\two.5} \nn \\
&+& \frac{\spab5.\two.3\* \spaba5.\two.(3+5).6}{\spab5.\two.5\* \spabab5.\two.(\one+\two).(4+6).3\* \spaba4.(3+5).\two.5} \nn\\
         &\times& \big( \spb3.4\* \spabab5.(\two+3).(3+5).\two.4
           -\spb4.6 \* (\spa6.5\* \spb5.3\* \spab5.\two.4+\spab5.\two.3\* \spab6.\two.4)\big)\nn \\
&-& \frac{m^2\* s_{35}}{s_{345}}
         \* \frac{\spb3.4\* \spa3.5\* \spa4.5\* \spb4.6\* \spaba5.(\one+\two).\one.6}{\spa3.4\* \spab5.(3+4).6\* \spab6.\one.6\* \spaba4.(3+5).\two.5} \Big] \nn \\
&-& \frac{m\* \spb\one.\two\* \spab5.(3+6).4^3\* \spab5.(4+6).3}{s_{346}\* s_{3456}\* \spb3.4\* \spb4.6\* \spab5.(3+4).6\* \spabab5.\two.(\one+\two).(4+6).3}
\,  .
\eeqn
\beqn
&& -i \Asix(\one,\two,5_g^+,3_q^-,4_{\qb}^+,6_g^+)=
 \nonumber \\ &&
   \frac{\spa\one.\two\* \spb4.6\* \spab3.\two.5^3\* \spaba3.(4+6).\one.4}{
     \spa3.4\* \spa3.5\* \spa4.6\* \spab5.\two.5\* \spabab3.(4+6).(\one+\two).\two.5\* \spbab4.(3+5).\two.5\* \mt}
 \nonumber \\ &&
 + \frac{\spa\one.3\* \spa3.\two\* \spab3.\two.5\* \spab4.\one.6}{
    \spa3.4\* \spa3.5\* \spa4.6\* \spab5.\two.5\* \spab6.\one.6\* \mt}
 - \frac{\spa\one.\two\* \spab4.\one.6\* \spab3.\two.5^2\* \spab3.(\two+5).4}{
    \spa3.4\* \spa3.5\* \spa4.6\* \spab5.\two.5\* \spab6.\one.6\* \spbab4.(3+5).\two.5\* \mt}
 \nonumber \\ &&
 - \frac{\mt\* \spa\one.\two\* \spb4.5^3\* \spbab5.(\one+\two).\one.6}{
    \spb3.4\* \spab6.\one.6\* \spab6.(3+4).5\* \spbab4.(3+5).\two.5\* s_{345}}
 \nonumber \\ &&
 + \frac{\mt\* \spab3.(4+6).5^2\* \spa\one.\two\* \spab3.(4+6).5\* \spab4.(3+6).5}{
    \spa3.4\* \spa4.6\* \spab6.(3+4).5\* \spabab3.(4+6).(\one+\two).\two.5\* s_{346}\* s_{3456}}
\,  .
\eeqn
 \beqn
&& -i \Asix(\one,\two,5_g^-,3_q^-,4_{\qb}^+,6_g^+)= \nonumber \\ &&
   \frac{\spab4.\one.6}{\spa4.6\* \spab5.\two.5\* \spab6.\one.6\* \spaba4.(3+5).\two.5\* \mt}  \* \Big(
      \Big(\spa\one.\two\* \spab5.\two.3 - \spa\one.5\* \spa5.\two\* \spb3.5 \Big) \*
      \frac{\Big((s_{235}-\mt^2)\* \spab5.\two.4-\mt^2\* \spa5.3\* \spb3.4 \Big)}{\spb3.4\* \spb3.5} \nonumber \\ && \qquad \qquad
    -  \spab5.\two.3\* \Big(\spa\one.3\*\spa5.\two\* \spa3.5-\frac{\spa\one.3\*\spa3.\two\* \spab5.\two.3}{\spb3.5} \Big)
    \Big)
 \nonumber \\ &&
 + \frac{(s_{235}-\mt^2)\* \spb4.6\* \spab5.\two.3}{
    \spb3.4\* \spa4.6\* \spab5.\two.5\* \spabab5.\two.(\one+\two).(4+6).3\* \mt} \* \Big(
      \frac{\spa\one.\two\* \spab5.\two.3}{\spb3.5} - \spa\one.5\* \spa5.\two
    \Big)
 \nonumber \\ &&
 + \frac{\spb4.6\* \spab5.\two.3 \*
    \Big(\spa\one.\two\* \spab5.\two.3 - \spa\one.5\* \spa5.\two\* \spb3.5 \Big) \*
      \Big((s_{235}-\mt^2)\* \spab5.\two.6-\mt^2\* \spa5.3\* \spb3.6 \Big)}{
    \spb3.4\* \spb3.5\* \spab5.\two.5\* \spabab5.\two.(\one+\two).(4+6).3\* \spaba4.(3+5).\two.5\* \mt}
 \nonumber \\ &&
 \nonumber \\ &&
 - \frac{\Big(\spa\one.5 \* \spb6.\two+\spb\one.6 \* \spa5.\two\Big) \* \spb3.6 \* \spb4.6 \* \spab5.(3+6).4 \* \spaba5.\one.\two.5}{
    \spb3.4 \* \spab5.(3+4).6 \* \spabab5.\two.(\one+\two).(4+6).3 \* s_{346} \* s_{3456}}
 - \frac{\spa3.5^2 \* \spa4.5 \* \spab5.\two.6
      \* \Big( \spb\one.\two \* \spbab6.\one.\two.6 + \spb\one.6 \* \spb6.\two \* s_{3456} \Big)}{
    \spa3.4 \* \spab5.(3+4).6 \* \spaba4.(3+5).\two.5 \* m \* s_{345} \* s_{3456}}
 \nonumber \\ &&
 + \frac{(\spa\one.5 \* \spb4.\two+\spb\one.4 \* \spa5.\two) \* \spb3.6 \* \spb4.6 \* \spaba5.\one.\two.5}{
    \spb3.4 \* \spabab5.\two.(\one+\two).(4+6).3 \* s_{346} \* s_{3456}} 
 + \frac{\spab4.\one.6\*  \Big(
      (\spa\one.3\* \spa5.\two + \spa\one.5\* \spa3.\two )\* \spab5.\two.3
    +  \spa\one.5\* \spa5.\two\* \spab5.(\two+3).5 \Big)}{
    \spa4.6\* \spb3.5\* \spab6.\one.6\* \spaba4.(3+5).\two.5\* \mt}
 \nonumber \\ &&
 + \frac{m \* \spa3.5^2 \* \spa4.5 \* (\spb\one.\two \* \spab4.\one.6+\spb\one.6 \* \spb6.\two \* \spa4.6)}{
    \spa3.4 \* \spa4.6 \* \spab6.\one.6 \* \spaba4.(3+5).\two.5 \* s_{345}}
 - \frac{\mt\* \spa3.5^2\* \spa4.5\* \spb\one.\two\* \spab4.(3+5).6}{
     \spa3.4\* \spa4.6\* \spaba4.(3+5).\two.5\* s_{345}\* s_{3456}}
\,  .
\eeqn
 \beqn
&& -i \Asix(\one,\two,5_g^+,3_q^-,4_{\qb}^+,6_g^-)=
 \nonumber \\ &&
   \frac{\spab3.\two.5\* \spab6.\one.4}{\spb4.6\* \spab5.\two.5\* \spab6.\one.6\* \spbab4.(3+5).\two.5\* \mt} \* \Big(
      \frac{\spab3.(\two+5).4\* \spab3.\two.5\* \spb\one.\two}{\spa3.4\* \spa3.5}
    - \frac{\spab3.(\two+5).4\* \spb\one.5\* \spb5.\two}{\spa3.4} \nonumber \\ && \qquad \qquad
    - \frac{\spab3.\two.5\* \spb\one.4\* \spb4.\two}{\spa3.5}
    + \spb\one.4\* \spb5.\two\* \spb4.5
    \Big)
 \nonumber \\ &&
 + \frac{\spa3.6\* \spab3.\two.5\* \spab3.(\two+5).4\* \Big( \spab3.\two.5\* \spb\one.\two-\spa3.5\* \spb\one.5\* \spb5.\two \Big)}{\spa3.4\* \spa3.5\* \spb4.6\* \spab5.\two.5\* \spabab3.(4+6).(\one+\two).\two.5\* \mt}
 \nonumber \\ &&
 + \frac{\spa3.6\* \spab3.\two.5 \* \Big( \spab3.\two.5\* \spab6.\two.4 -\spab6.\two.5\*\spa3.5\* \spb4.5 \Big)
   \* \Big( \spab3.\two.5\* \spb\one.\two - \spa3.5\* \spb\one.5\* \spb5.\two \Big)}{
   \spa3.4\* \spa3.5\* \spab5.\two.5\* \spabab3.(4+6).(\one+\two).\two.5\* \spbab4.(3+5).\two.5\* \mt} 
 \nonumber \\ &&
 - \frac{\spa3.6^3\* \spbab5.\one.\two.5\* \Big(\spb\one.5 \* \spba5.(3+4+6).\two -\spab\one.(3+4+6).5\* \spb5.\two\Big)}{\spa3.4\* \spab6.(3+4).5\* \spabab3.(4+6).(\one+\two).\two.5\* s_{346}\* s_{3456}}
 \nonumber \\ &&
 + \frac{\mt\* \spb4.5^3\* \Big(\spa\one.\two\* \spab6.\one.4+\spa\one.6\* \spa6.\two\* \spb4.6 \Big)}{\spb3.4\* \spb4.6\*
         \spab6.\one.6\* \spbab5.\two.(3+5).4\* s_{345}}
 - \frac{\spa\one.6\* \spa6.\two\* \spb4.5^3\* \spab6.\two.5}{\spb3.4\* \spab6.(3+4).5\* \spbab5.\two.(3+5).4\* \mt\* s_{345}}
 \nonumber \\ &&
 - \frac{\spa\one.\two\* \spb4.5^3\* \spab6.\two.5\* \spaba6.\one.\two.6}{\spb3.4\* \spab6.(3+4).5\* \spbab5.\two.(3+5).4\* \mt\* s_{345}\* s_{3456}}
 - \frac{\mt\* \spa\one.\two\* \spb4.5^3\* \spab6.(3+5).4}{\spb3.4\* \spb4.6\* \spbab5.\two.(3+5).4\* s_{345}\* s_{3456}}
\,  .
\eeqn

We note that the $(5^-,6^-)$ amplitude can be obtained from the
$(5^+,6^+)$ one by the operation,
$1 \leftrightarrow 2$, $3 \leftrightarrow 4$, $5 \leftrightarrow 6$,
$\langle\rangle \leftrightarrow []$.  On the other hand, the
$(5^-,6^+)$ and $(5^+,6^-)$ amplitudes are symmetric under this operation,
although this is not manifest in the forms given above.
This can be understood from the charge conjugation properties of this
color-ordered amplitude.

\subsubsection{Results for $\Asix(\one_Q,\two_{\Qb},3_q,5_g,4_{\qb},6_g,)$}
\label{sec:qagg3546}
\beqn
&& -i \Asix(\one,\two,3_q^-,5_g^-,4_{\qb}^+,6_g^-)=
\frac{1}{m\* \spb3.5\* \spb4.5\* \spb4.6}\* \Big[
               \frac{-\spb\one.4\* \spb4.\two\* \spab6.\one.4}{\spab6.\one.6} \nn \\
              &-&\frac{\spb\one.\two\* \spab6.\one.4\* \spbab4.\two.(3+5).4}{\spab6.\one.6\* s_{345}}
              -\frac{\spb\one.\two\* \spab6.(3+5).4\* \spbab4.\two.\one.4}{s_{345}\* s_{3456}}\Big]
\,  .
\eeqn
 \beqn
&& -i \Asix(\one,\two,3_q^-,5_g^-,4_{\qb}^+,6_g^+)=
-\frac{1}{m\* \spb3.5\* \spb4.5}\* \Big[
     \frac{\spa\one.\two\* \spb4.6}{s_{3456}}
     \* \Big(\frac{\spbab6.\two.(3+5).4}{s_{345}}+\frac{\spbab3.\two.(3+5).4}{\spab6.(4+5).3}\Big)\nn\\
&+&\frac{\big(\spab\one.(3+5).4\* \spab\two.(3+5).4-\spa\one.\two\* \spbab4.\two.(3+5).4\big)\* \spbab3.(4+5).\one.6}{\spab6.\one.6\* \spab6.(4+5).3\* s_{345}}\Big] \nn\\
    &-&\frac{\big(\spab5.(4+6).\one\* \spaba\two.(\one+\two).(4+6).5
                 +\spab5.(4+6).\two\* \spaba\one.(\one+\two).(4+6).5\big)}{\spab6.(4+5).3\* \spa4.6\* s_{3456}\* s_{456}}
\,.    
\eeqn
 
The remaining amplitudes are obtained from those in section~\ref{sec:qagg6354} as follows,
\begin{eqnarray}
A(\one_Q,\two_{\Qb},3_q^-,5_g^+,4_{\qb}^+,6_g^+) &=&
 \left. A(\two_{\Qb},\one,6_g^-,4_q^-,5_g^-,3_{\qb}^+) \right|_{\langle\rangle \leftrightarrow []}
\\
A(\one_Q,\two_{\Qb},3_q^-,5_g^+,4_{\qb}^+,6_g^-) &=&
 \left. A(\two_{\Qb},\one,6_g^+,4_q^-,5_g^-,3_{\qb}^+) \right|_{\langle\rangle \leftrightarrow []}
\end{eqnarray}

\subsubsection{Results for $\Asix(\one_Q,\two_{\Qb},6_g,3_q,5_g,4_{\qb})$}
\label{sec:qagg6354}
\beqn
&& -i \Asix(\one,\two,6_g^-,3_q^-,5_g^-,4_{\qb}^+)=
-\frac{1}{m\* \spb3.5\* \spb3.6\* \spb4.5}
    \* \Big[\frac{-\spb\one.4\* \spb4.\two\* \spab6.\two.3}{\spab6.\two.6} \nn \\
    &+&\frac{\spb\one.\two\* \spab6.\two.3\* \spbab4.\one.(3+5).4}{\spab6.\two.6\* s_{345}} 
       +\frac{\spb\one.\two\* \spab6.(3+5).4\* \spbab4.\one.(\one+\two).3}{s_{345}\* s_{3456}}\Big]
\, .
\eeqn
 \beqn
&& -i \Asix(\one,\two,6_g^+,3_q^-,5_g^-,4_{\qb}^+)=
-\frac{1}{m \* \spb3.5 \* \spb4.5 \* \spab6.(3+5).4\* s_{345}}
       \* \Big( \frac{\spa\one.\two \* \spb4.6 \* \spbabab4.(3+5).\one.(\one+\two).(3+5).4}{s_{3456}}
 \nonumber \\ &&
      +\frac{(\spa\one.\two \* \spbab4.\one.(3+5).4
            +\spab\one.(3+5).4 \* \spab\two.(3+5).4) \* \spbab6.\two.(3+5).4}{
       \spab6.\two.6} \Big)
 \nonumber \\ &&
     -\frac{\spa3.5^2 \* \spa4.6}{m \* \spa3.6 \* \spa4.6 \* \spab6.(3+5).4\* s_{356}} \* \Big(
       \spb\one.4 \* \spb4.\two
      +\frac{\spb\one.\two \* \spbab4.\one.\two.4}{s_{3456}} \Big)
\,  .
\eeqn
 
The remaining amplitudes are obtained from those in section~\ref{sec:qagg3546} as follows,
\begin{eqnarray}
A(\one_Q,\two_{\Qb},6_g^+,3_q^-,5_g^+,4_{\qb}^+) &=&
 \left. A(\two_{\Qb},\one,4_q^-,5_g^-,3_{\qb}^+,6_g^-) \right|_{\langle\rangle \leftrightarrow []}
\\
A(\one_Q,\two_{\Qb},6_g^-,3_q^-,5_g^+,4_{\qb}^+) &=&
 \left. A(\two_{\Qb},\one_Q,4_q^-,5_g^-,3_{\qb}^+,6_g^+) \right|_{\langle\rangle \leftrightarrow []}
\end{eqnarray}

\subsection{Six quark amplitudes}
\subsubsection{Color structure for three quark pairs}
The Feynman diagram evaluation containing six possible color factors is easily reduced
to the Melia basis through commutation relations. One then arrives at the four
possible color structures,
\beqn
  A(\one_Q, \two_{\Qb},3^{h_3}_q,4^{-h_3}_{\qb},5^{h_5}_{q^\prime},6^{-h_5}_{\qb^\prime}) &=& g^4
   \Big( 
   (t^A t^B)_{x_1 x_2} t^A_{x_3 x_4} t^B_{x_5 x_6} A^{(1)}
  +(t^B t^A)_{x_1 x_2} t^A_{x_3 x_4} t^B_{x_5 x_6} A^{(2)} \nn \\
  &+& 
  (t^B t^A)_{x_3 x_4} t^A_{x_5 x_6} t^B_{x_1 x_2} A^{(4)} 
  +(t^A t^B)_{x_5 x_6} t^A_{x_1 x_2} t^B_{x_3 x_4} A^{(5)} 
   \Big)\, .
\eeqn
The color-summed and squared amplitude then takes the form,
\beqn
  && \sum_{\rm colors}|A(\one_Q,\two_{\Qb},3_q,4_{\qb},5_{q^\prime},6_{\qb^\prime})|^2 = g^8 \,
   VN \bigg( 
   |A^{(1)}+A^{(5)}|^2
  +|A^{(2)}+A^{(4)}|^2\nn \\
  && \qquad +\frac{1}{N^2} \Big(
   |A^{(1)}+A^{(2)}|^2 + |A^{(4)}|^2 + |A^{(5)}|^2 - 2\, |A^{(1)}+A^{(2)}+A^{(4)}+A^{(5)}|^2 
   \Big) \bigg) \,.
\eeqn
For identical quarks ($q^\prime = q$) the amplitude can be obtained by forming
the combination (needed when $h_3=h_5$),
\beqn
A(\one_Q, \two_{\Qb},3^{h_3}_q,4^{-h_3}_{\qb},5^{h_5}_{q},6^{-h_5}_{\qb}) &=&
 A(\one_Q, \two_{\Qb},3^{h_3}_q,4^{-h_3}_{\qb},5^{h_5}_{q^\prime},6^{-h_5}_{\qb^\prime}) 
-A(\one_Q, \two_{\Qb},5^{h_3}_q,4^{-h_3}_{\qb},3^{h_5}_{q^\prime},6^{-h_5}_{\qb^\prime}) \,. 
\nn \\
\eeqn
With the shorthand notation $B^{(i)} = A^{(i)} (3 \leftrightarrow 5)$,
the color-summed and squared amplitude for identical quarks is then,
\beqn
 && \sum_{\rm colors}|A(\one_Q,\two_{\Qb},3^{h_3}_q,4^{-h_3}_{\qb},5^{h_5}_{q},6^{-h_5}_{\qb})|^2 = \nn \\
 &&\sum_{\rm colors}|A(\one_Q,\two_{\Qb},3^{h_3}_q,4^{-h_3}_{\qb},5^{h_5}_{q^\prime},6^{-h_5}_{\qb^\prime})|^2
 +\sum_{\rm colors}|A(\one_Q,\two_{\Qb},3^{h_3}_q,4^{-h_3}_{\qb},5^{h_5}_{q^\prime},6^{-h_5}_{\qb^\prime})|^2
  \big(A^{(i)} \rightarrow B^{(i)}\big) \nn \\
&&+\delta_{h_3 h_5} \, g^8 \,
   2V \, \mathbb{Re} \Bigg( 
   A^{(1)} \left( B^{(1)}+B^{(2)}+B^{(4)}\right)^*
  +A^{(2)} \left( B^{(1)}+B^{(2)}+B^{(5)}\right)^*
   \nn \\ && \qquad \qquad \qquad
  +A^{(4)} \left( B^{(1)}+B^{(5)}-B^{(4)}\right)^*
  +A^{(5)} \left( B^{(2)}+B^{(4)}-B^{(5)}\right)^* \nn \\
  && \qquad \qquad -\frac{1}{N^2} \bigg[
   \left(3A^{(1)}+3A^{(2)}+2A^{(4)}+2A^{(5)}\right) \left( B^{(1)}+B^{(2)}\right)^*
   \nn \\ && \qquad \qquad \qquad
  +\left(2A^{(1)}+2A^{(2)}+A^{(4)}+A^{(5)}\right) \left( B^{(4)}+B^{(5)}\right)^*
   \bigg] \Bigg) \,.
\eeqn

\subsubsection{Results for six quark amplitudes}
We present results only for the case of distinct flavors of massless quarks.
The amplitudes for the case of massless quarks of the same flavor are obtained
in an obvious way by imposing Fermi statistics. Detailed results are given above.

All amplitudes can be constructed from the following five, in which
the helicity of quark 3 has been fixed to be negative ($h_3=-$).
The first two correspond to $h_5=-$,
\beqn
&& -i A^{(5)}(\one_Q, \two_{\Qb},3^-_q,4^+_{\qb},5^-_{q^\prime},6^+_{\qb^\prime})=\nn \\
&& \frac{1}{s_{34}\* s_{3456}}\* \Bigg(
         \frac{(\spa4.3\* \spb4.6\* (\spa\one.5\* \spb4.\two+\spb\one.4\* \spa5.\two)
          + \spa6.3\* \spb4.6\* (\spb\one.6\* \spa5.\two+\spa\one.5\* \spb6.\two))}{s_{346}} \nn \\
       &-& \frac{(\spa5.3\* \spb4.3\* (\spb\one.6\* \spa3.\two+\spa\one.3\* \spb6.\two)
          + \spa5.3\* \spb4.5\* (\spb\one.6\* \spa5.\two+\spa\one.5\* \spb6.\two))}{s_{345}}
	  \Bigg)
\,  .
\eeqn
 \beqn
&& -i A^{(1)}(\one_Q, \two_{\Qb},3^-_q,4^+_{\qb},5^-_{q^\prime},6^+_{\qb^\prime})  = \nn \\
&&  \frac{1}{s_{34}\* s_{56}\* s_{3456}} \* \Big[
            \spa3.5\* \spb6.4\* (\spab\one.(3+4).\two+\spab\two.(3+4).\one)
          + \spab3.(5+6).4\* (\spb\one.6\* \spa5.\two+\spa\one.5\* \spb6.\two) \nn \\
     && \qquad \qquad \qquad 
          - \spab5.(3+4).6\* (\spb\one.4\* \spa3.\two+\spa\one.3\* \spb4.\two) \Big] \nn \\
&-&  \frac{(m\* \spa3.5\* \spb\one.4\* \spb6.\two
          + m\* \spb4.6\* \spa\one.3\* \spa5.\two
          + \spab5.(\one+3).4\* \spa\one.3\* \spb6.\two
          + \spab3.(\one+4).6\* \spb\one.4\* \spa5.\two)}{s_{34}\* s_{56}\* (s_{134}-m^2)} \nn \\
&+&  \frac{(\spa5.3\* \spb4.3\* (\spb\one.6\* \spa3.\two+\spa\one.3\* \spb6.\two)
          + \spa5.3\* \spb4.5\* (\spb\one.6\* \spa5.\two+\spa\one.5\* \spb6.\two))}{s_{34}\* s_{345}\* s_{3456}} \nn \\
&-&   \frac{(\spa3.5\* \spb6.3\* (\spb\one.4\* \spa3.\two+\spa\one.3\* \spb4.\two)
          + \spa3.5\* \spb6.5\* (\spa\one.5\* \spb4.\two+\spb\one.4\* \spa5.\two))}{s_{56}\* s_{356}\* s_{3456}}
\,  .
\eeqn
 
The other three have $h_5=+$,
\beqn
&& -i A^{(5)}(\one_Q, \two_{\Qb},3^-_q,4^+_{\qb},5^+_{q^\prime},6^-_{\qb^\prime}) = \nn \\ 
&& \frac{1}{s_{34}\* s_{3456}}\*  \Bigg(
          \frac{\spa3.6\* \spb3.4\* (\spb\one.5\* \spa3.\two+\spa\one.3\* \spb5.\two)
          + \spa3.6\* \spb6.4\* (\spa\one.6\* \spb5.\two+\spb\one.5\* \spa6.\two)}{s_{346}}\nn \\
       &-&  \frac{\spa3.4\* \spb5.4\* (\spa\one.6\* \spb4.\two+\spb\one.4\* \spa6.\two)
          + \spa3.5\* \spb5.4\* (\spa\one.6\* \spb5.\two+\spb\one.5\* \spa6.\two)}{s_{345}}
	   \Bigg)
\,  .
\eeqn
 \beqn
&& -i A^{(1)}(\one_Q, \two_{\Qb},3^-_q,4^+_{\qb},5^+_{q^\prime},6^-_{\qb^\prime}) =
 \nn \\
&&  \frac{1}{s_{34}\* s_{56}\* s_{3456}} \* \Big[
     \spa3.6\* \spb5.4\* (\spab\one.(3+4).\two+\spab\two.(3+4).\one)
          + \spab3.(5+6).4\* (\spa\one.6\* \spb5.\two+\spb\one.5\* \spa6.\two) \nn \\
     && \qquad \qquad \qquad 
          - \spab6.(3+4).5\* (\spa\one.3\* \spb4.\two+\spb\one.4\* \spa3.\two) \Big] \nn \\
&-& \frac{(m\* \spa3.6\* \spb\one.4\* \spb5.\two
          + m\* \spb4.5\* \spa\one.3\* \spa6.\two
          + \spab3.(1+4).5\* \spb\one.4\* \spa6.\two
          + \spab6.(1+3).4\* \spa\one.3\* \spb5.\two)}{s_{34}\* s_{56}\* (s_{134}-m^2)}\nn \\
&+& \frac{(\spa3.4\* \spb5.4\* (\spa\one.6\* \spb4.\two+\spb\one.4\* \spa6.\two)
          + \spa3.5\* \spb5.4\* (\spa\one.6\* \spb5.\two+\spb\one.5\* \spa6.\two))}{s_{34}\* s_{345}\* s_{3456}}\nn \\
          &-&\frac{(\spa3.6\* \spb5.3\* (\spa\one.3\* \spb4.\two+\spb\one.4\* \spa3.\two)
          + \spa3.6\* \spb5.6\* (\spa\one.6\* \spb4.\two+\spb\one.4\* \spa6.\two))}{s_{56}\* s_{356}\* s_{3456}}
\,  .
\eeqn
 \beqn
&& -i A^{(2)}(\one_Q, \two_{\Qb},3^-_q,4^+_{\qb},5^+_{q^\prime},6^-_{\qb^\prime}) =\nn \\
&&  \frac{1}{s_{34}\* s_{56}\* s_{3456}} \* \Big[
      \spab6.(3+4).5\* (\spa\one.3\* \spb4.\two+\spb\one.4\* \spa3.\two)
          - \spa3.6\* \spb5.4\* (\spab\one.(3+4).\two+\spab\two.(3+4).\one) \nn \\
&& \qquad \qquad \qquad 
          - \spab3.(5+6).4\* (\spa\one.6\* \spb5.\two+\spb\one.5\* \spa6.\two) \Big] \nn \\
&-&\frac{(m\* \spa6.3\* \spb\one.5\* \spb4.\two
          + m\* \spb5.4\* \spa\one.6\* \spa3.\two
          + \spab3.(1+6).5\* \spa\one.6\* \spb4.\two
          + \spab6.(1+5).4\* \spb\one.5\* \spa3.\two)}{s_{34}\* s_{56}\* (s_{156}-m^2)}\nn \\
&-&\frac{(\spa3.4\* \spb5.4\* (\spa\one.6\* \spb4.\two+\spb\one.4\* \spa6.\two)
          + \spa3.5\* \spb5.4\* (\spa\one.6\* \spb5.\two+\spb\one.5\* \spa6.\two))}{s_{34}\* s_{345}\* s_{3456}}\nn \\
 &+&\frac{(\spa3.6\* \spb5.3\* (\spa\one.3\* \spb4.\two+\spb\one.4\* \spa3.\two)
          + \spa3.6\* \spb5.6\* (\spa\one.6\* \spb4.\two+\spb\one.4\* \spa6.\two))}{s_{56}\* s_{356}\* s_{3456}}
\,  .
\eeqn

The $A^{(2)}$ amplitude for the other helicity, and the $A^{(4)}$ amplitudes,
can be obtained by interchange of labels and spinor brackets,
\beqn
&& A^{(2)}(\one_Q, \two_{\Qb},3^-_q,4^+_{\qb},5^-_{q^\prime},6^+_{\qb^\prime}) =
 A^{(1)}(\one_Q, \two_{\Qb},5^-_q,6^+_{\qb},3^-_{q^\prime},4^+_{\qb^\prime}) \,,
 \nn \\
&& A^{(4)}(\one_Q, \two_{\Qb},3^-_q,4^+_{\qb},5^-_{q^\prime},6^+_{\qb^\prime}) =
 A^{(5)}(\one_Q, \two_{\Qb},5^-_q,6^+_{\qb},3^-_{q^\prime},4^+_{\qb^\prime}) \,,
 \nn \\
&& A^{(4)}(\one_Q, \two_{\Qb},3^-_q,4^+_{\qb},5^+_{q^\prime},6^-_{\qb^\prime}) =
 A^{(5)}(2^A_Q, 1^B_{\Qb},5^-_q,6^+_{\qb},3^+_{q^\prime},4^-_{\qb^\prime})
 |_{\spa{}.{} \leftrightarrow \spb{}.{}} \, .
\eeqn

Finally, the amplitudes for $h_3 = +$ are obtained by similar relabelings,
\beqn
&& A^{(1)}(\one_Q, \two_{\Qb},3^+_q,4^-_{\qb},5^{h_5}_{q^\prime},6^{-h_5}_{\qb^\prime}) =
 -A^{(2)}(2^A_Q, 1^B_{\Qb},3^-_q,4^+_{\qb},5^{-h_5}_{q^\prime},6^{h_5}_{\qb^\prime})
 |_{\spa{}.{} \leftrightarrow \spb{}.{}} \, ,
 \nn \\
&& A^{(2)}(\one_Q, \two_{\Qb},3^+_q,4^-_{\qb},5^{h_5}_{q^\prime},6^{-h_5}_{\qb^\prime}) =
 -A^{(1)}(2^A_Q, 1^B_{\Qb},3^-_q,4^+_{\qb},5^{-h_5}_{q^\prime},6^{h_5}_{\qb^\prime})
 |_{\spa{}.{} \leftrightarrow \spb{}.{}} \, ,
 \nn \\
&& A^{(4)}(\one_Q, \two_{\Qb},3^+_q,4^-_{\qb},5^{h_5}_{q^\prime},6^{-h_5}_{\qb^\prime}) =
  A^{(4)}(2^A_Q, 1^B_{\Qb},3^-_q,4^+_{\qb},5^{-h_5}_{q^\prime},6^{h_5}_{\qb^\prime})
 |_{\spa{}.{} \leftrightarrow \spb{}.{}} \, ,
 \nn \\
&& A^{(5)}(\one_Q, \two_{\Qb},3^+_q,4^-_{\qb},5^{h_5}_{q^\prime},6^{-h_5}_{\qb^\prime}) =
  A^{(5)}(2^A_Q, 1^B_{\Qb},3^-_q,4^+_{\qb},5^{-h_5}_{q^\prime},6^{h_5}_{\qb^\prime})
 |_{\spa{}.{} \leftrightarrow \spb{}.{}} \, .
\eeqn

\section{Relation to classic formalism}
\label{classic}
In order to evaluate amplitudes with massive fermions we need a definite representation for the massive spinors.
This we do by expressing the massive momenta as the sum of two light-like vectors; this approach meshes nicely
with our technique to introduce spin correlations in top decay, as illustrated below in
subsection~\ref{Tree-level-decays}. 
The states for the massive fermions are computed introducing
arbitrary light-like vectors $\eta_p$ and $\eta_q$ and decomposing
massive vectors $p,q$ into two light-like vectors, $(p^\flat,\eta_p),(q^\flat,\eta_q)$,~\cite{Kleiss:1986qc}
\beq \label{lightlikedecomposition}
  p=p^\flat+\frac{m^2}{\spa{p^\flat}.{\eta_p}\spb{\eta_p}.{p^\flat}} \eta_p,\;\;\;
  q=q^\flat+\frac{m^2}{\spa{q^\flat}.{\eta_q}\spb{\eta_q}.{q^\flat}} \eta_q \, ,
\eeq
\beqn
  \bar{u}_-(p) &=\frac{[{\eta_p}\big|(\slsh{p}+m)}{\spb{\eta_p}.{p^\flat}}=\frac{m}{\spb{\eta_p}.{p^\flat}} [\eta_p\big|_\da+ \langle p^\flat\big|^\alpha  \, ,\quad
  \bar{u}_+(p) &=\frac{\langle {\eta_p}\big|(\slsh{p}+m)}{\spa{\eta_p}.{p^\flat}}=[p^\flat\big|_{\da} + \frac{m}{\spa{\eta_p}.{p^\flat}} \langle \eta_p\big|^\alpha\, , \nn\\
    v_+(q) &=\frac{(\slsh{q}-m)\big| \eta_q \rangle}{\spa{q^\flat}.{\eta_q}} =\big|q^\flat]^\da - \big|\eta_q\rangle_\alpha \frac{m}{\spa{q^\flat}.{\eta_q}} \, , \quad
    v_-(q) &= \frac{(\slsh{q}-m)\big|\eta_q]}{\spb{q^\flat}.{\eta_q}}= -\big|\eta_q]^\da \frac{m}{\spb{q^\flat}.{\eta_q}} +\big|q^\flat\rangle_\alpha \, ,
\eeqn
where $p$ and $q$ are the heavy quark momenta and in the labels for the Dirac spinors we have suppressed the
dependence on their common mass, $m$.
Using the expressions for the Dirac spinors in Eq.~(\ref{eq:ubarandv}) we can
read off the spin-spinors as (suppressing SL$(2,\mathbb{C})$ components from now on),
\begin{align*}
   \langle p^{I=1}| &= \bra{p^{\flat}} \,,
    & [p^{I=1}| &= \frac{m}{[\eta_p p^\flat]}[\eta_p|
      \\
   \langle p^{I=2}| &=  \frac{m}{\spa{\eta_p}.{p^\flat}}\bra{\eta_p} \,, 
    & [p^{I=2}| &= [p^{\flat}|
      \\
   |p_{J=2}\rangle &= -\ket{\eta_q}\frac{m}{\spa{q^\flat}.{\eta_q}} \,,
    & |p_{J=2}] &= |q^{\flat}] 
      \\
   |p_{J=1}\rangle &= \ket{q^\flat} \,,
    & |p_{J=1}] &= -|\eta_q] \frac{m}{[q^\flat \eta_q]} 
\end{align*}
Note that we have associated $v_+(q)$ with the $J=2$ component and $v_-(q)$ with $J=1$.
With these definitions we then have that,
\beqn
\spa \one.\two &=&  \left( \begin{matrix}
  \spa{p_1^{I=1} \, | \,}.{p_{2,J=1}}  & \; \spa{p_1^{I=1} \, | \,}.{p_{2,J=2}} \\
  \spa{p_1^{I=2} \, | \,}.{p_{2,J=1}} & \; \spa{p_1^{I=2} \, | \,}.{p_{2,J=2}} &
  \end{matrix} \right) \\
  &=& \left( \begin{matrix}
   \spa{1^\flat}.{2^\flat}
   & \; -\frac{m \spa{1^\flat}.{\eta_2}}{\spa{2^\flat}.{\eta_2}} \\
    \frac{m \spa{2^\flat}.{\eta_1}}{\spa{1^\flat}.{\eta_1}}
   & \; \frac{m^2 \spa{\eta_1}.{\eta_2}}{\spa{1^\flat}.{\eta_1}\spa{2^\flat}.{\eta_2}}
  \end{matrix} \right)
\eeqn
and, for example,
\beqn \spa\one.{k} \spa{l}.\two &=& \left( \begin{matrix}
  \bra{p_1^{I=1}} {k} \rangle \langle{l} \ket{p_{2,J=1}} & \;\; \bra{p_1^{I=1}} {k} \rangle \langle{l} \ket{p_{2,J=2}} \\
  \bra{p_1^{I=2}} {k} \rangle \langle{l} \ket{p_{2,J=1}} & \;\; \bra{p_1^{I=2}} {k} \rangle \langle{l} \ket{p_{2,J=2}} &  
  \end{matrix} \right) \\
  &=& \left( \begin{matrix}
    \spa{1^\flat}.{k} \spa{l}.{2^\flat} & \;
      - \frac{m \spa{1^\flat}.{k} \spa{l}.{\eta_2}}{\spa{2^\flat}.{\eta_2}} \\
    - \frac{m \spa{\eta_1}.{k} \spa{l}.{2^\flat}}{\spa{1^\flat}.{\eta_1}} &
      \frac{m^2 \spa{\eta_1}.{k} \spa{l}.{\eta_2}}{\spa{1^\flat}.{\eta_1} \spa{2^\flat}.{\eta_2}}
           
  \end{matrix} \right)  
\eeqn

\subsection{Inclusion of tree-level decays}
\label{Tree-level-decays}
Kleiss and Stirling~\cite{Kleiss:1988xr} have provided a procedure for including tree-level
top quark decays\footnote{Note that Eq.~(4) of ref.~\cite{Kleiss:1988xr} should read $t \to b(p_6) + \nu (p_7) + e^+(p_8)$.}.
Consider the leptonic decays of on-shell top quarks and antitops,
\beq
\bar{t} \to  \bar{b}(p_3) + e^-(p_4) + \bar{\nu}(p_5) 
\, , \quad \quad t \to  b(p_6) + \nu(p_7)+ e^+(p_8) 
\eeq
If we denote the four momenta of the top quarks and their decay products by the symbols given above,
the contribution to the matrix element of the heavy quark line and subsequent decays will
be,
\beqn
M &\propto& \frac{g_W^4}{4} \bar{u}(p_6) \gamma^\mu \gamma_L (\slsh{t}+m)\; \ldots\; (-\bar{\slsh{t}} +m) \gamma_R \gamma^\nu v(p_3) \,
  \times \, \spab{7}.{\gamma^\mu}.{8} \, \spab{4}.{\gamma^\nu}.{5} \nn \\
     &=& g_W^4\, \langle \bar{u}(p_6) |7\rangle \; [8|(\slsh{t}+m)  \ldots (-\bar{\slsh{t}} +m) |4\rangle \; [5| v(p_3)] 
\eeqn
Thus the full spin correlations for the decay of the top and antitop can be included by using the decomposition in Eq.~(\ref{lightlikedecomposition})
with auxiliary vectors $e,\bar{e}$ and a single helicity combination, $h_8=h_4=+\frac{1}{2}$.
This approach has been followed at next-to-leading order in the parton-level Monte Carlo program MCFM~\cite{Campbell:2012uf}
using one-loop results for the top amplitudes from ref.~\cite{Badger:2011yu}. This approach has also been pursued for
the case of a top quark pair accompanied by one~\cite{Melnikov:2010iu,Melnikov:2011qx} or two jets~\cite{Bevilacqua:2022ozv}.
A necessary first step to extend these analyses to NNLO is the calculation of the amplitudes for top quark pair production
at the two loop level. Although this program is not yet complete, first steps have been taken in refs.~\cite{Badger:2022mrb,Badger:2021owl}.

\section{Conclusions}
\label{conclusions}

In this paper we have provided explicit analytic expressions for all four-,
five- and six-parton amplitudes needed for the calculation of $pp \to t \bar t$,
$pp \to t \bar t+j$ and $pp \to t \bar t + jj$ production at hadron colliders.
These amplitudes have been presented using the Spin-spinor approach,
that extends the usual spinor notation for massless particles
to the massive case.  It thus retains many of the advantages of the original
spinor formalism, in particular its ability to provide results in a
compact form.
The results, although not always simple, are considerably more compact than the results obtained using normal Feynman diagram calculation.
We have elucidated the application of BCFW recursion in this
approach and used lower-point amplitudes as buildings blocks to provide
new results for some 6-point amplitudes.  In addition we have summarized the
BCJ relations that apply in each case and shown how to construct the squared
matrix element, summed over colors, from the color-ordered amplitudes.
As well as their utility in tree-level calculations, we anticipate that the
simple form of some of the amplitudes presented in this paper will enable
new analytic one-loop calculations.  Unitarity methods exploit these amplitudes
in calculations of processes 
containing a loop of massive fermions.  
Machine readable forms of our results are available in a
Fortran code which evaluates and squares these amplitudes. The Fortran code is attached to the arXiv version of this
paper. They will also be distributed
in a future version of MCFM~\cite{MCFM}.

\section*{Acknowledgments}
We would like to thank Tom Melia for useful discussions.
RKE is grateful for hospitality at LBNL and Fermilab during the preparation of this paper.
This manuscript has been authored by Fermi Research Alliance, LLC
under Contract No. DE-AC02-07CH11359 with the U.S. Department of
Energy, Office of Science, Office of High Energy Physics.

\appendix
\section{Review of spinor techniques}
\label{SpinorReview}
\subsection{Conventions}
We introduce spinor techniques departing from the Dirac equation, since we believe that the reader may be more
familiar with $\gamma$-matrix technology than Weyl spinors.
We work in the metric given by ${\rm diag}(1,-1,-1,-1)$ and use the Weyl representation of the Dirac gamma matrices given by,
\beq \label{gammamatrices}
\gamma^\mu = \left(\begin{matrix}{\bf 0}&{\boldsigma}^\mu\cr
  {\bar{\boldsigma}}^\mu&{\bf 0}\cr \end{matrix}\right)\ ,(\mu=0,3),\,\,
\gamma_5 = i \gamma^0\gamma^1\gamma^2\gamma^3 =
\left(\begin{matrix}-{\bf 1}&{\bf 0}\cr{\bf 0}&+{\bf 1}\cr \end{matrix}\right)\ ,
\eeq
where $\boldsigma^\mu=({\bf 1},{\boldsigma}^i)$, ${\bar{\boldsigma}}^\mu=({\bf 1},-{\boldsigma}^i)$
where $\boldsigma$ are the Pauli matrices,
\beq
{\boldsigma}^1 =\left(\begin{matrix} 0 & 1 \\ 1 & 0 \end{matrix}\right),\;\;
{\boldsigma}^2 =\left(\begin{matrix} 0 & -i \\ i & 0 \end{matrix}\right),\;\;
{\boldsigma}^3 =\left(\begin{matrix} 1 & 0 \\ 0 & -1 \end{matrix}\right)\, .
\eeq
Contracting the four-momentum with the gamma matrices we find an expression for $\slsh{p}$,
\beq
\slsh{p} \equiv p_\mu \gamma^\mu = \left( \begin{matrix} {\bf 0} & p^{\da \beta}\\
  p_{\alpha \db} & {\bf 0} \end{matrix}\right),\;\;\;
p^{\da \beta}=p_\mu (\sigma^\mu)^{\da \beta},\,\,
p_{\alpha \db}=p_\mu (\bar{\sigma}^\mu)_{\alpha \db},
\eeq
Explicitly we find in terms of the components of $p^\mu=(p^0,p^1,p^2,p^3)$,
\beq \label{masslesscase}
p^{\da \beta}=\left(\begin{matrix}p^-& -\bar{p}_\perp\\
                                           -p_\perp& p^+\end{matrix}\right),\;\;
                                           p_{\alpha \db}=
                                           \left(\begin{matrix}p^+& \bar{p}_\perp\\
                                           p_\perp& p^-\end{matrix}\right),\;\;{\rm where}~p^\pm=p^0\pm p^3, \,p_\perp=p^1+i p^2, \,\bar{p}_\perp=p^1-i p^2\,.
\eeq

\subsection{Spinor techniques for massless particles}
For massless particles the $p_\mu p^\mu={\rm det}~p_{\alpha \db}=0$ and the matrices
can be expressed as bi-spinors
\beq \label{bispinors}
p^{\da \beta}=|p]^\da \langle p|^{\beta},\;\;
p_{\alpha \db}=|p\rangle_\alpha [ p|_{\db}\, .
\eeq
By convention in the calculation of amplitudes we take all particles to be outgoing.
Therefore the ingredients that we require are the wave functions associated with outgoing
fermions and anti-fermions. The wave functions satisfy the massless Dirac equation
for fermions
\beq
\bar{u}_\pm(p) \slsh{p}=0,\;{\rm where}~\bar{u}_-(p)=\big( \,\mathbf{0} \, , \,\langle p |^{\beta}\big),\;\;
\bar{u}_+(p)=\big(\, [ p |_\db \, ,\, \mathbf{0} \,\big)
\eeq
and anti-fermions
\beq
\slsh{p} v_\pm(p)=0,\;{\rm where}~v_-(p)=\left(\begin{matrix}
         \mathbf{0}  \\
         |p\rangle_{\alpha}\end{matrix} \right),\;\;
v_+(p)=\left( \begin{matrix}
     {|p]}^\da\\
    \mathbf{0}\end{matrix} \right)
\eeq

Since the charge conjugation relation for Dirac spinors is $v_{\pm}=-i \gamma^2 u_\pm^*$ so that
$v_{\pm}(p)=C \bar{u}_{\pm}^T$ with
\beqn
{ C} &=&
{ i \gamma^2 \gamma^0 = \ \left(\begin{matrix} i{\boldsigma}^2& {\bf0}
\cr                                      {\bf0}& -i{\boldsigma}^2\cr \end{matrix}\right)}
=   \left(\begin{matrix}
      0& 1 & & 0& 0 \cr
      -1& 0 & & 0& 0 \cr
      0& 0 & & 0& -1 \cr
      0& 0 & & 1& 0 \cr \end{matrix}\right)
=\left(\begin{matrix} \epsilon^{\da \db}
  & {\bf0}\\
  {\bf 0} & \epsilon_{\alpha\beta} \end{matrix}\right)
\eeqn
where the two dimensional antisymmetric tensor is,
\beq
\epsilon^{\alpha \beta} = \epsilon^{\da \db} = i \boldsigma^2 =\left(\begin{matrix} 0 & 1 \\
                                              -1 & 0 \end{matrix}\right),\;\;\;
\epsilon_{\alpha \beta}= \epsilon_{\da \db}= -i \boldsigma^2 =\left(\begin{matrix} 0 & -1 \\
                                               1 & 0 \end{matrix}\right)
\label{eq:epdef}
\eeq
Thus to raise or lower the index of a spinor quantity, adjacent spinor
indices are summed over when multiplied on the left by the appropriate epsilon symbol,
\beq
|p\rangle_{\alpha}=\epsilon_{\alpha \beta}\langle p|^{\beta}, \quad
|p]^\da=\epsilon^{\da \db}[p|_{\db},\quad
\eeq
and analogously,
\beq
\langle p|^{\alpha}=\epsilon^{\alpha \beta}|p\rangle_{\beta}, \quad
[p|_\da=\epsilon_{\da \db}|p]^{\db}\, ,
\eeq
and,
\beq
p_{\alpha \db}=\epsilon_{\alpha \beta} \epsilon_{\db \da}p^{\da \beta} \,.
\eeq
Using Eq.~(\ref{bispinors}) we see that the massless spinors satisfy the Weyl equations of motion,
\beq
p^{\da \beta} |p\rangle_\beta = p_{\alpha \db} |p]^\db =
\langle p|^{\alpha} p_{\alpha \db}=[ p|_{\da} p^{\da \beta}=0 \,.
\eeq

Part of the simplicity of the spinor calculus derives from the fact that we do not
need explicit expressions for the spinor solutions, until we arrive at the stage of numerical evaluation.
However we can derive solutions to the Weyl equations of motion using the results in Eq.~(\ref{masslesscase}),
\beqn
{|p\rangle}_{\alpha}
    & =  (\sqrt{p^+}, p_{\perp}/\sqrt{p^+}),\quad\quad
{[ p |}_\db
      &=(\sqrt{p^+}, \bar{p}_{\perp}/\sqrt{p^+}),\\
{\langle p|}^{\beta}
  &=   (-{p}_{\perp}/\sqrt{p^+}, \sqrt{p^+}),\quad\quad
{| p ]}^\da
    &= (-\bar{p}_{\perp}/\sqrt{p^+}, \sqrt{p^+} ).
\eeqn
We see that the angle (square) brackets automatically encode the north-west~$\to$~south-east
(south-west~$\to$~north-east) summation convention 
for the SL$(2,\mathbb{C})$ undotted (dotted) indices.
Thus in most circumstances these indices
can be dropped. The spinor products satisfy $\spa{i}.{j}=-\spa{j}.{i}$, $\spb{i}.{j}=-\spb{j}.{i}$.
For light-like vectors we can combine the Weyl spinors to form Dirac spinors as follows,
\beq
v(p)=\left(\begin{matrix} |p]^{\db} \\
                          |p\rangle_{\beta}
\end{matrix}\right)\, , \quad\quad
\bar{u}(p)=\left(\begin{matrix} [p|_\da & \langle p|^\alpha \end{matrix}\right)\, .
\eeq

\subsection{Spinor techniques for massive particles}
\subsubsection{Angle notation}
We now turn to consider particles with mass, $m$, so that $E^2-P^2=m^2$.
Now in terms of a four-vector $p^\mu=(E, P\sin\theta\cos\phi,P\sin\theta\sin\phi,P\cos\theta)$
we find using the Weyl representation for the gamma matrices, Eq.~(\ref{gammamatrices}),  that,
\beqn \label{explicitpslash} 
p^{\da \beta}=\left(\begin{matrix} c^2 P^- + ss^* P^+   & -2 \,c \, s^* P\\
                                   -2 \, c\, s P          & ss^* P^- + c^2 P^+\end{matrix}\right) \\
p_{\alpha \db}=\left(\begin{matrix} c^2 P^+ + ss^* P^- & 2 \, c\, s^* P\\
                                    2 \, c\, s P           & ss^* P^+ + c^2 P^-\end{matrix}\right) 
\eeqn
where $c=\cos(\frac{\theta}{2}),s=\sin(\frac{\theta}{2})\exp(i\phi),s^*=\cos(\frac{\theta}{2})\exp(-i\phi)$.
In this equation we have introduced the notation $P^\pm=E \pm P$, which we write in upper case (to distinguish
it from $p^\pm$ in the massless case Eq.~(\ref{masslesscase}) which was defined differently).
We can express components of the tensor $|\bm{p}\rangle^I_\alpha$ and $[\bm{p}_I|_{\da}$ where we let the label
$I$ run over the two values $1$ and $2$,
\beqn \label{matrixexpanded}
  \lambda_{\alpha}^{I} &= |\bm{p}^I\rangle_\alpha =
  \sqrt{P^-}\left(\begin{matrix}   - s^* \\
                                     c   \end{matrix} \right) = \sqrt{P^-} \zeta^{-}_\alpha,\;\;
  \tilde{\lambda}_{I\; \da} &=[\bm{p}_I|_{\da} =
  \sqrt{P^-} \left(\begin{matrix} - s \\
                                    c  \end{matrix} \right)=\sqrt{P^-} \tilde{\zeta}^{+}_\da\;\;{\rm for}~I=1 \nn \\
  \lambda_{\alpha}^{I} &= |\bm{p}^I\rangle_\alpha =
  \sqrt{P^+}\left(\begin{matrix}  c\\
                                  s \end{matrix} \right)= \sqrt{P^+} \zeta^{+}_\alpha,\;\;
  \tilde{\lambda}_{\da\;I} &=[\bm{p}_I|_{\da} =
    \sqrt{P^+} \left(\begin{matrix} c \\
                                    s^* \end{matrix} \right)= \sqrt{P^+} \tilde{\zeta}^{-}_\da,\;\;{\rm for}~I=2 \nn  \\
\eeqn
so that,
\beq
\sum_{I=\mp 1} |\bm{p}^I\rangle_\alpha [\bm{p}_I|_{\db} =
 P^-\left(\begin{matrix} s s^* & -cs^* \\
  -cs & c^2 \end{matrix} \right)
    +P^+\left(\begin{matrix} c^2 & cs^* \\
                            cs & ss^* \end{matrix} \right)
= p_{\alpha \db} \,,
\eeq
using the expression for $p_{\alpha \db}$ in Eq.~(\ref{explicitpslash}).
Note using expressions below we have,
\beq
-\sum_{I=\pm} |\bm{p}^I]^{\da} \langle\bm{p}_I|^{\db} =
 P^-\left(\begin{matrix}c^2 & cs^* \\
  cs & s s^* \end{matrix} \right)
    +P^+\left(\begin{matrix} ss^* & -cs^* \\
                            -cs & c^2 \end{matrix} \right)
= p^{\da \beta} \,.
\eeq

We can write Eq.~(\ref{matrixexpanded}) equivalently as~\cite{Arkani-Hamed:2017jhn} 
\beqn
\lambda_{\alpha}^I &= |\bm{p}^I\rangle_\alpha &=\sqrt{P^+} \; \zeta^+_\alpha(p)   \otimes \delta^{I}_{1} \, \zeta^{-}
                                              + \sqrt{P^-} \; \zeta_\alpha^{-}(p) \otimes \delta^{I}_{2}\, \zeta^{+} 
\nn \\ 
\tilde{\lambda}_{I\, \da} &= [\bm{p}_I|_{\da} &=\sqrt{P^+} \; \tilde{\zeta}^{-}_{\da}(p) \otimes\delta^{1}_{I}\, \zeta^{-}
                                                 + \sqrt{P^-} \; \tilde{\zeta}^{+}_{\da}(p) \otimes\delta^{2}_{I}\, \zeta^{+}
\eeqn
where $I$ runs over the values $1$ and $2$.
Here we have chosen a representation of the SU(2) algebra in which $\boldsigma_z$ is diagonal with eigenstates,
\beq
\zeta^+ =
\left(\begin{matrix} 1 \\ 0 \end{matrix}\right),\quad\quad
\zeta^- =
\left(\begin{matrix} 0 \\ 1 \end{matrix}\right).
\eeq
and the expression for the spinors with SL$(2,\mathbb{C})$ Lorentz indices is, 
\beqn
\zeta^{+}_{\alpha} &= \left(\begin{array}{c} c \\ s \end{array} \right) \equiv    +\tilde{\zeta}^{+\,\da}\,,\quad\quad\quad
\zeta^{-}_{\alpha} &= \left(\begin{array}{c} -s^* \\ c \end{array} \right)\equiv  -\tilde{\zeta}^{- \,\da}\, ,\nn \\
\tilde{\zeta}^-_{\da} &=  \left(\begin{array}{c} c \\ s^* \end{array} \right)\equiv +\zeta^{- \, \alpha} \,,\quad\quad\quad
\tilde{\zeta}^+_{\da} &=  \left(\begin{array}{c} -s \\ c \end{array} \right)\equiv  -\zeta^{+ \, \alpha} \, .
\eeqn
In terms of these Weyl spinors we have the following relations,
\begin{eqnarray} \label{followingrelations}
  p^{\da\beta} \, \zeta^{\pm}_{\beta}  &=&  \pm P^{\mp} \tilde{\zeta}^{\pm\,\da} \,, \nn\\
   p_{\alpha \db} \, \tilde{\zeta}^{\pm\,\db} &=& \pm P^{\pm} \zeta^{\pm}_{\alpha} \,, \nn \\
   \zeta^{\pm\,\alpha} \, p_{\alpha \db} &=& \mp P^{\mp} \tilde{\zeta}^{\pm}_{\db} \,, \nn \\
   \tilde{\zeta}_{\da}^{\pm}  \, p^{\da\beta} &=&\mp P^{\pm} \zeta^{\pm\,\beta} \,.
\end{eqnarray}

Raising and lowering the $SU(2)$ index is performed 
by multiplying by the two-dimensional totally antisymmetric tensor $\epsilon$ on the right,
$|\bm{p}_I\rangle_{\alpha} = |\bm{p}^J\rangle_{\alpha}\epsilon_{JI}$
and $\lbrack\bm{p}^I|_{\da} = \lbrack\bm{p}_J|_{\da}\epsilon^{JI}$.
To be completely explicit we write out a complete set\footnote{Eqs.~(\ref{C2},\ref{C3}) correct
Eqs.~(C.2) and (C.3) of AHH~\cite{Arkani-Hamed:2017jhn} which contain errors.},
\beqn \label{C2}
|\bm{p}^I\rangle_\alpha &=&+\sqrt{P^+} \; \zeta^+_\alpha(p) \otimes\delta^{I}_1 \, \zeta^{-} + \sqrt{P^-} \; \zeta_\alpha^{-}(p) \otimes \delta^{I}_2\, \zeta^{+} \nn \nn \\ 
|\bm{p}^I]^{\da} &=&  -\sqrt{P^-} \; \tilde{\zeta}^{+\, \da}(p) \otimes\delta^{I}_1 \, \zeta^{-} + \sqrt{P^+} \; \tilde{\zeta}^{-\, \da}(p) \otimes \delta^{I}_2\, \zeta^{+}\nn \\
\langle \bm{p}^I|^{\alpha} &=&+\sqrt{P^+} \; \zeta^{+\, \alpha}(p) \otimes\delta^{I}_1 \, \zeta^{-} + \sqrt{P^-} \; \zeta^{-\, \alpha}(p) \otimes \delta^{I}_2\, \zeta^{+} \nn \\
{[\bm{p}^I|}_{\da} &=&  -\sqrt{P^-} \; \tilde{\zeta}^{+}_{\da}(p) \otimes\delta^{I}_1 \, \zeta^{-} + \sqrt{P^+} \; \tilde{\zeta}^{-}_{\da}(p) \otimes \delta^{I}_2\, \zeta^{+}
\eeqn
\beqn
\label{C3}
|\bm{p}_I\rangle_{\alpha} &=&+\sqrt{P^-} \; \zeta^-_\alpha(p) \otimes\delta^1_{I} \, \zeta^{-} -\sqrt{P^+}  \; \zeta^+_\alpha(p) \; \otimes \delta^2_{I}\, \zeta^{+} \nn  \\
|\bm{p}_I]^{\da} &=&+\sqrt{P^+} \; \tilde{\zeta}^{-\,\da}(p) \otimes\delta^1_{I} \, \zeta^{-} + \sqrt{P^-} \; \tilde{\zeta}^{+\,\da}(p) \otimes \delta^2_{I}\, \zeta^{+} \nn \\
\langle \bm{p}_I|^{\alpha} &=& +\sqrt{P^-} \; \zeta^{-\, \alpha}(p) \otimes\delta^1_{I} \, \zeta^{-} -\sqrt{P^+} \; \zeta^{+\, \alpha}(p) \otimes \delta^2_{I}\, \zeta^{+} \nn \\
{[\bm{p}_I|}_{\da} &=&+\sqrt{P^+} \; \tilde{\zeta}^{-}_{\da}(p) \otimes\delta^1_{I} \, \zeta^{-} + \sqrt{P^-} \; \tilde{\zeta}^{+}_{\da}(p) \otimes \delta^2_{I}\, \zeta^{+} 
\eeqn
In our notation $I$ and $J$ taken on the values $1$ and $2$. The SU(2) little group indices are lowered and raised
by multiplying to the right by $\epsilon^{IJ}$ and $\epsilon_{IJ}$, c.f. Eq.~(\ref{eq:epdef}).
From these expressions for the spinors we can see that,
\beqn
\Big( \langle \bm{p}^I|^\alpha \Big)^* = |\bm{p}_I]^\da \, \qquad
\Big( | \bm{p}^I \rangle_\alpha \Big)^* = [\bm{p}_I|_\da \,,
\eeqn
but on the other hand,
\beqn
\Big( \langle \bm{p}_I|^\alpha \Big)^* = - |\bm{p}^I]^\da \, \qquad
\Big( | \bm{p}_I \rangle_\alpha \Big)^* = - [\bm{p}^I|_\da \,.
\eeqn
In other words, taking the complex conjugate of an angle spinor with a lowered spin index $I$, or a square
spinor with a raised spin index, introduces an additional minus sign.
This means that if we define the spinors for an outgoing quark and antiquark as,
\beq
\bar{u}(p) = \left(
\begin{matrix} {[} \bm{p}^I |_\da & \; \langle \bm{p}^I |^\alpha \end{matrix}
\right) \, \qquad
v(p) = \left(
\begin{array}{c} | \bm{p}_I ]^\da \\ | \bm{p}_I \rangle_\alpha \end{array}
\right) \,,
\label{eq:ubarandv}
\eeq
then we must also have,
\beq
u(p) = \left(
\begin{array}{c} | \bm{p}_I ]^\da \\ - | \bm{p}_I \rangle_\alpha \end{array}
\right) \, \qquad
\bar v(p) = \left(
\begin{matrix} \; -[ \bm{p}^I |_\da & \langle \bm{p}^I |^\alpha \end{matrix}
\right) \,.
\eeq

In the massless case the spinor-helicity states $|p_\alpha], |p_{\db}\rangle$ satisfy the Weyl equation and are independent of each other.
In the massive case dotted and undotted massive spinor states are related through the equation of motion for the Weyl fields.
In terms of this set of tensors, using the relations in Eq.~(\ref{followingrelations}),
we obtain the following equations of motion,
\beqn
p^{\da \beta} \, |\bm{p}^I\rangle_\beta &=-m |\bm{p}^I]^{\da}, \quad\quad
p_{\alpha \db} \, |\bm{p}^I]^{\db} &=-m |\bm{p}^I\rangle_{\alpha}, \nn\\
\langle \bm{p}^I|^{\alpha} \, p_{\alpha \db} &=+m {[\bm{p}^I|}_{\db}, \quad\quad
{[\bm{p}^I|}_{\da} \, p^{\da \beta} &= +m \langle \bm{p}^I|^{\beta}.
  \eeqn
Therefore the scattering amplitude involving massive particles
can be expressed either in terms of $|{\bm{p}}_I]^\alpha$ or $|{\bm{p}_I\rangle_\db}$.
In addition we have, 
\beqn
   \ket{\bm{p}^I}_{\alpha}[\bm{p}_I|_{\db}& =  p_{\alpha\db} \quad\quad |\bm{p}^I]^{\da}\:\bra{\bm{p}_I}^{\beta} &= -p^{\da\beta} \\
   \ket{\bm{p}^I}_{\alpha} \bra{\bm{p}_I}^{\beta} & =m\:\delta_\alpha^\beta  \quad \quad |\bm{p}^I]^{\da}\:[\bm{p}_I|_{\db} &=-m\:\delta^{\da}_{\db}\\
   \ket{\bm{p}_I}_{\alpha}[\bm{p}^I|_{\db}& =  -p_{\alpha\db} \quad\quad |\bm{p}_I]^{\da}\:\bra{\bm{p}^I}^{\beta} &= p^{\da\beta} \\
   \ket{\bm{p}_I}_{\alpha} \bra{\bm{p}^I}^{\beta} & =-m\:\delta_\alpha^\beta  \quad \quad |\bm{p}_I]^{\da}\:[\bm{p}^I|_{\db} &= m\:\delta^{\da}_{\db}
\eeqn
The massive fermion propagator is reconstructed as follows,
\beq
\gamma_\mu p^\mu +m =
\left(\begin{matrix}
  m \delta^\da_\db & p^{\da \beta}\\
  p_{\alpha \db} & m \delta^\beta_\alpha 
\end{matrix}\right) = 
\left(\begin{matrix}
  |p_I]^\da [p^I|_\db & \; |p_I]^\da \langle p^I|^\beta\\
  -|p_I\rangle_\alpha [p^I|_\db & \; -|p_I\rangle_\alpha \langle p^I|^\beta
\end{matrix}\right) = \left( \begin{matrix}  |p_I]^\da \\ - |p_I\rangle_\alpha) \end{matrix} \right) \otimes
 \left(\begin{matrix}[p^I|_\db & \; \langle p^I|^\beta \end{matrix}\right)
\eeq
Adopting the convention,
\beq
|-p\rangle = -|p\rangle\, ,\quad\quad
|-p] = |p]\, ,\quad\quad
\label{eq:fliprule}
\eeq
we have that,
\beq
\gamma_\mu p^\mu +m =
\left( \begin{matrix}  |-p_I]^\da \\ |-p_I\rangle_\alpha  \end{matrix} \right) \otimes
 \left(\begin{matrix}[p^I|_\db & \; \langle p^I|^\beta \end{matrix}\right)
 = v(-p) \otimes \bar{u}(p)
\eeq
Explicit representations for spinors that satisfy the rules in Eq.~(\ref{eq:fliprule}) are given in
refs.~\cite{Ochirov:2018uyq,Lazopoulos:2021mna}.
In a similar way,
\beq
\varepsilon^+_\mu(-k) \varepsilon^-_\nu(k)+\varepsilon^-_\mu(-k) \varepsilon^+_\nu(k)=
 \left(g_{\mu \nu}-\frac{(k_{\mu}b_{\nu}+k_{\nu}b_{\mu})}{b\cdot k}\right)
\eeq
For a review of the Spin-spinor formalism for massive particles, see also refs.~\cite{Christensen:2018zcq,Christensen:2019mch}.
\section{Melia basis for two quark pair + two gluon amplitudes}
\label{sec:Melia}

An alternative color-ordered basis for this process can be obtained following
Melia~\cite{Melia:2013xok}.  In this basis we have,
\beqn
  A(\one_Q, \two_{\Qb},3^{h_3}_q,4^+_{\qb},5^+_g,6^+_g) &=& g^4 \Big(
 A_{256143} \, C_{256143}+A_{215643} \, C_{215643}+A_{214563} \, C_{214563} \nn \\
&&\quad+A_{251643} \, C_{251643}+A_{251463} \, C_{251463}+A_{215463} \, C_{215463}\Big) \nn \\
&& + (5 \leftrightarrow 6) \, .
\eeqn
The color factors in this decomposition can be read off from the Feynman rules,
\beqn
C_{256143}&=&\left( t^{A} t^{C_5} t^{C_6} \right)_{i_1 i_2} \, t^{A}_{i_3 i_4} \, , \nn \\
C_{215643}&=&t^{A}_{i_1 i_2} \, t^{B}_{i_3 i_4} \, F^{A C_5 D} \, F^{D C_6 B} \, , \nn \\
C_{214563}&=&t^{A}_{i_1 i_2} \, \left( t^{C_6} t^{C_5} t^{A} \right)_{i_3 i_4} \, , \nn \\
C_{251643}&=&\left( t^{A} t^{C_5} \right)_{i_1 i_2} \, t^{B}_{i_3 i_4} \, F^{C_6 B A} \, , \nn \\
C_{251463}&=&\left( t^{A} t^{C_5} \right)_{i_1 i_2} \, \left( t^{C_6} t^{A} \right)_{i_3 i_4} \, , \nn \\
C_{215463}&=&t^{A}_{i_1 i_2} \, F^{C_5 B A} \, \left( t^{C_6} t^{B} \right)_{i_3 i_4} \, ,
\eeqn
where $F^{ABC} = i f^{ABC} \sqrt 2$, and similarly for $5 \leftrightarrow 6$.
Note that in this basis each color structure consists of a single term.
The amplitudes are related to the ones in the Feynman diagram decomposition by,
\beqn
A^{(1)}&=&A_{215643} \nn \\
A^{(2)}&=&-A_{216453}+A_{261453}-A_{261543}+A_{214653}+A_{216543}+A_{256143} \nn \\
A^{(3)}&=&A_{261543}-A_{216543}+A_{215463}-A_{215643} \nn \\
A^{(4)}&=&-A_{256143} \nn \\
A^{(5)}&=&-A_{214653} \nn \\
A^{(6)}&=&-A_{251463} \nn \\
B^{(1)}&=&A_{216543} \nn \\
B^{(2)}&=&A_{265143}-A_{215463}+A_{251463}-A_{251643}+A_{214563}+A_{215643} \nn \\
B^{(3)}&=&A_{216453}-A_{216543}+A_{251643}-A_{215643} \nn \\
B^{(4)}&=&-A_{265143} \nn \\
B^{(5)}&=&-A_{214563} \nn \\
B^{(6)}&=&-A_{261453}
\eeqn
\bibliography{TopAmp}       

\providecommand{\href}[2]{#2}\begingroup\raggedright\begin{thebibliography}{10}

\bibitem{DeCausmaecker:1981jtq}
P.~De~Causmaecker, R.~Gastmans, W.~Troost and T.T.~Wu, \emph{{Multiple
  Bremsstrahlung in Gauge Theories at High-Energies. 1. General Formalism for
  Quantum Electrodynamics}},
  \href{https://doi.org/10.1016/0550-3213(82)90488-6}{\emph{Nucl. Phys. B}
  {\bfseries 206} (1982) 53}.

\bibitem{Berends:1981uq}
F.A.~Berends, R.~Kleiss, P.~De~Causmaecker, R.~Gastmans, W.~Troost and T.T.~Wu,
  \emph{{Multiple Bremsstrahlung in Gauge Theories at High-Energies. 2. Single
  Bremsstrahlung}},
  \href{https://doi.org/10.1016/0550-3213(82)90489-8}{\emph{Nucl. Phys. B}
  {\bfseries 206} (1982) 61}.

\bibitem{Kleiss:1985yh}
R.~Kleiss and W.J.~Stirling, \emph{{Spinor Techniques for Calculating p anti-p
  $\to W^\pm/Z^0$ + Jets}},
  \href{https://doi.org/10.1016/0550-3213(85)90285-8}{\emph{Nucl. Phys. B}
  {\bfseries 262} (1985) 235}.

\bibitem{Xu:1986xb}
Z.~Xu, D.-H.~Zhang and L.~Chang, \emph{{Helicity Amplitudes for Multiple
  Bremsstrahlung in Massless Nonabelian Gauge Theories}},
  \href{https://doi.org/10.1016/0550-3213(87)90479-2}{\emph{Nucl. Phys. B}
  {\bfseries 291} (1987) 392}.

\bibitem{Mangano:1990by}
M.L.~Mangano and S.J.~Parke, \emph{{Multiparton amplitudes in gauge theories}},
  \href{https://doi.org/10.1016/0370-1573(91)90091-Y}{\emph{Phys. Rept.}
  {\bfseries 200} (1991) 301}
  [\href{https://arxiv.org/abs/hep-th/0509223}{{\ttfamily hep-th/0509223}}].

\bibitem{Dixon:1996wi}
L.J.~Dixon, \emph{{Calculating scattering amplitudes efficiently}},  in
  \emph{{Theoretical Advanced Study Institute in Elementary Particle Physics
  (TASI 95): QCD and Beyond}}, pp.~539--584, 1, 1996
  [\href{https://arxiv.org/abs/hep-ph/9601359}{{\ttfamily hep-ph/9601359}}].

\bibitem{Ochirov:2018uyq}
A.~Ochirov, \emph{{Helicity amplitudes for QCD with massive quarks}},
  \href{https://doi.org/10.1007/JHEP04(2018)089}{\emph{JHEP} {\bfseries 04}
  (2018) 089} [\href{https://arxiv.org/abs/1802.06730}{{\ttfamily
  1802.06730}}].

\bibitem{Lazopoulos:2021mna}
A.~Lazopoulos, A.~Ochirov and C.~Shi, \emph{{All-multiplicity amplitudes with
  four massive quarks and identical-helicity gluons}},
  \href{https://doi.org/10.1007/JHEP03(2022)009}{\emph{JHEP} {\bfseries 03}
  (2022) 009} [\href{https://arxiv.org/abs/2111.06847}{{\ttfamily
  2111.06847}}].

\bibitem{Britto:2004ap}
R.~Britto, F.~Cachazo and B.~Feng, \emph{{New recursion relations for tree
  amplitudes of gluons}},
  \href{https://doi.org/10.1016/j.nuclphysb.2005.02.030}{\emph{Nucl. Phys. B}
  {\bfseries 715} (2005) 499}
  [\href{https://arxiv.org/abs/hep-th/0412308}{{\ttfamily hep-th/0412308}}].

\bibitem{Britto:2005fq}
R.~Britto, F.~Cachazo, B.~Feng and E.~Witten, \emph{{Direct proof of tree-level
  recursion relation in Yang-Mills theory}},
  \href{https://doi.org/10.1103/PhysRevLett.94.181602}{\emph{Phys. Rev. Lett.}
  {\bfseries 94} (2005) 181602}
  [\href{https://arxiv.org/abs/hep-th/0501052}{{\ttfamily hep-th/0501052}}].

\bibitem{Bern:2008qj}
Z.~Bern, J.J.M.~Carrasco and H.~Johansson, \emph{{New Relations for
  Gauge-Theory Amplitudes}},
  \href{https://doi.org/10.1103/PhysRevD.78.085011}{\emph{Phys. Rev. D}
  {\bfseries 78} (2008) 085011}
  [\href{https://arxiv.org/abs/0805.3993}{{\ttfamily 0805.3993}}].

\bibitem{Bern:2019prr}
Z.~Bern, J.J.~Carrasco, M.~Chiodaroli, H.~Johansson and R.~Roiban, \emph{{The
  Duality Between Color and Kinematics and its Applications}},
  \href{https://arxiv.org/abs/1909.01358}{{\ttfamily 1909.01358}}.

\bibitem{Maltoni:2002qb}
F.~Maltoni and T.~Stelzer, \emph{{MadEvent: Automatic event generation with
  MadGraph}}, \href{https://doi.org/10.1088/1126-6708/2003/02/027}{\emph{JHEP}
  {\bfseries 02} (2003) 027}
  [\href{https://arxiv.org/abs/hep-ph/0208156}{{\ttfamily hep-ph/0208156}}].

\bibitem{Gleisberg:2008fv}
T.~Gleisberg and S.~Hoeche, \emph{{Comix, a new matrix element generator}},
  \href{https://doi.org/10.1088/1126-6708/2008/12/039}{\emph{JHEP} {\bfseries
  12} (2008) 039} [\href{https://arxiv.org/abs/0808.3674}{{\ttfamily
  0808.3674}}].

\bibitem{Actis:2016mpe}
S.~Actis, A.~Denner, L.~Hofer, J.-N.~Lang, A.~Scharf and S.~Uccirati,
  \emph{{RECOLA: REcursive Computation of One-Loop Amplitudes}},
  \href{https://doi.org/10.1016/j.cpc.2017.01.004}{\emph{Comput. Phys. Commun.}
  {\bfseries 214} (2017) 140}
  [\href{https://arxiv.org/abs/1605.01090}{{\ttfamily 1605.01090}}].

\bibitem{Buccioni:2019sur}
{\scshape OpenLoops 2} collaboration, \emph{{OpenLoops 2}},
  \href{https://doi.org/10.1140/epjc/s10052-019-7306-2}{\emph{Eur. Phys. J. C}
  {\bfseries 79} (2019) 866}
  [\href{https://arxiv.org/abs/1907.13071}{{\ttfamily 1907.13071}}].

\bibitem{Kleiss:1988xr}
R.~Kleiss and W.J.~Stirling, \emph{{Top quark production at hadron colliders:
  some useful formulae}}, \href{https://doi.org/10.1007/BF01548856}{\emph{Z.
  Phys. C} {\bfseries 40} (1988) 419}.

\bibitem{Badger:2005jv}
S.D.~Badger, E.W.N.~Glover and V.V.~Khoze, \emph{{Recursion relations for gauge
  theory amplitudes with massive vector bosons and fermions}},
  \href{https://doi.org/10.1088/1126-6708/2006/01/066}{\emph{JHEP} {\bfseries
  01} (2006) 066} [\href{https://arxiv.org/abs/hep-th/0507161}{{\ttfamily
  hep-th/0507161}}].

\bibitem{Badger:2005zh}
S.D.~Badger, E.W.N.~Glover, V.V.~Khoze and P.~Svrcek, \emph{{Recursion
  relations for gauge theory amplitudes with massive particles}},
  \href{https://doi.org/10.1088/1126-6708/2005/07/025}{\emph{JHEP} {\bfseries
  07} (2005) 025} [\href{https://arxiv.org/abs/hep-th/0504159}{{\ttfamily
  hep-th/0504159}}].

\bibitem{Ozeren:2006ft}
K.J.~Ozeren and W.J.~Stirling, \emph{{Scattering amplitudes with massive
  fermions using BCFW recursion}},
  \href{https://doi.org/10.1140/epjc/s10052-006-0007-7}{\emph{Eur. Phys. J. C}
  {\bfseries 48} (2006) 159}
  [\href{https://arxiv.org/abs/hep-ph/0603071}{{\ttfamily hep-ph/0603071}}].

\bibitem{Huang:2012gs}
J.-H.~Huang and W.~Wang, \emph{{Multigluon tree amplitudes with a pair of
  massive fermions}},
  \href{https://doi.org/10.1140/epjc/s10052-012-2050-x}{\emph{Eur. Phys. J. C}
  {\bfseries 72} (2012) 2050}
  [\href{https://arxiv.org/abs/1204.0068}{{\ttfamily 1204.0068}}].

\bibitem{Arkani-Hamed:2017jhn}
N.~Arkani-Hamed, T.-C.~Huang and Y.-t.~Huang, \emph{{Scattering amplitudes for
  all masses and spins}},
  \href{https://doi.org/10.1007/JHEP11(2021)070}{\emph{JHEP} {\bfseries 11}
  (2021) 070} [\href{https://arxiv.org/abs/1709.04891}{{\ttfamily
  1709.04891}}].

\bibitem{Berends:1987me}
F.A.~Berends and W.T.~Giele, \emph{{Recursive Calculations for Processes with n
  Gluons}}, \href{https://doi.org/10.1016/0550-3213(88)90442-7}{\emph{Nucl.
  Phys. B} {\bfseries 306} (1988) 759}.

\bibitem{MCFM}
J.~Campbell, R.K.~Ellis, T.~Neumann and C.~Williams, ``{MCFM}-10.4 (in
  preparation).'' {\url{https://mcfm.fnal.gov/}}, September, 2023.

\bibitem{Bern:1995db}
Z.~Bern and A.G.~Morgan, \emph{{Massive loop amplitudes from unitarity}},
  \href{https://doi.org/10.1016/0550-3213(96)00078-8}{\emph{Nucl. Phys. B}
  {\bfseries 467} (1996) 479}
  [\href{https://arxiv.org/abs/hep-ph/9511336}{{\ttfamily hep-ph/9511336}}].

\bibitem{Budge:2020oyl}
L.~Budge, J.M.~Campbell, G.~De~Laurentis, R.K.~Ellis and S.~Seth, \emph{{The
  one-loop amplitudes for Higgs + 4 partons with full mass effects}},
  \href{https://doi.org/10.1007/JHEP05(2020)079}{\emph{JHEP} {\bfseries 05}
  (2020) 079} [\href{https://arxiv.org/abs/2002.04018}{{\ttfamily
  2002.04018}}].

\bibitem{Kleiss:1986qc}
R.~Kleiss and W.J.~Stirling, \emph{{Cross-sections for the Production of an
  Arbitrary Number of Photons in Electron - Positron Annihilation}},
  \href{https://doi.org/10.1016/0370-2693(86)90454-5}{\emph{Phys. Lett. B}
  {\bfseries 179} (1986) 159}.

\bibitem{Christensen:2018zcq}
N.~Christensen and B.~Field, \emph{{Constructive standard model}},
  \href{https://doi.org/10.1103/PhysRevD.98.016014}{\emph{Phys. Rev. D}
  {\bfseries 98} (2018) 016014}
  [\href{https://arxiv.org/abs/1802.00448}{{\ttfamily 1802.00448}}].

\bibitem{Christensen:2019mch}
N.~Christensen, B.~Field, A.~Moore and S.~Pinto, \emph{{Two-, three-, and
  four-body decays in the constructive standard model}},
  \href{https://doi.org/10.1103/PhysRevD.101.065019}{\emph{Phys. Rev. D}
  {\bfseries 101} (2020) 065019}
  [\href{https://arxiv.org/abs/1909.09164}{{\ttfamily 1909.09164}}].

\bibitem{Kleiss:1988ne}
R.~Kleiss and H.~Kuijf, \emph{{Multi - Gluon Cross-sections and Five Jet
  Production at Hadron Colliders}},
  \href{https://doi.org/10.1016/0550-3213(89)90574-9}{\emph{Nucl. Phys. B}
  {\bfseries 312} (1989) 616}.

\bibitem{DelDuca:1999rs}
V.~Del~Duca, L.J.~Dixon and F.~Maltoni, \emph{{New color decompositions for
  gauge amplitudes at tree and loop level}},
  \href{https://doi.org/10.1016/S0550-3213(99)00809-3}{\emph{Nucl. Phys. B}
  {\bfseries 571} (2000) 51}
  [\href{https://arxiv.org/abs/hep-ph/9910563}{{\ttfamily hep-ph/9910563}}].

\bibitem{Ellis:2011cr}
R.K.~Ellis, Z.~Kunszt, K.~Melnikov and G.~Zanderighi, \emph{{One-loop
  calculations in quantum field theory: from Feynman diagrams to unitarity
  cuts}}, \href{https://doi.org/10.1016/j.physrep.2012.01.008}{\emph{Phys.
  Rept.} {\bfseries 518} (2012) 141}
  [\href{https://arxiv.org/abs/1105.4319}{{\ttfamily 1105.4319}}].

\bibitem{Johansson:2015oia}
H.~Johansson and A.~Ochirov, \emph{{Color-Kinematics Duality for QCD
  Amplitudes}}, \href{https://doi.org/10.1007/JHEP01(2016)170}{\emph{JHEP}
  {\bfseries 01} (2016) 170}
  [\href{https://arxiv.org/abs/1507.00332}{{\ttfamily 1507.00332}}].

\bibitem{Melia:2013bta}
T.~Melia, \emph{{Dyck words and multiquark primitive amplitudes}},
  \href{https://doi.org/10.1103/PhysRevD.88.014020}{\emph{Phys. Rev. D}
  {\bfseries 88} (2013) 014020}
  [\href{https://arxiv.org/abs/1304.7809}{{\ttfamily 1304.7809}}].

\bibitem{Melia:2013epa}
T.~Melia, \emph{{Getting more flavor out of one-flavor QCD}},
  \href{https://doi.org/10.1103/PhysRevD.89.074012}{\emph{Phys. Rev. D}
  {\bfseries 89} (2014) 074012}
  [\href{https://arxiv.org/abs/1312.0599}{{\ttfamily 1312.0599}}].

\bibitem{Melia:2015ika}
T.~Melia, \emph{{Proof of a new colour decomposition for QCD amplitudes}},
  \href{https://doi.org/10.1007/JHEP12(2015)107}{\emph{JHEP} {\bfseries 12}
  (2015) 107} [\href{https://arxiv.org/abs/1509.03297}{{\ttfamily
  1509.03297}}].

\bibitem{Mangano:1987kp}
M.L.~Mangano and S.J.~Parke, \emph{{Quark - Gluon Amplitudes in the Dual
  Expansion}}, \href{https://doi.org/10.1016/0550-3213(88)90368-9}{\emph{Nucl.
  Phys. B} {\bfseries 299} (1988) 673}.

\bibitem{Melia:2013xok}
T.~Melia, \emph{{Dyck words and multi-quark amplitudes}},
  \href{https://doi.org/10.22323/1.197.0031}{\emph{PoS} {\bfseries RADCOR2013}
  (2013) 031}.

\bibitem{Campbell:2012uf}
J.M.~Campbell and R.K.~Ellis, \emph{{Top-Quark Processes at NLO in Production
  and Decay}}, \href{https://doi.org/10.1088/0954-3899/42/1/015005}{\emph{J.
  Phys. G} {\bfseries 42} (2015) 015005}
  [\href{https://arxiv.org/abs/1204.1513}{{\ttfamily 1204.1513}}].

\bibitem{Badger:2011yu}
S.~Badger, R.~Sattler and V.~Yundin, \emph{{One-Loop Helicity Amplitudes for
  $t\bar{t}$ Production at Hadron Colliders}},
  \href{https://doi.org/10.1103/PhysRevD.83.074020}{\emph{Phys. Rev. D}
  {\bfseries 83} (2011) 074020}
  [\href{https://arxiv.org/abs/1101.5947}{{\ttfamily 1101.5947}}].

\bibitem{Melnikov:2010iu}
K.~Melnikov and M.~Schulze, \emph{{NLO QCD corrections to top quark pair
  production in association with one hard jet at hadron colliders}},
  \href{https://doi.org/10.1016/j.nuclphysb.2010.07.003}{\emph{Nucl. Phys. B}
  {\bfseries 840} (2010) 129}
  [\href{https://arxiv.org/abs/1004.3284}{{\ttfamily 1004.3284}}].

\bibitem{Melnikov:2011qx}
K.~Melnikov, A.~Scharf and M.~Schulze, \emph{{Top quark pair production in
  association with a jet: QCD corrections and jet radiation in top quark
  decays}}, \href{https://doi.org/10.1103/PhysRevD.85.054002}{\emph{Phys. Rev.
  D} {\bfseries 85} (2012) 054002}
  [\href{https://arxiv.org/abs/1111.4991}{{\ttfamily 1111.4991}}].

\bibitem{Bevilacqua:2022ozv}
G.~Bevilacqua, M.~Lupattelli, D.~Stremmer and M.~Worek, \emph{{A study of
  additional jet activity in top quark pair production and decay at the LHC}},
  \href{https://arxiv.org/abs/2212.04722}{{\ttfamily 2212.04722}}.

\bibitem{Badger:2022mrb}
S.~Badger, M.~Becchetti, E.~Chaubey, R.~Marzucca and F.~Sarandrea,
  \emph{{One-loop QCD helicity amplitudes for pp \textrightarrow{} $
  t\overline{t}j $ to O(\ensuremath{\varepsilon}$^{2}$)}},
  \href{https://doi.org/10.1007/JHEP06(2022)066}{\emph{JHEP} {\bfseries 06}
  (2022) 066} [\href{https://arxiv.org/abs/2201.12188}{{\ttfamily
  2201.12188}}].

\bibitem{Badger:2021owl}
S.~Badger, E.~Chaubey, H.B.~Hartanto and R.~Marzucca, \emph{{Two-loop leading
  colour QCD helicity amplitudes for top quark pair production in the gluon
  fusion channel}}, \href{https://doi.org/10.1007/JHEP06(2021)163}{\emph{JHEP}
  {\bfseries 06} (2021) 163}
  [\href{https://arxiv.org/abs/2102.13450}{{\ttfamily 2102.13450}}].

\end{thebibliography}\endgroup
\bibliographystyle{JHEP}       
\end{document}